\documentclass{sage}

\usepackage[bookmarks=false]{hyperref}
\usepackage[brazil,american]{babel}
\usepackage[T1]{fontenc}
\usepackage{indentfirst}
\usepackage{natbib}
\usepackage{graphicx,url}
\usepackage[utf8]{inputenc}
\usepackage{amsmath,amssymb,amsthm, natbib}
\usepackage{footnote}
\usepackage{tablefootnote} 
\usepackage{verbatim}

\usepackage[table,xcdraw]{xcolor}
\usepackage{pgfplots} 

\usepackage{listings} 
\usepackage[algoruled,linesnumbered]{algorithm2e}
\SetKwFor{ParaTodo}{para todo}{faça}{fim para todo}
\SetKwFor{ParaCada}{para cada}{faça}{fim para cada}
\SetKwFor{Varredura}{varredura $D_I$ na ordem}{faça}{fim varredura}
\SetKwFor{VarreduraIJ}{varredura $D_I$ em I e J na ordem}{faça}{fim varredura}
\SetAlgorithmName{Algoritmo}{algoritmo}{Lista de algoritmos}
\usepackage{textcomp} 

\usepackage{pdfpages} 

\usepackage{pgfplots}
\pgfplotsset{compat=newest}
\pgfplotscreateplotcyclelist{my chart colors}{
    black,  postaction={pattern color=blue,     pattern=vertical lines}\\%
    black,  postaction={pattern color=orange,   pattern=horizontal lines}\\%
    black,  postaction={pattern color=green,    pattern=grid}\\%
    black,  postaction={pattern color=red,      pattern=north east lines}\\%
    black,  postaction={pattern color=purple,   pattern=crosshatch dots}\\%
}
\usetikzlibrary{patterns}
\usepgfplotslibrary{groupplots}

\begin{document}

\title{Efficient Execution of Irregular Wavefront Propagation Pattern on Many
Integrated Core Architecture}

\author{Jeremias Moreira Gomes, George Luiz Medeiros Teodoro}

\maketitle

\selectlanguage{american}

\begin{abstract}
The efficient execution of image processing algorithms is an active area of
Bioinformatics.  In image processing, one of the classes of algorithms or
computing pattern that works with irregular data structures is the Irregular
Wavefront Propagation Pattern (IWPP).  In this class, elements propagate
information to neighbors in the form of wave propagation. This propagation
results in irregular access to data and expansions.  Due to this irregularity,
current implementations of this class of algorithms requires atomic operations,
which is very costly and also restrains implementations with Single
Instruction, Multiple Data (SIMD) instructions in Many Integrated Core (MIC)
architectures, which are critical to attain high performance on this processor.
The objective of this study is to redesign the Irregular Wavefront Propagation
Pattern algorithm in order to enable the efficient execution on processors with
Many Integrated Core architecture using SIMD instructions. In this work, using
the Intel\textsuperscript{\textregistered} Xeon
Phi\textsuperscript{\texttrademark} coprocessor, we have implemented a vector
version of IWPP with up to $5.63\times$ gains on non-vectored version, a
parallel version using First In, First Out (FIFO) queue that attained speedup
up to $55\times$ as compared to the single core version on the coprocessor, a
version using priority queue whose performance was $1.62\times$ better than the
fastest version of GPU based implementation available in the literature, and a
cooperative version between heterogeneous processors that allow to process
images bigger than the Intel\textsuperscript{\textregistered} Xeon
Phi\textsuperscript{\texttrademark} memory and also provides a way to utilize
all the available devices in the computation.
\end{abstract}


\vspace*{1cm}

\selectlanguage{brazil}

\hyphenation{au-xi-li-ar}
\hyphenation{pa-ra-le-la}

\begin{abstract}
A execução eficiente de algoritmos de processamento de imagens é uma área ativa
da Bioinformática. Uma das classes de algoritmos em processamento de imagens ou
de padrão de computação comum nessa área é a \textit{Irregular Wavefront
Propagation Pattern} (IWPP). Nessa classe, elementos propagam informações para
seus vizinhos em forma de ondas de propagação. Esse padrão de propagação
resulta em acessos a dados e expansões irregulares. Por essa característica
irregular, implementações paralelas atuais dessa classe de algoritmos
necessitam de operações atômicas, o que acaba sendo muito custoso e também
inviabiliza a implementação por meio de instruções \textit{Single Instruction,
Multiple Data} (SIMD) na arquitetura \textit{Many Integrated Core} (MIC), que
são fundamentais para atingir alto desempenho nessa arquitetura. O objetivo
deste trabalho é reprojetar o algoritmo \textit{Irregular Wavefront Propagation
Pattern}, de forma a possibilitar sua eficiente execução em processadores com
arquitetura \textit{Many Integrated Core} que utilizem instruções SIMD.\ Neste
trabalho, utilizando o Intel\textsuperscript{\textregistered} Xeon
Phi\textsuperscript{\texttrademark}, foram implementadas uma versão vetorizada,
apresentando ganhos de até $5.63\times$ em relação à versão não-vetorizada; uma
versão paralela utilizando fila \textit{First In, First Out} (FIFO) cuja
escalabilidade demonstrou-se boa com \textit{speedups} em torno de $55\times$
em relação à um núcleo do coprocessador; uma versão utilizando fila de
prioridades cuja velocidade foi de $1.62\times$ mais veloz que a versão mais
rápida em GPU conhecida na literatura, e uma versão cooperativa entre
processadores heterogêneos que permitem processar imagens que ultrapassem a
capacidade de memória do Intel\textsuperscript{\textregistered} Xeon
Phi\textsuperscript{\texttrademark}, e também possibilita a utilização de
múltiplos dispositivos na execução do algoritmo.
\end{abstract}


  \section{Introdução}%

A Ciência da Computação tem sido amplamente utilizada nos diversos níveis de
cuidado a saúde, incluindo aqueles ligados a Patologia Clínica. A Patologia é a
área da Medicina que estuda os desvios de um organismo em relação ao que é
considerado normal~\cite{dictionary1989oxford}. Essa área tem como objetivo
analisar e diagnosticar doenças, estabelecendo estágios e verificando fatores
de risco para a saúde. Nos dias atuais essa análise advém, consideravelmente,
do processamento de imagens e de sistemas de apoio a decisão.

O processamento de imagens é uma das áreas da computação que tem concebido
numerosos trabalhos em conjunto com a
Medicina~\cite{6732495,Saltz01082013,5289193,Teodoro:2010:ROR:1851476.1851479}.
Tal fato tem ganhado destaque e, em parte, é consequência da constante evolução
dos dispositivos de captura de imagens médicas, vinculados a formação de
grandes bases de dados públicas.  Imagens capturadas por \textit{scanners}
modernos chegam a ter resoluções de até 100K x
100K~\cite{Kurc2015,teodoro2013efficient,Teodoro2014589,DBLP:journals/cluster/TeodoroHCF12}.
Em decorrência do uso dessas tecnologias, há um significativo crescimento no
volume dessas bases de dados, e a utilização das mesmas cria uma grande demanda
por tarefas de armazenamento, de análise, de visualização e de compressão
desses dados. Uma maneira de atender aos requisitos computacionais necessários
para se extrair informação desses dados é a utilização de processamento
paralelo, principalmente pelo uso de aceleradores.

Aceleradores, em especial, vem ganhando atenção da comunidade devido seu alto
poder de processamento e bom custo benefício. Um dos processadores dessa classe
de arquiteturas promissoras é o Intel\textsuperscript{\textregistered} Xeon
Phi\textsuperscript{\texttrademark}~\cite{rahman2013xeon}. O
Intel\textsuperscript{\textregistered} Xeon Phi\textsuperscript{\texttrademark}
é baseado na microarquitetura Larrabee~\cite{seiler2008larrabee} e é o primeiro
produto a utilizar a arquitetura Intel\textsuperscript{\textregistered}
\textit{Many Integrated Core} (MIC). Essa arquitetura suporta vários núcleos em
um único processador altamente escalável~\cite{intelmanycore}, com
características semelhantes às unidades de processamento gráfico e
compatibilidade com a arquitetura x86~\cite{jeffers2013intel}.

Por outro lado, as aplicações de análise de imagens médicas são construídas
utilizando um conjunto de operações
básicas~\cite{10.1371/journal.pone.0081049,DBLP:conf/isbi/KongWTLZTB15,DBLP:conf/miccai/LiangWTMTZK15},
que são combinadas de formas diferentes conforme a análise desejada. Grande
parte dessas operações consistem de transformações morfológicas dos dados, tais
como Reconstrução Morfológica~\cite{vincent1993morphological}, Transformada de
Distância~\cite{vincent1991exact}, Transformação
Watershed~\cite{vincent1991watersheds}, entre outros. Assim, é importante
implementar essas operações de forma otimizada nesses novos processadores
(aceleradores) buscando beneficiar aplicações que utilizam as mesmas operações. 

Muitos algoritmos morfológicos populares podem ser implementados eficientemente
por meio de um padrão de computação, chamado Propagação de Ondas Irregulares -\
do inglês \textit{Irregular Wavefront Propagation Pattern}
(IWPP)~\cite{teodoro2013efficient}. IWPP é um padrão de computação no qual um
conjunto de elementos formam ondas iniciais que irão se expandir de forma
irregular a partir das suas frentes de onda. Essas ondas são dinâmicas, possuem
dependência de dados e são computadas ao longo de suas expansões. Cada elemento
da frente de onda, dada uma condição de propagação, pode propagar informação ao
seu conjunto de elementos vizinhos. 

Como apenas os elementos na frente de onda contribuem para o resultado das
operações, o IWPP pode ser implementado eficientemente utilizando alguma
estrutura de dados auxiliar, que identifique esses elementos, tal como uma
fila.

Além dos algoritmos morfológicos previamente citados, diversos outros métodos
em processamento de imagens tem seu funcionamento baseado no IWPP, como
Esqueletos Euclidianos~\cite{meyer1990digital}, Esqueletos por Zona de
Influência~\cite{lantuejoul1980skeletonization}, ou a Transformação
Watershed~\cite{vincent1993morphological}.

Devido ao seu extenso uso aplicado a grandes volumes de dados e ao seu
variado emprego, a construção de uma implementação eficiente para
algoritmos IWPP pode beneficiar diversas ferramentas ou aplicações que são
amplamente utilizadas em Bioinformática. Uma das formas de se alcançar a
eficiência desses algoritmos é por meio do paralelismo, usando arquiteturas
especializadas em \textit{Central Processing Unit} (CPU) ou coprocessadores
como \textit{Graphic Processing Unit} (GPU)~\cite{teodoro2012fast} e
\textit{Many Integrated Core} (MIC)~\cite{jeffers2013intel}. 

Uma particularidade sobre essa classe de algoritmos é que implementações atuais
exploram o uso de paralelismo com apoio em operações atômicas, para garantir o
resultado final do processamento. Essas operações atômicas geram um custo
adicional para as aplicações, e não permitem uma execução eficiente em
arquiteturas cuja caraterística principal se baseia no uso do conjunto de
instruções \textit{Single Instruciton, Multipe Data} (SIMD). Dessa maneira,
este trabalho busca reprojetar o algoritmo IWPP de forma a possibilitar sua
eficiente execução na arquitetura \textit{Many Integrated Core} utilizando
instruções \textit{Single Instruction, Multiple Data}.

\subsection{Problema}
Algoritmos e implementações atuais do IWPP executam ineficientemente no
Intel\textsuperscript{\textregistered} Xeon
Phi\textsuperscript{\texttrademark}~\cite{teodoro2015performance}, pois
requerem instruções atômicas que não são suportadas nesse coprocessador ao
utilizar o conjunto de instruções \textit{Single Instruction, Multiple Data}
(SIMD).\ O uso desse conjunto de instruções é fundamental para o desempenho do
Intel Phi, onde o tamanho do registrador é uma das características fundamentais
dessa arquitetura.

\subsection{Objetivo} 
O objetivo deste trabalho é reprojetar o algoritmo IWPP de forma a possibilitar
sua eficiente execução na arquitetura \textit{Many Integrated Core} utilizando
instruções \textit{Single Instruction, Multiple Data} (SIMD). 

\subsection{Objetivos Específicos}
\begin{enumerate}
    \item Reprojetar o algoritmo IWPP;\
    \item Implementar uma versão vetorial eficiente do algoritmo reprojetado;
    \item Paralelizar em nível de \textit{threads} o algoritmo vetorizado;
	\item Implementar casos de uso do algoritmo;
    \item Implementar uma versão cooperativa para ambientes com processadores

\end{enumerate}

Os primeiros 4 itens dos objetivos já foram publicados~\cite{gomes2015efficient}.

\subsection{Organização do Texto}
O restante deste trabalho segue a seguinte organização:
\begin{description} 
    \item[Seção 2. \text{[Referencial Teórico]}] 
        Faz uma revisão teórica dos conceitos, métodos e técnicas apresentadas
        em trabalhos anteriores envolvendo Arquiteturas Paralelas e Algoritmos
        Morfológicos.  
    \item[Seção 3. \text{[\textit{Irregular Wavefront Propagation Pattern} (IWPP)]}] 
        Detalha algoritmos da classe IWPP e discute trabalhos relacionados.
    \item[Seção 4. \text{[IWPP Paralelo Eficiente Utilizando MIC]}]
        Apresenta uma implementação eficiente para o algoritmo IWPP detalhando as
        estratégias utilizadas.  
    \item[Seção 5. \text{[Análise de Resultados]}]
        Descreve os resultados a partir do emprego das estratégias
        apresentadas na seção anterior.
    \item[Seção 6. \text{[Conclusão]}] 
        Apresenta as conclusões do trabalho por meio de uma discussão sobre
        objetivos alcançados, e indica sugestões de continuação do trabalho.
\end{description}

  \section{Referencial Teórico}\label{sec:referencial}

Nesta seção serão abordados Algoritmos Morfológicos apresentando notações,
exemplos e estratégias para implementação desses algoritmos. Também serão
detalhados alguns conceitos básicos sobre arquiteturas paralelas, processadores
vetoriais, uma exposição do coprocessador
Intel\textsuperscript{\textregistered} Xeon Phi\textsuperscript{\texttrademark}
e definições e classificações envolvendo \textit{data races} e condições de
corrida. 

\subsection{Algoritmos Morfológicos}\label{subsec:algoritmosmorf}

A Morfologia Matemática é uma abordagem não linear para análise espacial de
estruturas. A base da morfologia consiste em extrair de uma imagem desconhecida
a sua geometria através de transformações em uma outra imagem completamente
definida, ou seja, extrair informações associadas à geometria dos objetos e a
topologia de um conjunto desconhecido pela transformação através de outro
conjunto bem definido, chamado elemento
estruturante~\cite{wangenheim2001seminario}. Inicialmente, ela foi concebida
para manipular imagens binárias, sendo estendida, mais tarde, para imagens em
tons de cinza utilizando teoria de
reticulados~\cite{serra1983image}~\cite{sternberg1986grayscale}.

Transformações morfológicas são funções complexas que compartilham
características simples e intuitivas. Utilizar uma aplicação de análise de
imagens envolve, na maioria das vezes, a concatenação de diversas
transformações de baixo nível. Por esta razão, a implementação de cada uma
dessas transformações deve ser a mais otimizada possível.

Os algoritmos oriundos dessas transformações podem ser classificados em quatro
famílias, que serão detalhadas na
Seção~\nameref{subsubsec:estrategiasalgoritmosparalelos}: Algoritmos Paralelos,
Algoritmos Sequenciais, Algoritmos Baseados em Filas de Pixels e Algoritmos
Híbridos~\cite{vincent1992morphological}.

\subsubsection{Notações em Algoritmos Morfológicos}\label{subsubsec:notacoes}
Consideram-se imagens binárias ou em tons de cinza \textit{I} o mapeamento de
um domínio retangular $D_I \subset \mathbb{Z}^2$ em
$\mathbb{Z}$~\cite{vincent1992morphological}. Em uma imagem, cada pixel $p$
pode assumir o valor $0$ ou $1$ para imagens binárias, e valores de um conjunto
$\{0,\dots,L-1\}$ para imagens em tons de cinza, onde cada valor de $L$ denota
a tonalidade de cinza. Diversos algoritmos podem estender suas funcionalidades
para espaços $n$-dimensionais, mas para efeitos deste trabalho, serão
descritas, unicamente, imagens bidimensionais.

Uma grade subjacente $G \subset \mathbb{Z} \times \mathbb{Z}$ define uma 
relação de vizinhança entre pixels. $G$ diz respeito a conectividade, e define 
o número de vizinhos acessíveis a um determinado pixel. A conectividade pode 
ser dada por uma grade quadrada ou hexagonal. $N_G(p)$ representa o conjunto de 
vizinhos $q$ de um pixel $p \in \mathbb{Z}^2$, conforme uma grade $G$, de
acordo com o apresentado na Equação~\ref{eq:vizinhanca}.
\begin{equation}
	N_G(p)=\{q \in \mathbb{Z}^2 \mid (p,q) \in G\}
    \label{eq:vizinhanca}
\end{equation}

\begin{figure}[!htb]
\hspace*{8mm}
\begin{minipage}{.4\textwidth}
  \centering
  \includegraphics[width=.7\linewidth]{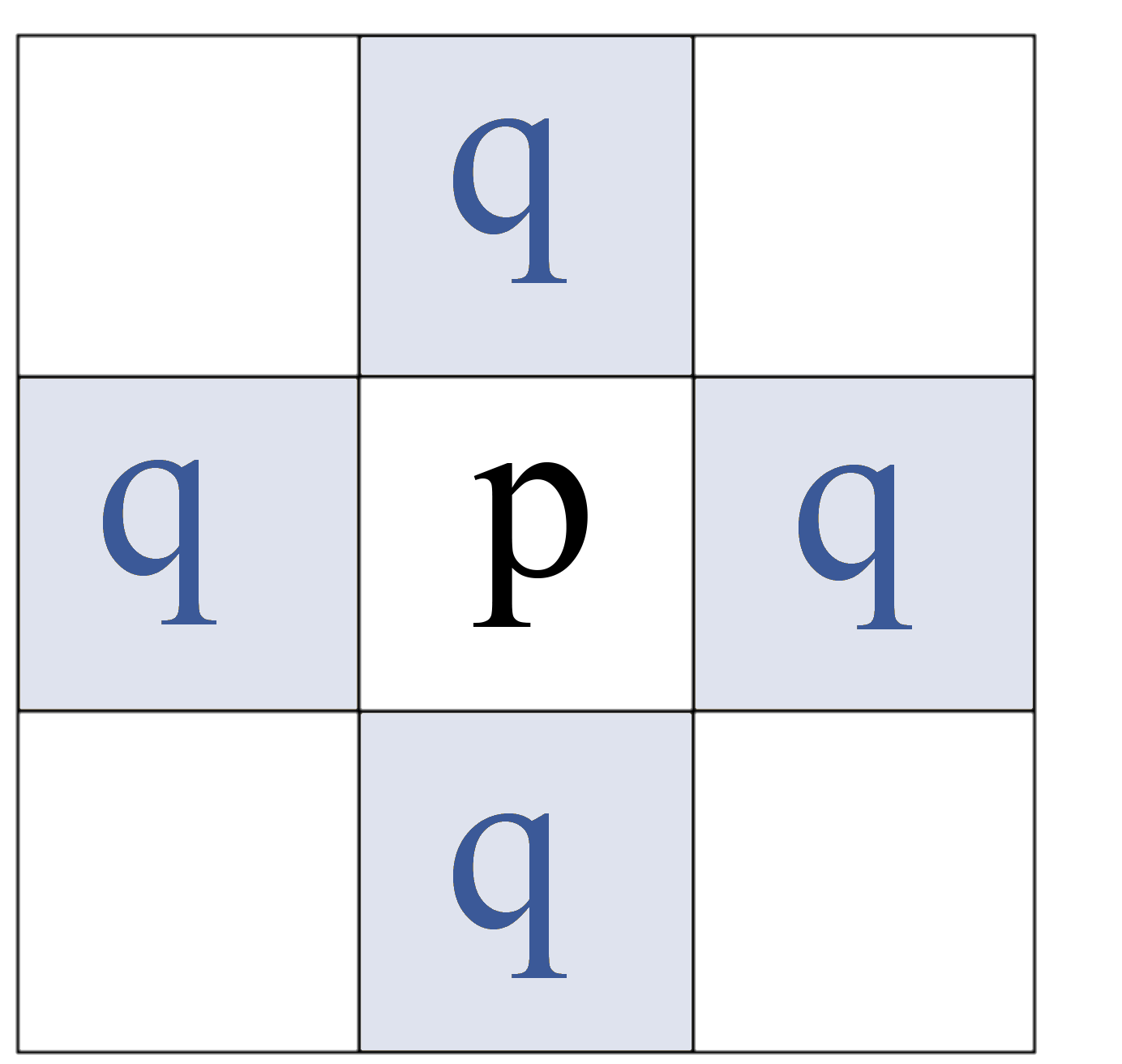}
  \caption{Representação de um Conjunto de Vizinhos 4-Conectado.}
  \label{fig:4conectado}
\end{minipage}
\hspace*{1cm}
\begin{minipage}{.4\textwidth}
  \centering
  \includegraphics[width=.7\linewidth]{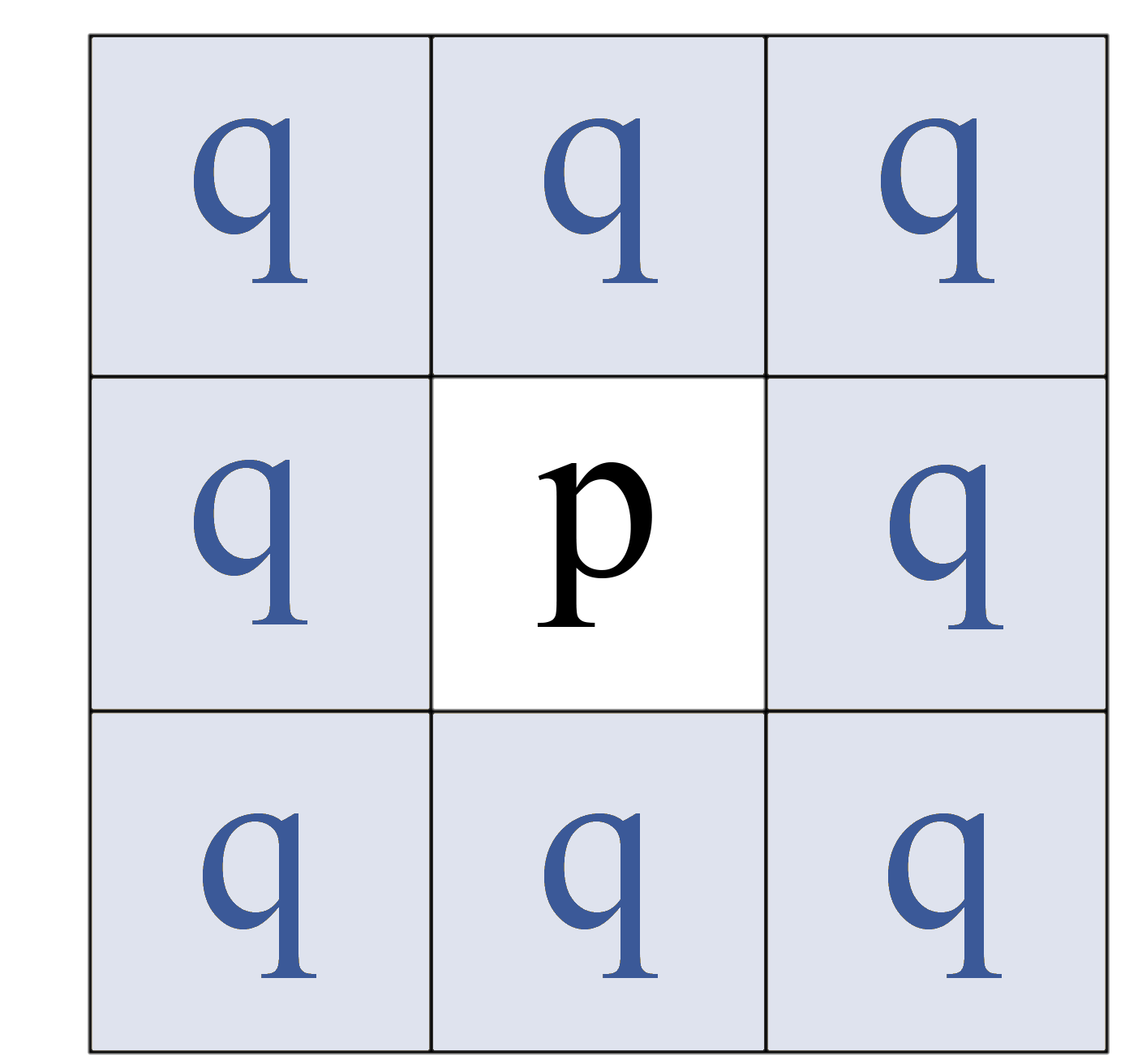}
  \caption{Representação de um Conjunto de Vizinhos 8-Conectado.}
  \label{fig:8conectado}
\end{minipage}
\end{figure}

Um pixel $p$ é vizinho de um pixel $q$ se $(p,q) \in G$. Se $G$ 
é 4-conectado, então os pixels acessíveis a $p$ são os quatro pixels adjacentes 
por borda (Figura~\ref{fig:4conectado}), e se $G$ é 8-conectado, $p$ acessa 
pixels que são adjacentes por borda e por vértice (Figura~\ref{fig:8conectado}). 

\subsubsection{Exemplos de Algoritmos Morfológicos}\label{subsubsec:exemplos}
Esta seção apresenta alguns exemplos de algoritmos morfológicos conhecidos da
literatura.

\textbf{\textit{Watershed}}

\begin{figure}[!htb]
\label{fig:watershed01}
\centering
\includegraphics[width=.6\linewidth]{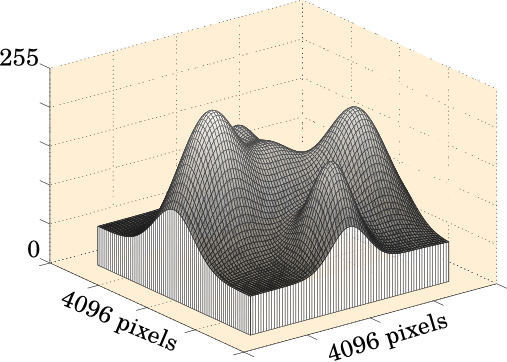}
\caption{Representação Tridimensional de uma Imagem em Tons de Cinza.}
\end{figure}

\textit{Watershed} é um método para segmentação de uma imagem em tons de cinza
em regiões nas quais, de um ponto de vista tridimensional, cada pixel
corresponde a uma posição e os níveis de cinza determinam a altitude
(Figura~\ref{fig:watershed01}).  

\begin{figure}[!htb]
\begin{minipage}{.5\textwidth}
  \centering
  \includegraphics[width=\linewidth]{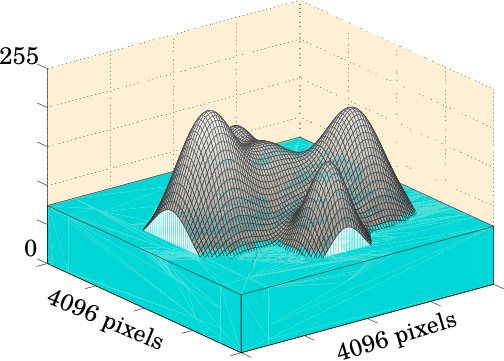}
  \caption{Início da Inundação.}
  \label{fig:watershed02}
\end{minipage}
\begin{minipage}{.5\textwidth}
  \centering
  \includegraphics[width=\linewidth]{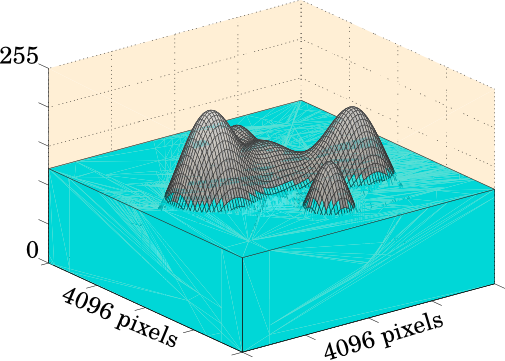}
  \caption{Avanço da Inundação.}
  \label{fig:watershed03}
\end{minipage}
\end{figure}

\begin{figure}[!htb]
\label{fig:watershed04}
\centering
\includegraphics[width=.6\linewidth]{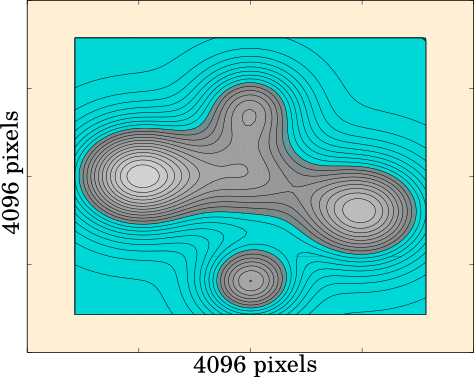}
\caption{Vista Superior da Segmentação.}
\end{figure}

Esse algoritmo utiliza dois passos: ordenação dos pixels e inundação do relevo.
A ordenação determina a distribuição cumulativa por níveis de cinza e guarda os
endereços únicos de acesso de cada um dos pixels. A inundação é o
relacionamento da estrutura topográfica formada, com a inserção progressiva de
água em suas regiões mais baixas. Essas inserções progressivas provocam o
surgimento de bacias de captação. Com o avanço da inundação e uma vez que
bacias de captação estão para se misturar, são formadas regiões de barragem que
consistirão na imagem segmentada da Figura~\ref{fig:watershed04} em regiões com
formatos descontínuos~\cite{gonzalez2002digital}. 

Para a inundação do relevo utiliza-se uma fila FIFO e busca em largura onde,
para cada nível, a inundação continua em seus novos mínimos locais ou nas
bacias de captação mais baixas. Todos os novos mínimos do nível analisados são,
então, descobertos e tratados separadamente.

\textbf{Esqueleto Morfológico}

\begin{figure}[!htb]
\begin{minipage}{.5\textwidth}
  \centering
  \includegraphics[width=.75\linewidth]{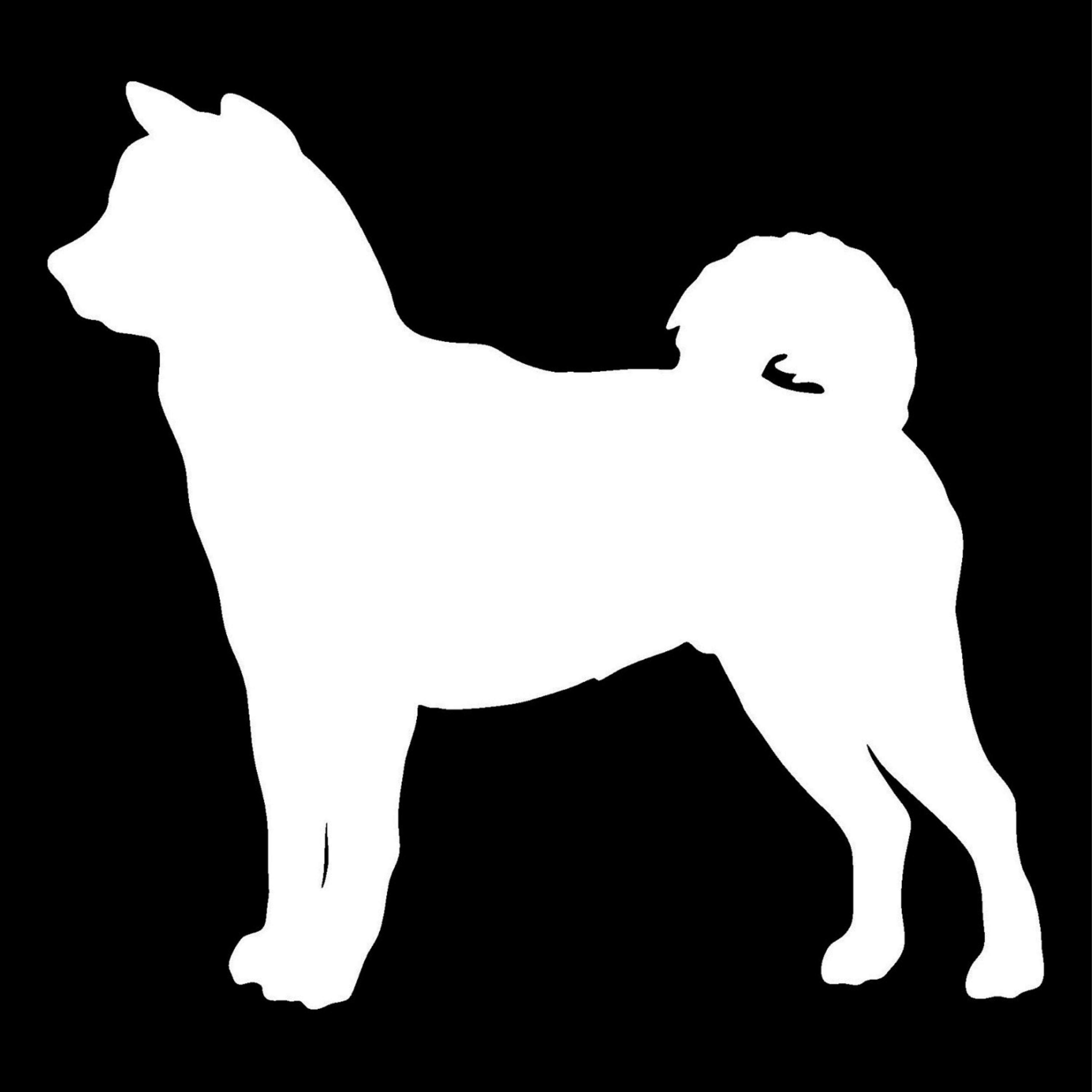}
  \caption{Imagem Original.}
  \label{fig:shibainu}
\end{minipage}
\begin{minipage}{.5\textwidth}
  \centering
  \includegraphics[width=.75\linewidth]{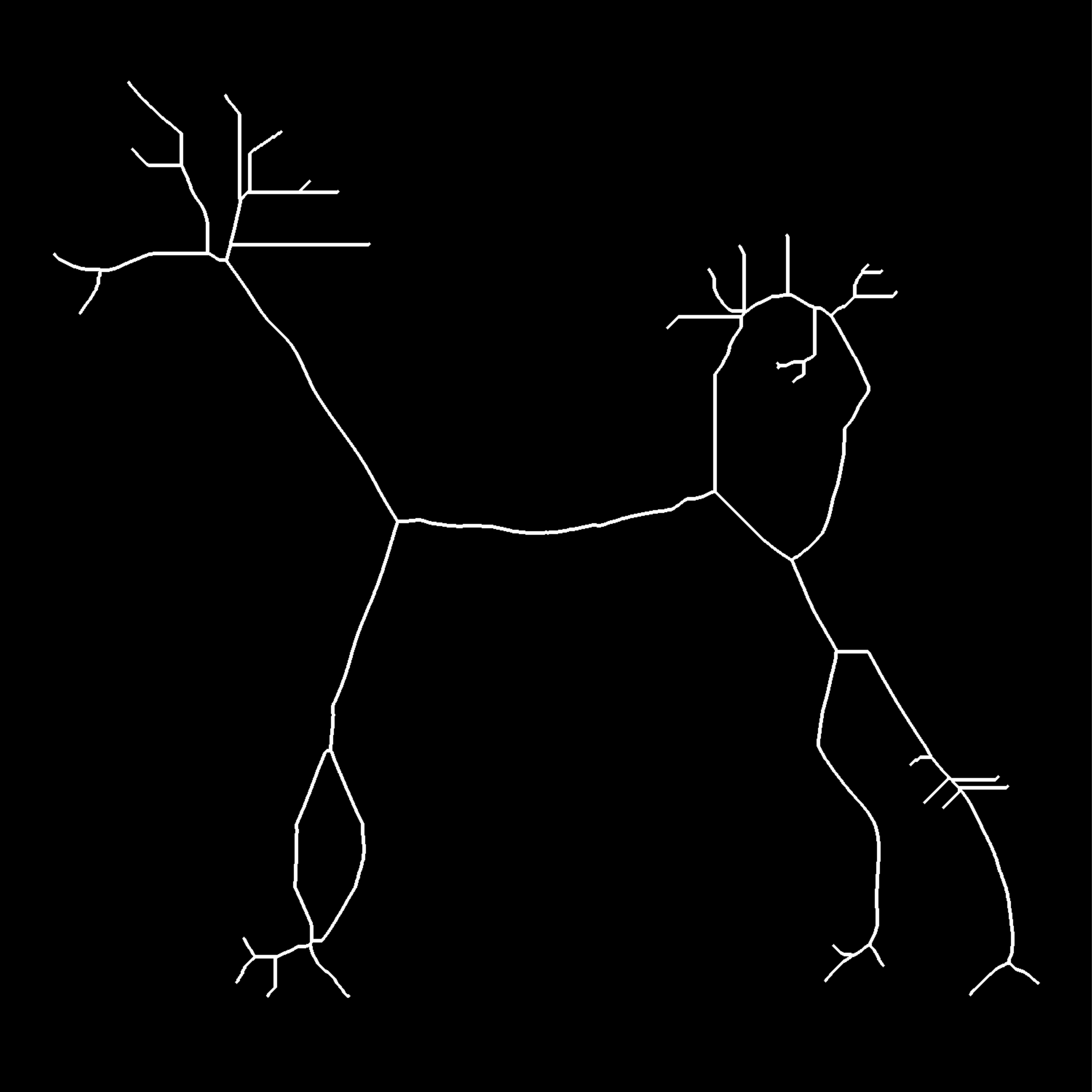}
  \caption{Esqueleto Morfológico.}
  \label{fig:shibainuskeleton}
\end{minipage}
\end{figure}

Na extração das características de um objeto, o esqueleto consiste no
afinamento do referido, com intenção de formar um conjunto de retas ou linhas
que sintetizem a informação do objeto original preservando a sua
homotopia\footnote{Funções contínuas que transformam um espaço topológico em
outro se uma puder ser ``continuamente deformada'' na outra.}
(Figuras~\ref{fig:shibainu} e~\ref{fig:shibainuskeleton}). Assim, em um
conjunto $X \subset \mathbb{Z}^2$, o esqueleto $S(X)$ é o conjunto de pixels
nos quais diferentes afinamentos, chamados de \textit{wavefronts}, se
encontram~\cite{blum1967transformation}. Pela necessidade da propriedade de
homotopia e para sua utilidade prática, esse processo precisa preservar o
número de componentes conectados e o número de buracos no conjunto original,
assim como as relações entre esses componentes e os buracos.

\begin{figure}
  \centering
  \includegraphics[width=.4\linewidth]{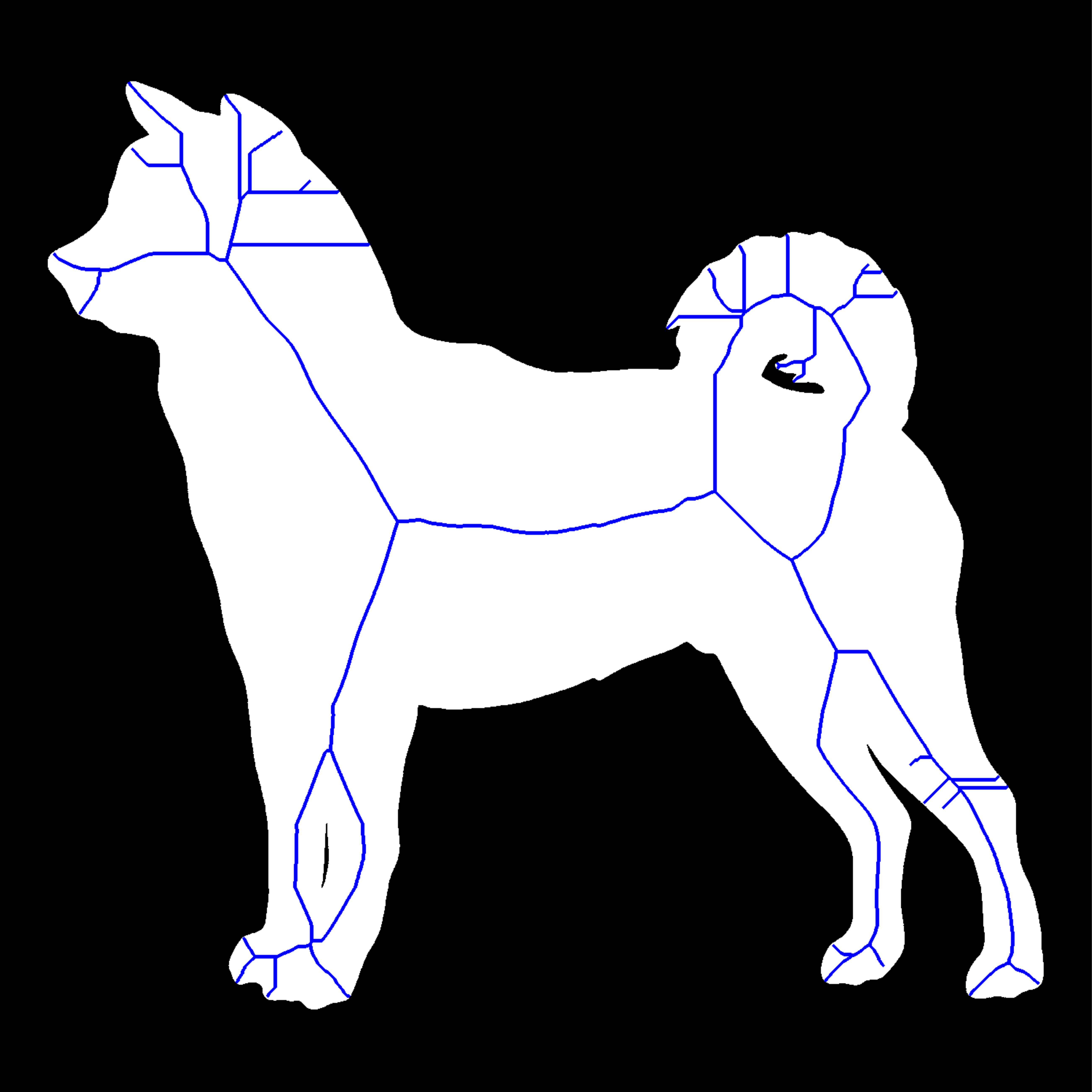}
  \caption{Visualização do Resultado.}
  \label{fig:shibainuresultado}
\end{figure}

Em uma definição mais formal~\cite{vincent1991efficient}, o esqueleto $S(X)$ de
um conjunto $X \subset \mathbb{Z}^2$ é o conjunto de centros de \textit{maximal
balls} $B$ dentro da forma dada pela Equação~\ref{eq:balls}:

\begin{equation}
    S(X)=\{p \in X,\exists r \geq 0, B(p,r) \text{ \textit{maximal ball} de } X \}
\label{eq:balls}
\end{equation}

A primeira proposta de solução para este método apresentada na
literatura~\cite{blum1967transformation} realizava sucessivas remoções até que
a estabilidade fosse alcançada. Propostas posteriores utilizaram uma abordagem
relacionada a uma sequência definida de pixels~\cite{rosenfeld1966sequential}.
Essas sequências são chamadas de varredura. Utilizando um número fixo de
varreduras, chegou-se a uma abordagem precisa, mais eficiente e que poderia ser
estendida a outros tipos de algoritmos. Contudo, esses algoritmos requeriam
custosa análise de vizinhança e possuíam baixa flexibilidade. Dessa forma,
outros métodos surgiram na
literatura~\cite{sternberg1986grayscale}~\cite{tzionas1995parallel}, porém os
mesmos apresentavam alto grau de complexidade e especificidade.

Baseada nas obervações anteriores, o algoritmo de esqueleto morfológico
apresentado em 1991~\cite{vincent1991efficient} utilizava remoções homotópicas,
\textit{crest points} e contornos. Esse algoritmo inicia-se pelas fronteiras de
X, onde são realizadas remoções até que a estabilidade seja alcançada. Tais
remoções são implementadas utilizando-se filas de pixels. A cada passo, o pixel
\textit{p} pode ser removido se, e somente se, uma das seguintes condições
forem atendidas: $p$ não pertence ao esqueleto de \textit{maximal balls}, ou
$p$ não modifica a localidade homotópica. A primeira condição garante a
precisão do algoritmo, e a segunda a propriedade de homotopia. Assim como
outros algoritmos que utilizam filas FIFO, é um método eficiente uma vez que
utiliza somente aqueles pixels ativos do passo de propagação. 

\textbf{Reconstrução Morfológica}

\begin{figure}[!htb]
	\centering
	\includegraphics[width=\columnwidth]{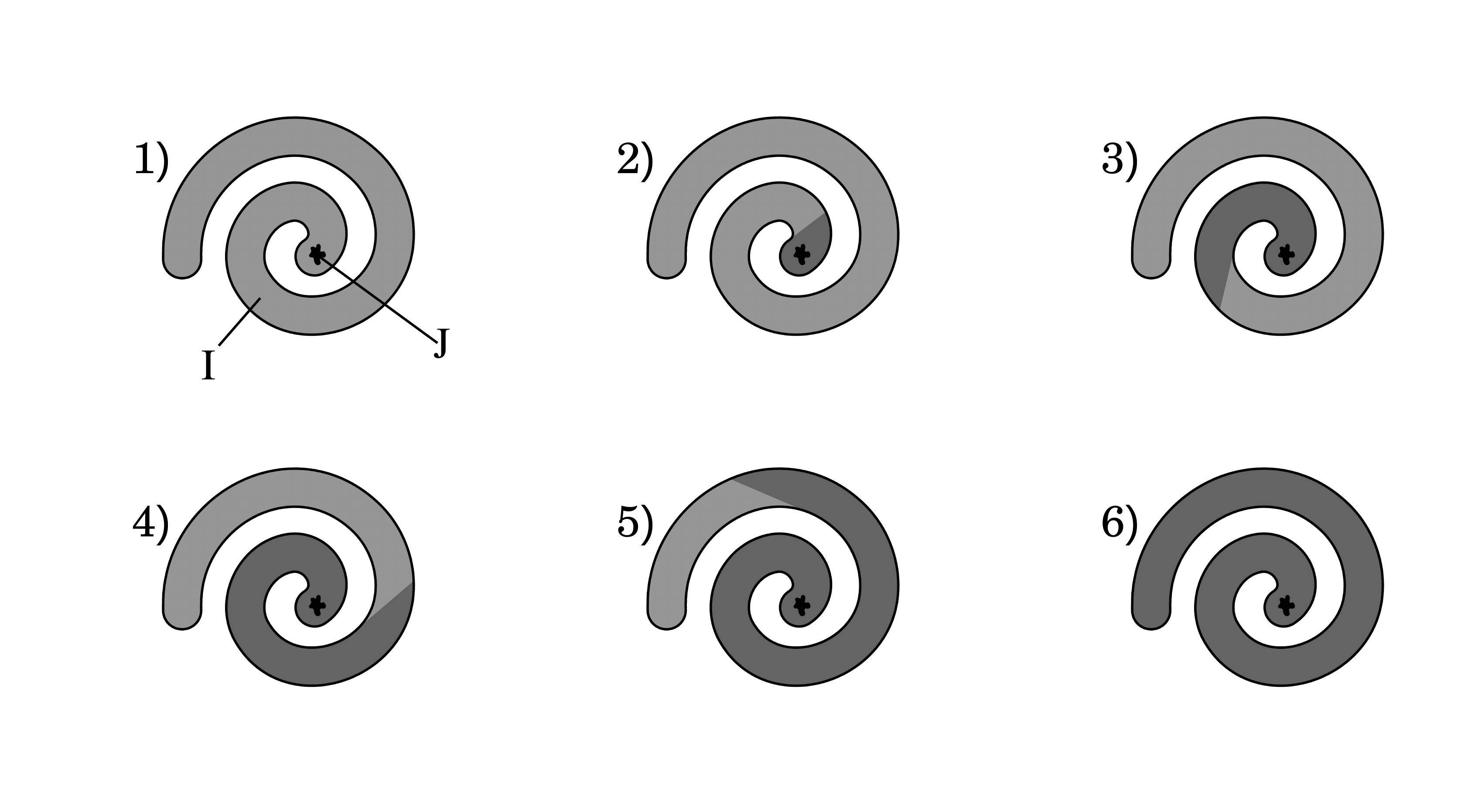}
    \caption{Reconstrução Morfológica. Adaptado
    de~\cite{vincent1992morphological}.}
	\label{fig:reconbinaria}
\end{figure}

Na reconstrução, dadas duas imagens (binárias ou em tons de cinza) $I$ e $J$
onde $J \leq I$ (ou seja, para cada pixel $p$, no domínio das imagens $J(p)
\leq I(p)$), a reconstrução $R_I(J)$ de $I$ em $J$ é obtida por sucessivas
dilatações em $J$ utilizando $I$ como limite, até que a estabilidade seja
alcançada (Figura~\ref{fig:reconbinaria}). $I$ é chamada de imagem máscara e
$J$ é chamada imagem marcadora. As conectividades elementares utilizadas na
reconstrução são a hexagonal (conectividade-6), quadrada $S_1$ (conectividade-4
com 5 pixels) ou quadrada $S_2$ (conectividade-8 com 9 pixels). Sejam
$\delta_G$ as dilatações com respeito a $G$, e $\wedge$ a operação de
comparação mínima pixel-a-pixel. 

A reconstrução de $I$ em $J$ é obtida aplicando, até que a estabilidade seja 
alcançada, a operação apresentada pela Equação~\ref{eq:recon}:

\begin{equation}
	J \longleftarrow \delta_G(J) \wedge I.
    \label{eq:recon}
\end{equation}

Observando a Equação~\ref{eq:recon} é possível obter o
Algoritmo~\ref{alg:recon} para a reconstrução. Esse algoritmo realiza
sucessivas propagações verificando a vizinhança de cada pixel $p$ (linhas 3 e
4), e esse procedimento continua até que a imagem reconstruída $J$ não seja
alterada por toda uma iteração do laço (linhas 1 a 6). Esse algoritmo é simples
em sua construção, porém é ineficiente devido a irregularidade de suas
atualizações.

\begin{algorithm}
	\label{alg:recon}
	\footnotesize
	\Entrada{I, imagem binária ou em tons de cinza}
	\Entrada{J, imagem binária ou em tons de cinza}
    \tcp{$\forall p \in D_I, J \leq I$}
	\Saida{J, imagem binária ou em tons de cinza reconstruída}
	\Repita{alcançar a estabilidade}{
        Copia $J$ para $J_{temp}$\\
        \ParaCada{pixel $p \in D_{J_{temp}}$}{
            $J(p) \leftarrow max\{J_{temp}(p), N_G(p)\} \wedge I(p)$
		}
	}
	\caption{Reconstrução Morfológica}
\end{algorithm}

Recorrendo a um algoritmo que realize alterações na própria imagem, é possível
realizar as atualizações por meio de varreduras \textit{raster} e
\textit{anti-raster}~\cite{vincent1992morphological}, porém tais varreduras
precisam ser repetidas diversas vezes até que a estabilidade seja alcançada. 

A reconstrução é uma ferramenta morfológica muito expressiva e possui diversas 
aplicações, principalmente, no caso de imagens em tons de cinza. Como exemplo, 
pode-se citar a extração das áreas de maior interesse (picos) que determinam 
objetos relevantes na imagem.

\textbf{Transformada de Distância}

\begin{figure}[!htb]
	\centering
	\includegraphics[width=\columnwidth]{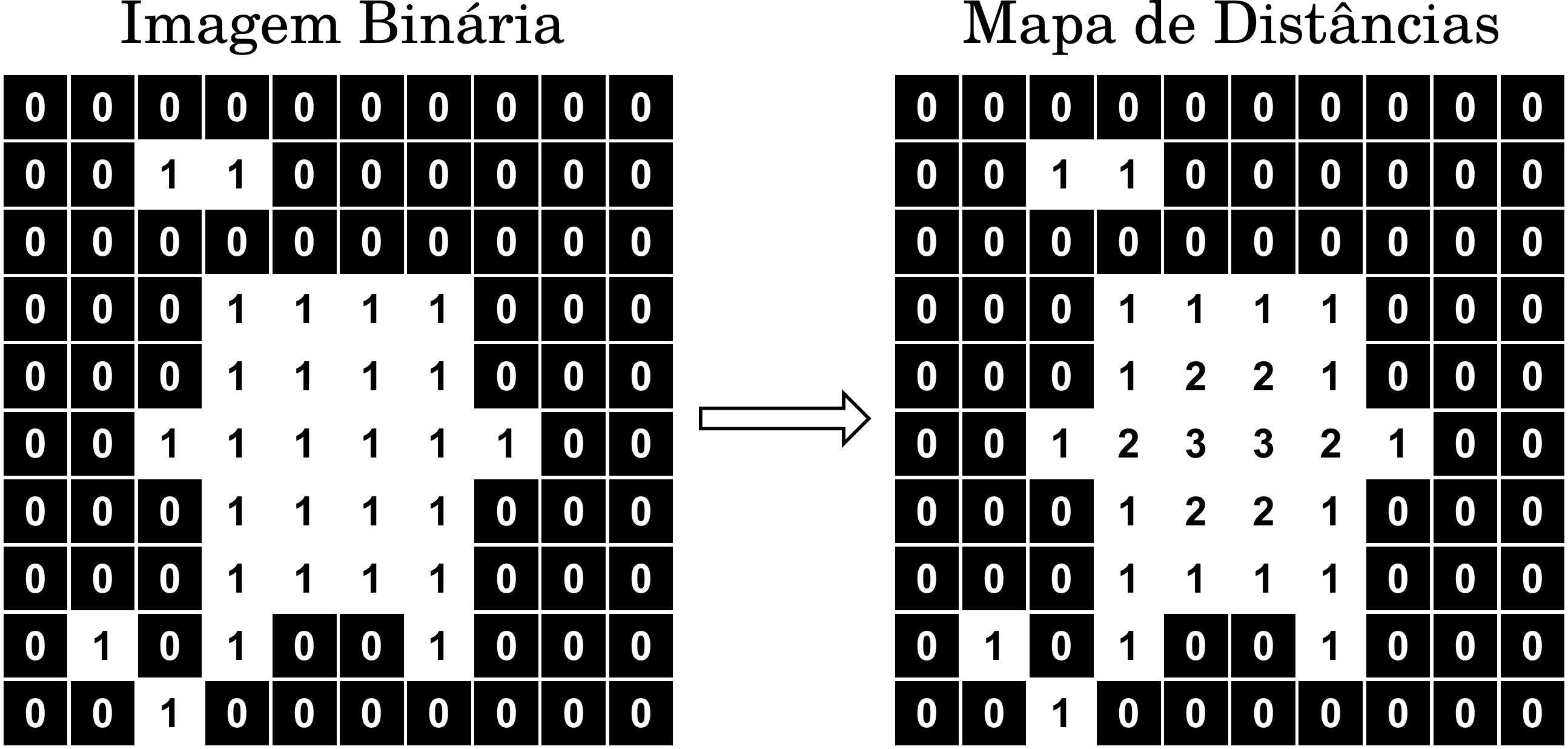}
    \caption{Exemplo Numérico da Transformada de Distância \textit{City-block}.}
	\label{fig:tde}
\end{figure}

A operação Transformada de Distância (DT) calcula o mapa de distâncias $M$ de
uma imagem binária de entrada $I$, onde para cada pixel $p \in I$, que faz
parte do objeto (\textit{foreground}), seu valor em $M$ é a menor distância a
partir de $p$ até o pixel de fundo (\textit{background}) mais próximo da
imagem. A Figura~\ref{fig:tde} apresenta um exemplo de execução da Transformada
de Distância. Inicialmente é dada uma imagem binária como entrada para o
algoritmo no qual o fundo da imagem possui valor 0 (zero) e os objetos, cuja
distância será calculada, possuem valor inicial de 1 (um). O algoritmo irá
identificar para cada elemento ou pixel do objeto qual é o
elemento do fundo mais próximo a ele, seguindo alguma métrica. Após a
identificação, o valor da distância é calculado e inserido na posição do
elemento do objeto. Após o cálculo da distância de todos elementos dos objetos,
a imagem constituirá um mapa com a distância de cada um dos objetos da imagem
binária.

A Transformada de Distância possui diversas aplicações, dentre elas podem ser
citadas a base para separação de objetos sobrepostos por meio da Transformação
\textit{Watershed}, a navegação robótica para escolha do menor caminho, a
computação de representações geométricas como diagramas de Voronói e
triangulações de Delaunay, análise comparativa da forma de objetos, e análise
de interação entre estruturas biológicas~\cite{fabbri20082d}.

Implementações para a Transformada de Distância podem ser classificadas de
diversas formas: pela complexidade, pela eficiência, pela ordem de varredura ou
pela métrica de distância utilizada. Quanto a métrica, a Transformada de
Distância pode ser chanfrada como nas \textit{city-block}, \textit{chess-board}
ou \textit{octogonal}~\cite{breu1995linear}; euclidiana ou euclidiana
aproximada~\cite{danielsson1980euclidean} devido as dificuldades de construções
eficientes para a sua forma exata.

\subsubsection{Estratégias para Implementação de Algoritmos 
Morfológicos}\label{subsubsec:estrategiasalgoritmosparalelos}
Esta seção apresenta estratégias de implementação para algoritmos morfológicos
que podem ser classificadas em quatro grandes grupos: Algoritmos Paralelos,
Algoritmos Sequenciais, Algoritmos Baseados em Filas de Pixels e Algoritmos
Híbridos.

\textbf{Algoritmos Paralelos}

São os algoritmos clássicos e mais comuns no campo da morfologia. Eles
funcionam da seguinte forma: dada uma imagem $I$ como entrada, todos os pixels
são escaneados e o novo valor de cada pixel $p$ é determinado a partir da
vizinhança $N(p)$. O êxito deste algoritmo reside na restrição de que a
atualização dos pixels deverá ser realizada em uma imagem $J$ diferente de $I$.
$J$ é então copiada para $I$ e é realizada mais uma rodada de escaneamento dos
pixels em $I$. Tais passos são executados até que nenhuma modificação em $J$
seja feita a partir dos escaneamentos de $I$. Os passos descritos podem ser
observados na Transformada de Distância Paralelo, demonstrados no
Algoritmo~\ref{alg:paralelo}.

\begin{algorithm}
	\label{alg:paralelo}
	\footnotesize
	\Entrada{I, imagem binária}
	\Saida{J, imagem em tons de cinza $D_I; J \neq I$}
	\Repita{alcançar a estabilidade} {
		\ParaCada {pixel $p \in D_I$}{
			\Se{$I(p) = 1$}{
				$J(p) \leftarrow min\{I(q),q\in N_G(p)\}+1$
			}
		}
		Copia $J$ para $I$
	}
	\caption{Transformada de Distância Paralelo}
\end{algorithm}

Como $I$ é diferente de $J$, vários pixels $p$ de uma rodada de escaneamento
podem ser processados em paralelo. O número de escaneamentos deste algoritmo é
proporcional à maior distância calculada. Devido a essa característica,
algoritmos paralelos requerem grande número de varreduras na imagem completa, o
que o torna inadequado para computadores convencionais, mesmo em algoritmos
paralelos mais básicos.

\textbf{Algoritmos Sequenciais}

\begin{figure}[b]
\begin{minipage}{.5\textwidth}
  \centering
  \includegraphics[width=.6\linewidth]{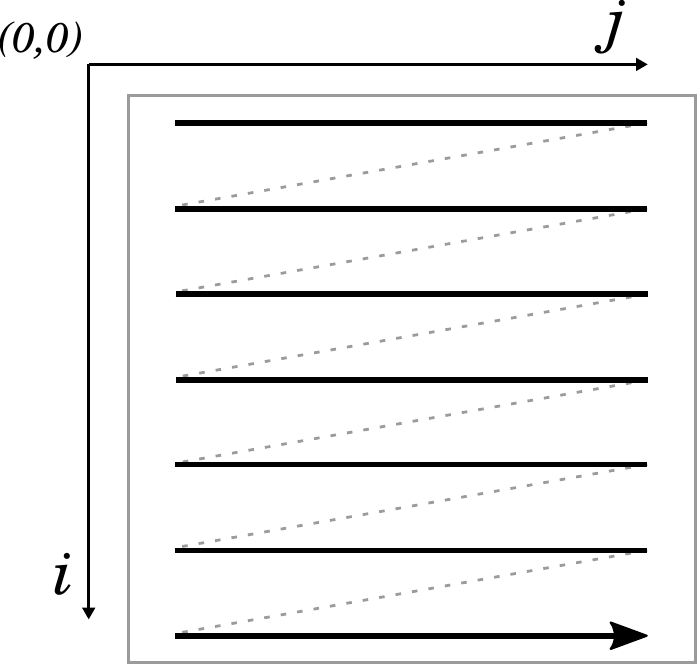}
  \caption{Varredura \textit{Raster}.}
  \label{fig:rastersimples}
\end{minipage}
\begin{minipage}{.5\textwidth}
  \centering
  \includegraphics[width=.6\linewidth]{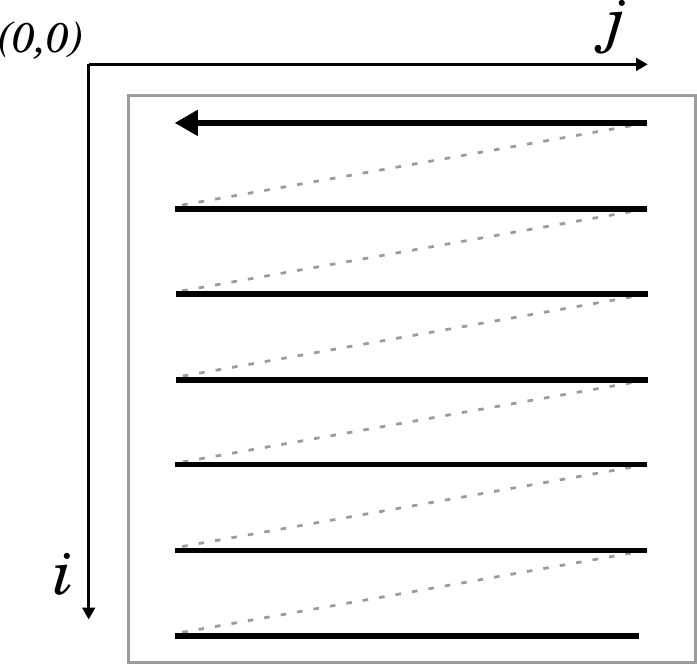}
  \caption{Varredura \textit{Anti-raster}.}
  \label{fig:antirastersimples}
\end{minipage}
\end{figure}

São algoritmos que foram desenvolvidos como forma de evitar o alto número de
escaneamentos. Essa classe de algoritmos remete aos seguintes preceitos: os
pixels devem ser escaneados em uma ordem pré-definida (chamado de varredura); e
o valor atualizado do pixel, determinado a partir da computação com seus
valores vizinhos, é escrito na própria imagem $I$
analisada~\cite{vincent1992morphological}. De maneira oposta aos algoritmos
paralelos, a ordem de execução dos algoritmos sequencias é essencial ao
resultado.

\begin{figure}[!htb]
	\centering
	\includegraphics[width=.25\columnwidth]{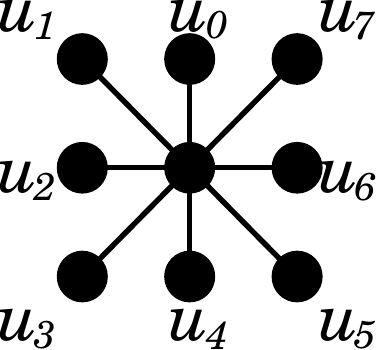}
	\caption{Pixels Utilizados nas Varreduras.}
	\label{fig:sequencialpixel}
\end{figure}

Para calcular a Transformada de Distância de uma imagem $I$, os escaneamentos
são suficientemente realizados na forma de duas varreduras: \textit{raster} que
é realizada do início para o fim da imagem (Figura~\ref{fig:rastersimples}); e
uma \textit{anti-raster} que realiza a varredura na imagem do fim para o início
(Figura~\ref{fig:antirastersimples}). Cada uma dessas varreduras irá utilizar
um seguimento específico do seu conjunto de vizinhos $u$ de um pixel $p \in
\mathbb{Z}^2$ para realizar a respectiva atualização
(Figura~\ref{fig:sequencialpixel}). Tais procedimentos podem ser observados no
Algoritmo~\ref{alg:dtsequencial}.

\begin{algorithm}
	\label{alg:dtsequencial}
	\footnotesize
	\Entrada{I, imagem binária}
	\Saida{I, imagem em tons de cinza}
	\Varredura{raster}{
		Sendo $p$ o pixel atual
		\Se {$I(p) \neq 0$}{
			$I(p) \leftarrow min\{I(p+u_0)+1, 
			I(p+u_1)+1,I(p+u_2)+1,I(p+u_3)+1\}$		
		}
	}
	\Varredura{anti-raster}{
		Sendo $p$ o pixel atual
		\Se {$I(p) \neq 0$}{
			$I(p) \leftarrow 
			min\{I(p),I(p+u_4)+1,I(p+u_5)+1,I(p+u_6)+1,I(p+u_7)+1\}$		
		}
	}
	\caption{Transformada de Distância Sequencial}
\end{algorithm}

A varredura \textit{raster} (linhas 1 a 5 do Algoritmo~\ref{alg:dtsequencial})
gera uma imagem em tons de cinza intermediária, onde os maiores valores estão
concentrados abaixo e a esquerda da imagem $I$. Cada pixel $p$ possuirá o menor
caminho $P$ entre $p$ e o \textit{background} da imagem, seguindo a regra: cada
caminho $xy$ de $P$ tem um componente vertical estritamente positivo ou um
componente vertical zero, e um componente positivo à esquerda.
Dessa forma, a varredura do tipo \textit{anti-raster} (linhas 6 a 9 do
Algoritmo~\ref{alg:dtsequencial}), após a varredura \textit{raster}, irá
realizar os ajustes de distância a partir dos componentes inferiores e a
direita~\cite{vincent1992morphological}.

Como duas varreduras são suficientes na execução do
algoritmo~\cite{rosenfeld1966sequential}, há uma evidente melhoria em relação
ao algoritmo paralelo. Assim, algoritmos sequenciais constituem uma das
melhores alternativas em termos de algoritmos morfológicos e, também, podem ser
facilmente expandidos para trabalhar com
$n$-dimensões~\cite{borgefors1984distance}.

\textbf{Algoritmos Baseados em Filas de Pixels}

Esta classe considera toda a imagem como um grafo, onde cada pixel é um vértice
e as arestas são as conexões entre os mesmos, admitidas pela grade
(conectividade). Os limites de dilatação são estruturas
isotrópicas\footnote{Estruturas que apresentam as mesmas propriedades em todas
as direções}, e é utilizada uma fila de pixels que executa uma busca em largura
no grafo.

\begin{algorithm}
	\label{distanciacomfila}
	\footnotesize
	\Entrada{I, imagem binária}
	\Saida{I, imagem em tons de cinza}
	\ParaCada{pixel $p \in D_I$}{
		\Se{$I(p)=1$ \textbf{ E } $\exists p' \in N_G(p),I(p')=0$}{
			fila.insere($p$); \\
			$I(p) \leftarrow 2$
		}
	}
	\Enqto{fila.vazia() $=$ \textbf{falso}}{
		$p \leftarrow fila.retira()$ \\
		\ParaCada {$p' \in N_G(p)$}{
			\Se{$I(p')=1$}{
				$I(p') \leftarrow I(p)+1$ \\
				fila.insere($p'$)
			}
		}
	}
	\caption{Transformada de Distância com Filas de Pixels}
\end{algorithm}

A estrutura de dados utilizada é a fila \textit{First In, First Out} (FIFO), na
qual os primeiros pixels inseridos, serão os primeiros a serem retirados para
processamento.  Dessa forma, cada pixel representa um vértice e a fila
constitui-se de uma determinada quantidade de ponteiros que assinalam o
endereço de um determinado pixel. As operações utilizadas na fila são
\textit{insere}, na qual é colocado um pixel $p$ no final da fila; $retira$ que
realiza a leitura do pixel que se encontra no início da fila (leitura do seu
endereço) e o retira da mesma; e \textit{vazia} que retorna verdadeiro se a
fila estiver vazia ou falso caso contrário. Dadas as operações do parágrafo
anterior, a Transformada de Distância é apresentada no
Algoritmo~\ref{distanciacomfila}.

Algoritmos baseados em filas de pixels são extremamente eficientes e simples.
Eles são, também, mais adequados para transformações mais complexas como a
reconstrução em tons de cinza~\cite{teodoro2012fast}, esqueleto morfológico e a
transformação \textit{watershed}.

\textbf{Algoritmos Híbridos}

Algoritmos Híbridos são algoritmos projetados para utilizar segmentos das
outras técnicas já citadas. Tal abordagem busca prover um algoritmo cuja
combinação de técnicas seja mais eficiente do que a utilização individual de
qualquer uma dessas.

A construção de um algoritmo híbrido é feita pela identificação de segmentos
que podem ser aproveitados de forma isolada. A divisão mais comum separa
algoritmos morfológicos nas fases de Inicialização e Processamento. Essa
separação pode ser observada no Algoritmo~\ref{distanciahibrida}. Na fase de
Inicialização podem ser realizadas varreduras sequenciais ou paralelas, como
aquelas já descritas nas seções anteriores. Essa fase tem como objetivo
realizar um processamento mais grosseiro, com a possibilidade de inicializar
uma estrutura de dados que terá continuidade na fase de Processamento. A fase
de Processamento executa tarefas que processam em elementos específicos dentro
da imagem, auxiliados por alguma estrutura de dados que indicará o ponto de
partida e a continuação para tal execução.

\begin{algorithm}
	\label{distanciahibrida}
	\footnotesize
	\Entrada{I, imagem binária}
	\Saida{I, imagem em tons de cinza}
	\Varredura{raster}{
		Sendo $p$ o pixel atual
		\Se {$I(p) \neq 0$ \textbf{ E } $\exists p' \in N_G(p),I(p')=0$}{
			fila.insere($p$); \\
			$I(p) \leftarrow 2$
		}
	}
	\Enqto{fila.vazia() $=$ \textbf{falso}}{
		$p \leftarrow fila.retira()$ \\
		\ParaCada {$p' \in N_G(p)$}{
			\Se{$I(p')=1$}{
				$I(p') \leftarrow I(p)+1$ \\
				fila.insere($p'$)
			}
		}
	}
	\caption{Transformada de Distância Híbrida}
\end{algorithm}

Uma implementação possível híbrida da Transformada de Distância utiliza
varreduras do método sequencial (apresentado no
Algoritmo~\ref{alg:dtsequencial}), em conjunto com uma estrutura de dados que
irá receber aqueles elementos identificados na varredura inicial. Utilizando a
estratégia baseada em filas de pixels, a estrutura de dados consome e insere
elementos a medida que seus valores são verificados e atualizados, conforme
condição de propagação.

\subsection{Arquiteturas Paralelas}\label{subsec:arquiteturas}

Arquitetura paralela é uma forma de fornecer estruturas explícitas e de alto
nível para o desenvolvimento de soluções utilizando processamento paralelo,
dado pelo uso de múltiplos processadores cooperando para resolver problemas
através de execução concorrente~\cite{duncan1990survey}. Dessa forma, busca-se
quebrar problemas em partes independentes de forma que cada um dos
processadores disponíveis seja responsável por uma parcela específica do
processamento.

O paralelismo pode ser realizado de diversas formas (bit, instrução, dado ou
tarefa), e computadores paralelos são classificados de acordo com o suporte de
paralelismo dado pelo hardware. 

\subsubsection{Classificação de Arquiteturas Paralelas}\label{subsubsec:classif}

Diversas arquiteturas paralelas já foram sugeridas e implementadas na
literatura, e também diversas classificações foram propostas. Dentre as mais
conhecidas é possível citar a proposta por Flynn~\cite{flynn1972some} e a
proposta por Duncan~\cite{duncan1990survey}.

\textbf{Taxonomia de Flynn}

A classificação proposta por Flynn~\cite{flynn1972some} agrupa as arquiteturas
segundo os seus fluxos de instruções e de dados. A classificação é a seguinte:
\begin{itemize}
    \item \textit{Single Instruction, Single Data} (SISD) - Um único fluxo de
        instrução para um único fluxo de dados. São as máquinas não paralelas
        da arquitetura de von Neumann~\cite{pacheco2011introduction}; 
    \item \textit{Múltiple Instruction, Single Data} (MISD) - Múltiplos fluxos
        de instrução para um único fluxo de dados; 
    \item \textit{Single Instruction, Multiple Data} (SIMD) - Um único fluxo de
        instrução para múltiplos fluxos de dados, sendo ideal para a
        paralelização de laços simples que operam sobre grandes vetores de
        dados~\cite{pacheco2011introduction}. Nesta classe podem ser citados,
        como exemplo, os processadores vetoriais; e 
    \item \textit{Multiple Instruction, Multiple Data} (MIMD) - Múltiplos
        fluxos de instrução para múltiplos fluxos de dados. Nesta classe,
        encontram-se, por exemplo, sistemas multicomputadores
        (\textit{clusters}) e multiprocessadores com memória compartilhada.
\end{itemize}

\textbf{Taxonomia de Duncan}

A Taxonomia de Duncan~\cite{duncan1990survey} surgiu de maneira a classificar
de forma detalhada e hierárquica as arquiteturas paralelas. Essa classificação
usa como base as principais arquiteturas paralelas e pode ser visualizada na
Figura~\ref{fig:duncan}. A classificação é descrita como:

\begin{figure}[!htb]
	\centering
	\includegraphics[width=\columnwidth]{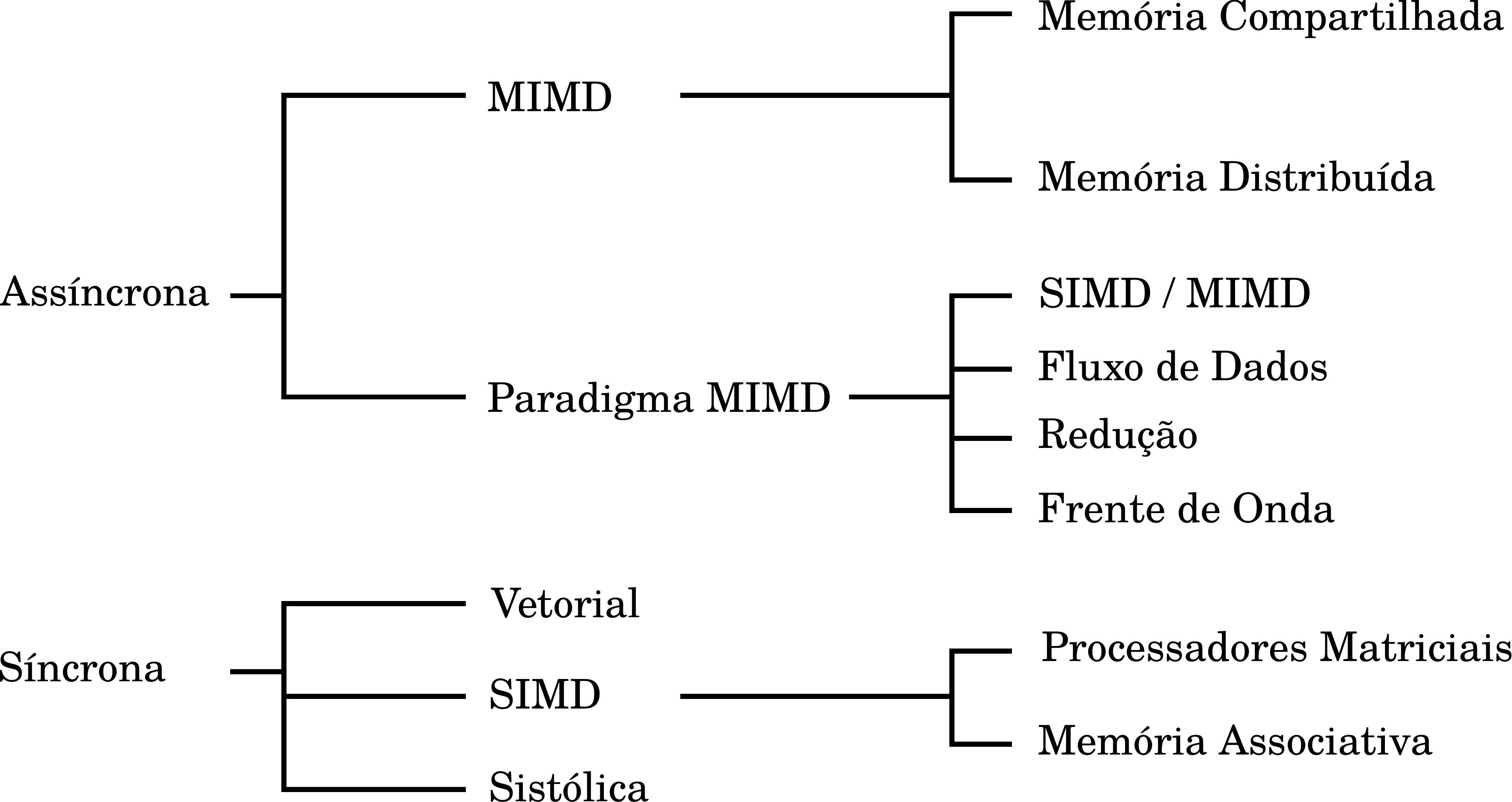}
    \caption{Taxonomia de Duncan. Adaptado de~\cite{duncan1990survey}.}
	\label{fig:duncan}
\end{figure}

\begin{itemize}
    \item Arquiteturas Assíncronas: Possuem controle descentralizado com
        unidades de processamento independentes que operam em conjuntos
        diferentes de dados. Elas são divididas em MIMD e paradigma MIMD.
    \begin{itemize}
        \item MIMD: São arquiteturas formadas por múltiplas unidades de
            processamento independentes, executando conjuntos de instruções
            diferentes sobre conjuntos de dados distintos. São divididas em
            memória compartilhada e memória distribuída.
        \begin{itemize}
            \item Memória Compartilhada: Possui uma única memória global
                utilizada para comunicação e sincronização entre as unidades de
                processamento.
            \item Memória Distribuída: Cada unidade de processamento possui sua
                própria memória local e sua comunicação ocorre por meio de
                troca de mensagens.
        \end{itemize}
        \item Paradigma MIMD: São arquiteturas com características peculiares
            além daquelas que as classificam como MIMD.
        \begin{itemize}
            \item SIMD / MIMD: Possui segmentos de arquitetura MIMD que
                funcionam sob a arquitetura SIMD.
            \item Fluxo de Dados: Instruções são executadas de acordo com a
                disponibilidade dos operandos utilizados.
            \item Redução: Instruções são executadas somente quando seus
                respectivos resultados são requeridos como operandos de outra
                instrução em execução.
            \item Frente de Onda: Combinam vários processadores em
                \textit{pipeline}, nos quais apenas alguns realizam comunicação
                com a memória, trabalhando de acordo com a disponibilidade dos
                operandos utilizados por cada instrução.
        \end{itemize}
    \end{itemize}
    \item Arquiteturas Síncronas: Possuem controle centralizado dado pelo
        \textit{clock}, unidade de controle central ou controlador de unidade
        vetorial.
        \begin{itemize}
            \item Vetorial: Executam operações em vetores de dados.
            \item SIMD: Um único fluxo de instrução para múltiplos fluxos de
                dados, sendo dividido em Arranjo de Processadores e Memória
                Associativa.  
            \begin{itemize}
                \item Processadores Matriciais: Nesta arquitetura as estruturas
                    de processamento são arranjadas na forma de matrizes de 
                    dados para operação.
                \item Memória Associativa: É a arquitetura com acesso a memória
                    vinculado ao seu conteúdo, ao invés do uso de endereço.
            \end{itemize}
            \item Sistólica: Esta arquitetura busca suprir necessidades de
                computação intensiva em operações de Entrada e Saída (E/S),
                combinando vários processadores em pipeline, onde apenas alguns
                desses realizam comunicação com a memória.
        \end{itemize}
\end{itemize}

A Taxonomia de Duncan detalha Arquiteturas Síncronas do tipo Vetorial no qual
uma mesma operação é executada em todo um vetor de uma só vez. Essa mesma
característica se apresenta na Taxonomia de Flynn em um contexto mais genérico
como SIMD, observando um fluxo de instrução único para diversos fluxos de
dados. O objeto de pesquisa deste trabalho é o coprocessador
Intel\textsuperscript{\textregistered} Xeon Phi\textsuperscript{\texttrademark}
que possui as características citadas, por ser um processador com instruções do
tipo vetorial. Dessa forma, as seções seguintes irão detalhar tanto processadores
vetoriais (Seção~\nameref{subsubsec:procvetorial}) quanto o coprocessador 
Intel\textsuperscript{\textregistered} Xeon Phi\textsuperscript{\texttrademark}
(Seção~\nameref{subsec:xeonphi}).

\subsubsection{Processadores Vetoriais}\label{subsubsec:procvetorial}

Processador Vetorial é um dos tipos de sistema SIMD que operam diretamente em 
matrizes ou vetores de dados, enquanto processadores convencionais operam em 
elementos de dados individuais ou escalares. Processadores Vetoriais possuem 
as seguintes características~\cite{pacheco2011introduction}:
\begin{enumerate}
    \item Registradores Vetoriais - Registradores capazes de guardar um vetor
        de operandos e processar, simultaneamente, múltiplos operandos nesse
        vetor.
	
    \item Unidades Funcionais Vetorizadas - As operações são aplicadas a cada
        elemento do vetor ou, então, a cada conjunto de elementos.
	
    \item Instruções Vetoriais - Instruções que operam em todo o vetor ao invés
        da forma escalar. Por exemplo, em um processador vetorial onde a
        largura do vetor é \textit{vector\_lenght}, um simples laço como
\begin{verbatim}
for (i = 0; i < n; i++) {
    x[i] += y[i];
}
\end{verbatim}
        requer uma instrução de \textit{load}, uma de \textit{add} e uma de
        \textit{store} para cada bloco de \textit{vector\_lenght} elementos,
        enquanto um sistema convencional (SISD) requer um \textit{load}, um
        \textit{add} e um \textit{store} para cada elemento.
	
    \item Memória Intercalada - A memória é dividida em bancos e, após o acesso
        a determinado local, é necessária uma espera antes de acessar esse
        local novamente. Então, os elementos de um vetor são distribuídos em
        vários bancos como forma de mitigação de atraso das instruções
        \textit{load} e \textit{store}.
	
    \item Acesso espaçado a memória e \textit{scatter}/\textit{gather} - O
        acesso espaçado a memória permite acessar elementos de um vetor que
        possuam intervalos fixos. Já o \textit{scatter}/\textit{gather} escreve
        ou lê elementos localizados em posições irregulares da memória em um
        registrador vetorial. 
\end{enumerate}

Em aplicações regulares que manipulam matrizes, processadores vetoriais são
fáceis e rápidos de se utilizar. Além disso, compiladores auxiliam na
identificação de códigos que podem ser vetorizados e também fornecem
informações úteis quando isso não é possível~\cite{pacheco2011introduction}. A
próxima seção discutirá o Intel\textsuperscript{\textregistered} Xeon
Phi\textsuperscript{\texttrademark} que é um exemplo específico de processador
vetorial, além de ser o foco deste trabalho.

\subsection{Intel\textsuperscript{\textregistered} Xeon 
Phi\textsuperscript{\texttrademark}}\label{subsec:xeonphi}

O coprocessador Intel\textsuperscript{\textregistered} Xeon
Phi\textsuperscript{\texttrademark} foi lançado em novembro de 2012 e é baseado
na arquitetura \textit{Intel Many Integrated
Core} (Intel\textsuperscript{\textregistered} MIC), suportando execução
simultânea de até 244 \textit{threads} em seus 61 núcleos para atingir um
desempenho teórico de até 1.2 \textit{teraflops} com baixo consumo
energético~\cite{jeffers2013intel}\footnote{Informações baseadas no modelo
7120P.}. 

Ele funciona acoplado a um \textit{host}, porém possui espaço de memória
independente com comunicação através do barramento \textit{Peripheral Component
Interconnect Express} 3.0 (\textit{PCI-Express}). Comparado com CPUs
convencionais, o Intel\textsuperscript{\textregistered} Xeon
Phi\textsuperscript{\texttrademark} possui características singulares que são
essenciais para se atingir alto desempenho. As principais características são:

\begin{figure}[!htb]
	\centering
	\includegraphics[width=1\columnwidth]{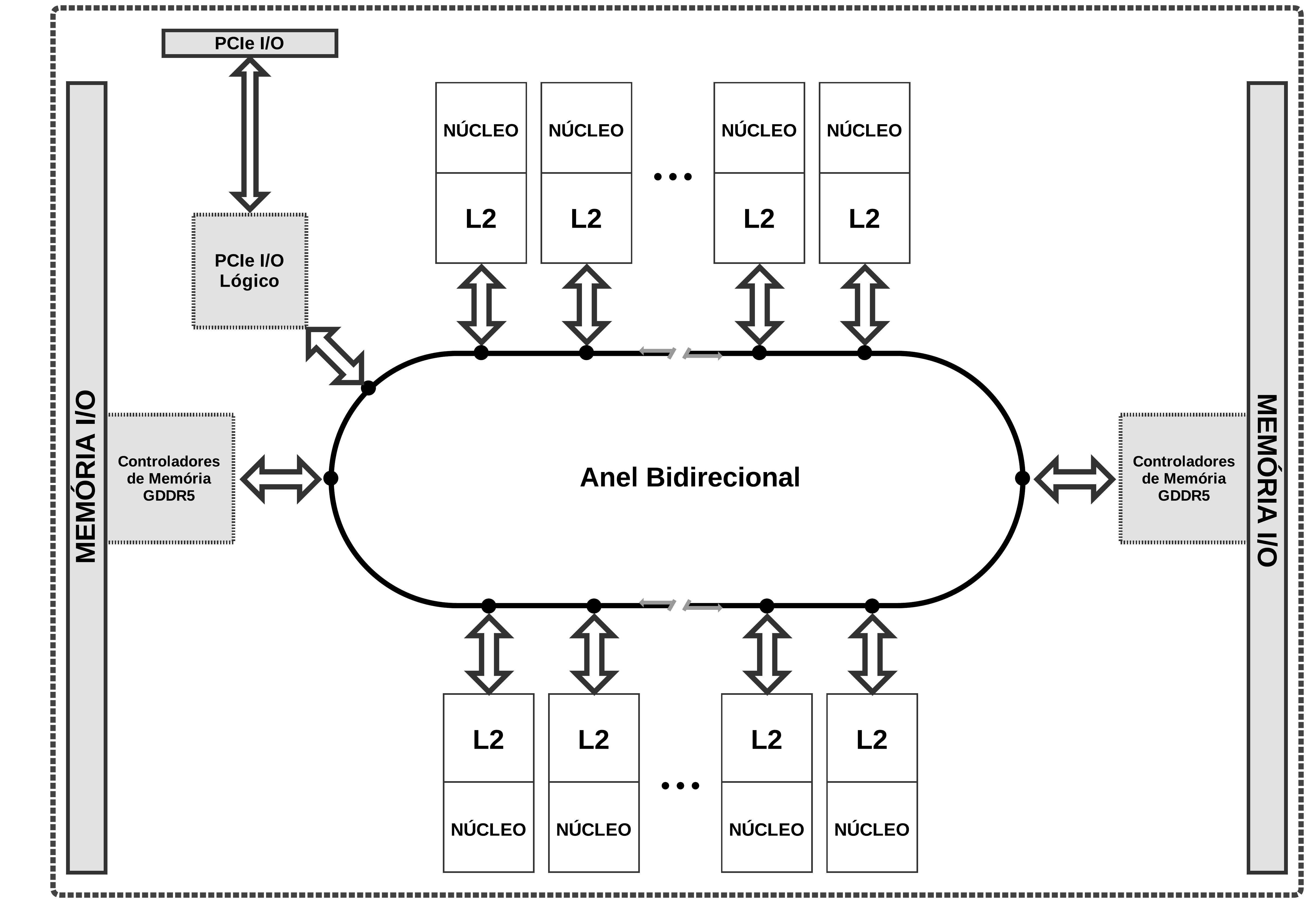}
	\caption{Diagrama de Bloco do Intel\textsuperscript{\textregistered} Xeon 
    Phi\textsuperscript{\texttrademark}. Adaptado de~\cite{phiSite}.}
	\label{fig:phiblock}
\end{figure}

\begin{itemize}
    \item A largura dos vetores utilizados pela Unidade de Processamento
        Vetorial (VPU) é de 512 bits. Sendo assim, uma unidade pode processar
        até 8 elementos de dados de precisão dupla ou 16 elementos de
        precisão simples. Comparado com CPUs mais recentes como o
        Intel\textsuperscript{\textregistered}
        Core\textsuperscript{\texttrademark} i7-6700K ou o AMD
        Opteron\textsuperscript{\texttrademark} 6386 SE, seus registradores
        possuem, pelo menos, o dobro da largura. Assim, a utilização das VPUs
        de forma efetiva é um do elementos essenciais para se obter alta
        performance nesse processador. As VPUs podem ser exploradas por
        implementações manuais, utilizando instruções SIMD, ou por
        autovetorização através do compilador
        Intel\textsuperscript{\textregistered}. A autovetorização é um módulo
        que tenta identificar \textit{loops} ou segmentos que podem ser
        vetorizados para utilizar VPUs SIMD em tempo de compilação.  
    \item Cada núcleo do Intel\textsuperscript{\textregistered} Xeon
        Phi\textsuperscript{\texttrademark} suporta até quatro \textit{threads}
        de hardware, totalizando 244 \textit{threads}. Isso possibilita uma
        melhor utilização do processador em casos nos quais uma \textit{thread}
        não é capaz de explorar completamente cada núcleo.
    \item Cada núcleo possui uma cache L2 unificada de 512 KB, que é coerente
        nos 61 núcleos. Tais caches são interconectados por um barramento em
        anel de largura de 512 bits (veja a Figura~\ref{fig:phiblock}). Se
        ocorrer um \textit{chache miss} em alguma cache L2, através de
        \textit{Distributed Tag Directories} (DTD), as solicitações são
        encaminhadas para os outros núcleos através da rede em anel. 
\end{itemize}

Plataformas MIC combinam características de CPUs de uso geral e aceleradores,
como GPUs. O Intel\textsuperscript{\textregistered} Xeon
Phi\textsuperscript{\texttrademark} é baseado no conjunto de instruções x86 e
suporta modelos de programação em memória compartilhada para a comunicação
entre núcleos. Essas características minimizam o esforço de programação para o
caso da portabilidade de código CPU para o Xeon Phi, em particular porque
linguagens de programação como C/C++ e Fortran, e modelos de programação como
\textit{Open Multi-Processing} (OpenMP), \textit{Posix Threads} (PThreads) e
\textit{Message Passing Interface} (MPI) podem ser utilizados sem modificação de
código~\cite{rahman2013xeon}.

\begin{figure}[t]
	\centering
	\includegraphics[width=\columnwidth]{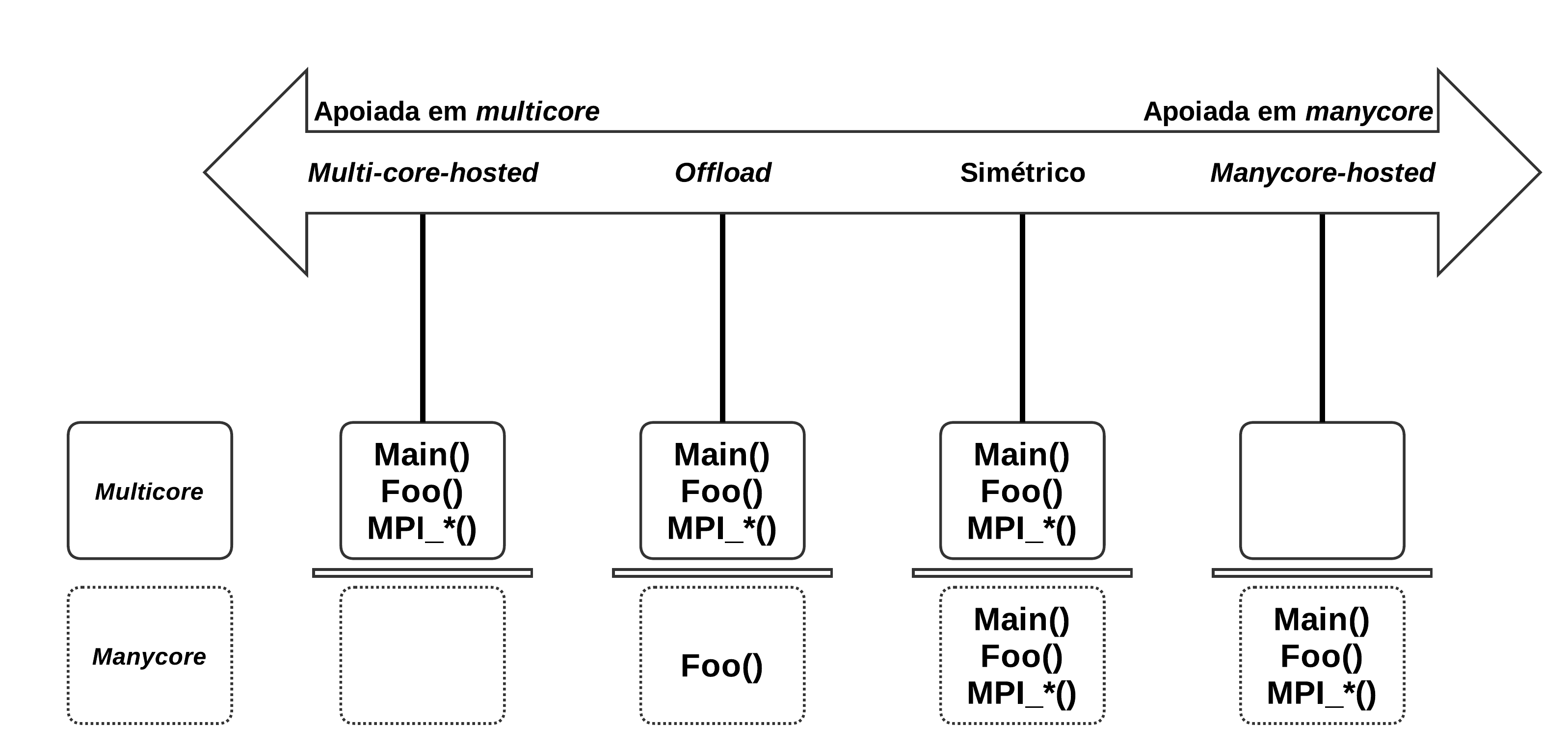}
	\caption{Abordagens de Programação Utilizando o 
	Intel\textsuperscript{\textregistered} Xeon 
	Phi\textsuperscript{\texttrademark}.}
	\label{fig:abordagem}
\end{figure}

Além disso, o Intel\textsuperscript{\textregistered} Xeon
Phi\textsuperscript{\texttrademark} possui 8GB de memória \textit{Graphics
Double Data Rate} (GDDR) local com 320Gb/s de transmissão, possui 32KB de
memória cache L1 e 512KB de cache L2, executa um \textit{kernel} do Linux em um
de seus núcleos e cada núcleo possui sua própria Unidade de Processamento
Vetorial (VPU). Assim, a partir das suas características, o
Intel\textsuperscript{\textregistered} Xeon Phi\textsuperscript{\texttrademark}
permite as seguintes abordagens de execução (veja a Figura~\ref{fig:abordagem}): 

\begin{itemize}
    \item Modo ``Nativo'' (\textit{Many-core-hosted}): Dado que o
        Intel\textsuperscript{\textregistered} Xeon
        Phi\textsuperscript{\texttrademark} executa seu próprio sistema
        operacional, é possível compilar uma aplicação de forma a permitir sua
        transferência e a execução do arquivo binário direto no coprocessador.
    \item Modo \textit{Offload}: Nesse modo uma aplicação inicia sua execução
        no \textit{host} e pode transferir partes da computação para o
        Intel\textsuperscript{\textregistered} Xeon
        Phi\textsuperscript{\texttrademark}.
    \item Modo Simétrico: Modo onde um aplicativo está executando tanto no
        \textit{host} quanto no Intel\textsuperscript{\textregistered} Xeon
        Phi\textsuperscript{\texttrademark}, e processos específicos podem
        executar em diferentes dispositivos.
\end{itemize}

\subsubsection{\textit{Data Race} e Condição de Corrida}\label{subsubsec:corrida}

Em arquiteturas paralelas, diversos erros ou \textit{bugs} de programação estão
diretamente ligados a \textit{data races} e condições de
corrida~\cite{narayanasamy2007automatically}.

\textit{Data race} ocorre quando dois acessos a mesma posição de memória entram
em conflito e pelo menos um desses acessos é uma
escrita~\cite{attiya1993shared}.  Já condição de corrida é definida como uma
situação na qual múltiplas \textit{threads} realizam leitura e escrita em uma
área compartilhada de memória e o resultado final depende da ordem de
execução~\cite{carr2001race}.

Em uma análise detalhada desse contexto, \textit{data races} podem ou não levar
a condições de corrida, e diversos mecanismos tem sido desenvolvidos para
localizar automaticamente \textit{data races} e classificá-los de acordo com
suas potencialidades dentro do código.  \textit{Data races} podem ser
potencialmente prejudiciais ou potencialmente
benignos~\cite{narayanasamy2007automatically}. \textit{Data races}
potencialmente prejudiciais afetam a corretude de um algoritmo e necessitam ser
corrigidos. Já \textit{data races} potencialmente benignos possuem
comportamentos que dificultam a sua identificação em análises estáticas, porém
eles não afetam a corretude do algoritmo, mas precisam de uma análise dinâmica
para facilitar a sua identificação. Alguns motivos para a ocorrência de
\textit{data races} potencialmente benignos são: 

\begin{itemize}
    \item Sincronização Construída pelo Usuário: São primitivas de
        sincronização construídas pelo próprio usuário sem a utilização de
        barreiras ou operações atômicas providas pelo conjunto de instruções da
        arquitetura.
    \item Dupla Checagem: Utiliza mais de uma checagem em determinada
        condição, com a finalidade de otimizar a necessidade de sincronização.
    \item Ambos os Valores Válidos: Nesse caso, independente de leitura ou
        escrita para valores diferentes, a execução do algoritmo continua
        correta. 
    \item Escritas Redundantes: Caso onde uma operação de escrita apenas
        reescreve o valor já contido na posição de memória.
    \item Manipulação Disjunta de Bits: Nesse caso, apesar de \textit{threads}
        escreverem valores diferentes em uma mesma posição de memória
        compartilhada, cada \textit{thread} modifica apenas bits específicos
        nessa posição.
\end{itemize}

A correta identificação e classificação de \textit{data races} em um código
permite direcionar o modelo de construção de algoritmos paralelos. Dessa forma,
a existência \textit{data races} benignos não exige tratamento específico para
garantir a corretude do algoritmo, o que possibilita a aplicação de
estratégias, principalmente, vetoriais que serão apresentadas na
Seção~\ref{sec:IWPPphi}.

\subsection{Sumário}

Nesta seção foram apresentados conceitos básicos sobre Algoritmos
Morfológicos e Arquiteturas Paralelas. A Seção~\ref{subsec:algoritmosmorf} tratou
sobre Algoritmos Morfológicos apresentando notações que serão utilizadas ao
longo deste trabalho. Também foram mostrados alguns exemplos de algoritmos
morfológicos buscando evidenciar a utilidade e a capacidade da Morfologia
Matemática, e foram discutidas as principais estratégias de implementação para
algoritmos morfológicos: Algoritmos Paralelos, Algoritmos Sequenciais,
Algoritmos Baseados em Filas de Pixels e Algoritmos Híbridos.

Em seguida, a Seção~\ref{subsec:arquiteturas} argumentou sobre
Arquiteturas Paralelas, abordando as classificações propostas por Flynn e
Duncan, detalhando as características de processadores vetoriais, uma visão
acurada do Intel\textsuperscript{\textregistered} Xeon
Phi\textsuperscript{\texttrademark} e uma categorização de \textit{data races}
e condições de corrida.

A próxima seção irá apresentar o \textit{Irregular Wavefront Propagation
Pattern} que generaliza as implementações atuais para Algoritmos Morfológicos,
apresentando casos de uso utilizando esse padrão.

  \section{\textit{Irregular Wavefront Propagation Pattern}
(IWPP)}\label{sec:IWPP}

\textit{Irregular Wavefront Propagation Pattern}~\cite{teodoro2013efficient} é
um padrão de computação encontrado em diversos algoritmos morfológicos, como
descritos anteriormente na Seção~\nameref{subsubsec:exemplos}. IWPP são estruturas
que, dados espaços multidimensionais $D_I \in \mathbb{Z}^n$, e um conjunto de
pixels ativos $S$, propagam os valores de pixels ativos a elementos em
\textit{wavefront} de forma irregular, seguindo uma condição de propagação.
Esses conjuntos de pixels ativos $S$ são denominados ondas e, durante a
execução, se expandem de forma irregular ao longo do espaço $D$. As principais
características dessas ondas são: 

\begin{itemize}
	\item Dinamismo - Não é possível antever as direções de suas expansões;
	\item Dependência - Cada expansão depende diretamente da anterior e 
	indiretamente da expansão de outras ondas.
\end{itemize}

O IWPP divide-se em duas fases: Identificação das Frentes de Onda e Propagação
Irregular. A Identificação das Frentes de Onda tem por finalidade identificar
as frentes de onda iniciais e alimentar a estrutura de dados que permitirá a
execução da fase de Propagação Irregular. Já na tarefa de Propagação Irregular,
cada elemento ativo constitui uma frente de onda que se expande enquanto a
estrutura de dados contiver elementos e a condição de propagação for
verdadeira. Essas duas tarefas podem ser observadas no
Algoritmo~\ref{alg:IWPPpadrao}.

\begin{algorithm}
    \label{alg:IWPPpadrao}
    \scriptsize
    \Entrada{$D$: \emph{Conjunto de elementos de um espaço multidimensional}}
    \Saida{$D$: \emph{Conjunto estável com todas as propagações realizadas}}
    $D \leftarrow$ \emph{conjunto de elementos de um espaço multidimensional}\\
    \{{\bf Fase de Inicialização}\}\\
    $S \leftarrow$ \emph{subconjunto de elementos ativos de $D$}\\
    \{{\bf Fase de Propagação Irregular}\}\\
    \Enqto{$S \neq \emptyset$}{
    Extrai $e_i$ de $S$\\
    $Q \leftarrow$ $N_G(e_i)$\\
    \Enqto{$Q \neq \emptyset$}{
        Extrai $e_j$ de $Q$\\
        \Se{$PropagationCondition$($D(e_i)$,$D(e_j)$) $=$ verdadeiro}{
            $D(e_j) \leftarrow$ $max$/$min(D(e_i), D(e_j))$\\
            Insere $e_j$ em $S$\\
        }
    }
}
\caption{Irregular Wavefront Propagation Pattern (IWPP)}
\end{algorithm}

Nesse algoritmo, a Fase de Inicialização identifica os elementos ativos de $D$
que formarão o subconjunto de elementos ativos $S$ (linhas 1 a 3). A partir
desse subconjunto, na Fase de Propagação Irregular, enquanto ele não
estiver vazio (linha 5), é extraído um elemento $e_i$ de $S$ (linha 6). Desse
elemento é gerado um subconjunto $Q$ com a vizinhança $N_G(e_i)$ de $e_i$
(linha 7) e, para cada elemento $e_j$ de Q, é verificado se condição de
propagação é satisfeita para esse elemento (linhas 9 e 10). Quando a condição
de propagação for satisfeita, $D(e_j)$ recebe a propagação (linha 11), e $e_j$
passa a fazer parte do subconjunto de elementos ativos $S$ que podem propagar
suas características aos seus elementos vizinhos. Então, $e_j$ é inserido em
$S$. A Fase de Propagação Irregular continua até que não hajam mais
elementos no subconjunto $S$, indicando a estabilidade de $D$.

As seções a seguir exemplificam e detalham casos de uso de algoritmos
morfológicos utilizando o padrão IWPP: a Reconstrução Morfológica
(Seção~\ref{subsec:reconstrucaomorfologica}), a Transformada de Distância
Euclidiana (Seção~\ref{subsec:tde}) e o Algoritmo \textit{Fill Holes}
(Seção~\ref{subsec:fillholes}).

\subsection{Reconstrução Morfológica}\label{subsec:reconstrucaomorfologica}
Reconstrução é uma transformação que utiliza operações morfológicas envolvendo
duas imagens e um elemento estruturante. As operações morfológicas utilizadas
na reconstrução são operações básicas adotadas em um conjunto amplo de
algoritmos de processamento. Essas operações são aplicadas a pixels individuais
e são processadas baseadas no valor atual do pixel e nos valores dos pixels em
sua respectiva vizinhança. 

No processo de reconstrução, as imagens são identificadas por marcadora ($J$), 
utilizada como ponto de partida para a transformação; e máscara ($I$), que 
representa o limite da transformação. O elemento estruturante utilizado define 
a conectividade do pixel $p$ com sua vizinhança $N_G(p)$. A reconstrução 
morfológica $\rho(J)$ de uma máscara $I$, a partir de uma imagem marcadora 
$J$, é feita através de dilatações elementares de tamanho 1 em $J$ por $G$. Uma 
dilatação elementar de um pixel $p$ corresponde a propagação de $p$ à sua 
vizinhança $G$. 

\begin{figure}[h]
	\centering
	\includegraphics[width=\columnwidth]{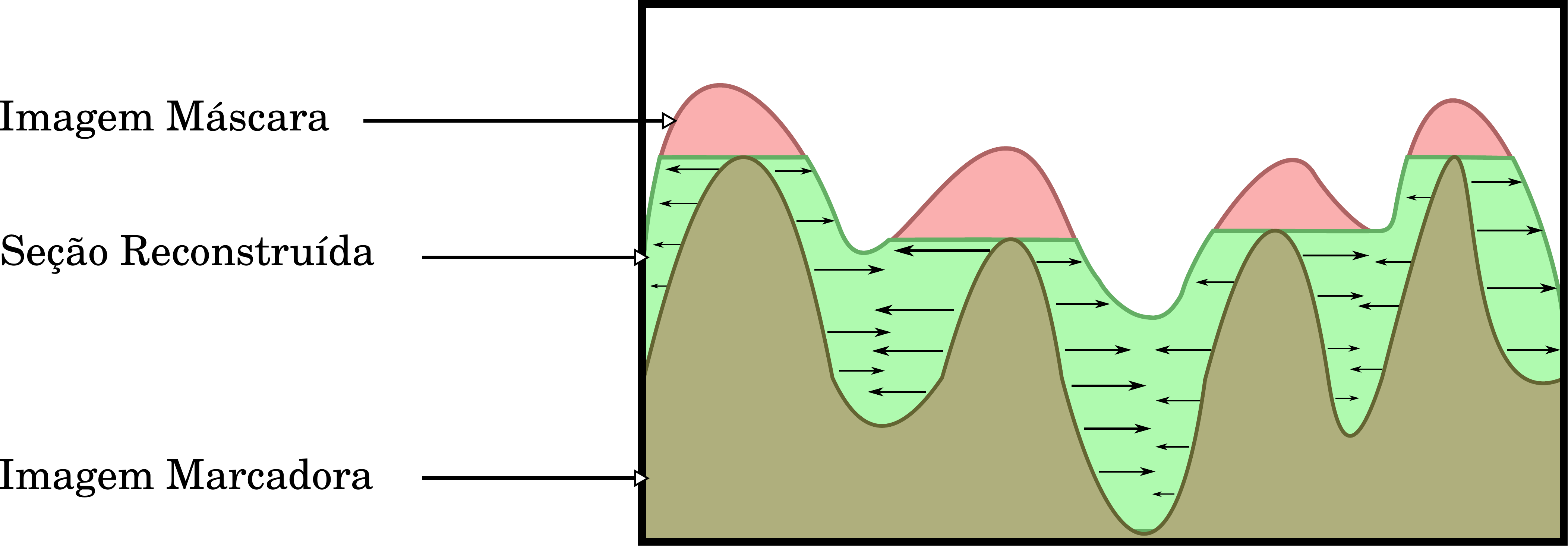}
	\caption{Reconstrução Morfológica.}
	\label{fig:reconstrucao}
\end{figure}

O algoritmo básico executa suas dilatações elementares sucessivamente ao longo
de toda a imagem $J$, atualizando cada pixel com uma comparação pixel-a-pixel
mínima do resultado de suas dilatações e o pixel correspondente em $I$, dado
por $J(p) \leftarrow (max\{J(q),q \in N_G(p)\cup \{p\}\}) \wedge I(p)$, até que
a estabilidade seja alcançada, quando não há mais modificações de valores em
qualquer pixel (conforme mostrado na Figura~\ref{fig:reconstrucao}). 

Baseado nessa técnica, o Algoritmo~\ref{alg:iwpp-recon} demonstra a
reconstrução morfológica utilizando o padrão IWPP para imagens em tons de
cinza. Na Fase de Varreduras desse algoritmo (linhas 3 a 14), o pixel é
propagado na imagem marcadora alternando varreduras \textit{raster} e
\textit{anti-raster}.  A varredura \textit{raster} inicia a partir do pixel
$(0,0)$ e desloca-se até o final da imagem $(i-1,j-1)$
(Figura~\ref{fig:rastersimples}), enquanto a varredura \textit{anti-raster}
realiza o deslocamento no caminho contrário, do final $(i-1,j-1)$ para o início
da imagem $(0,0)$ (Figura~\ref{fig:antirastersimples}). Em cada varredura, a
metade da vizinhança dos pixels vizinhos - superiores a esquerda ou inferiores
a direita - são propagados quando a condição de propagação é satisfeita. Na
passagem \textit{anti-raster}, além da varredura,  é utilizada uma fila de
pixels \textit{First In, First Out} (FIFO) que dará continuidade a execução do
algoritmo na Fase de Propagação Irregular. A fila é inicializada com pixels que
satisfazem a condição de propagação e a computação continua na Fase de
Propagação Irregular removendo elementos da fila, varrendo seus elementos
vizinhos, identificando aqueles pixels cuja condição de propagação ocorre, e
inserindo na fila os modificados pela propagação. Assim, o algoritmo termina
quando a fila está vazia, indicando que a estabilidade foi alcançada. 

\begin{figure}[!htb]
\begin{minipage}{.5\textwidth}
    \centering
    \includegraphics[width=\linewidth]{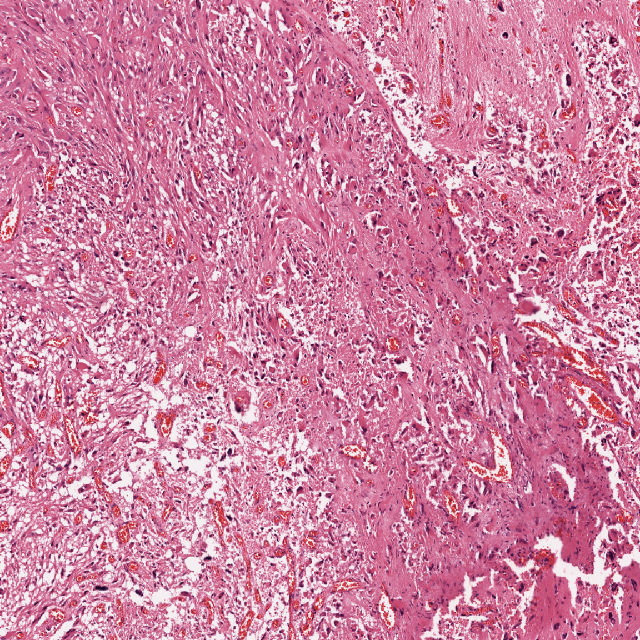}
    \caption{Imagem de Tecido Original}
    \label{img:recon-original}
\end{minipage}
\begin{minipage}{.5\textwidth}
    \centering
    \includegraphics[width=\linewidth]{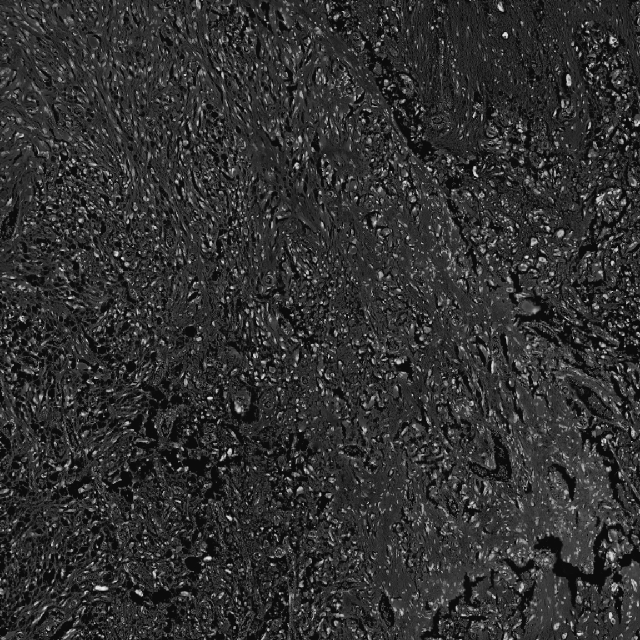}
    \caption{Imagem Máscara}
    \label{img:recon-mascara}
\end{minipage}
\end{figure}
\begin{figure}[!htb]
\begin{minipage}{.5\textwidth}
    \centering
    \includegraphics[width=\linewidth]{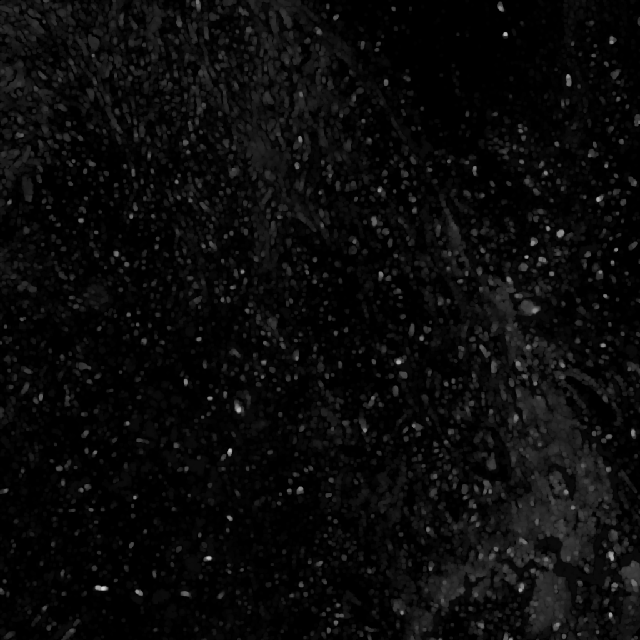}
    \caption{Imagem Marcadora}
    \label{img:recon-marcadora}
\end{minipage}
\begin{minipage}{.5\textwidth}
    \centering
    \includegraphics[width=\linewidth]{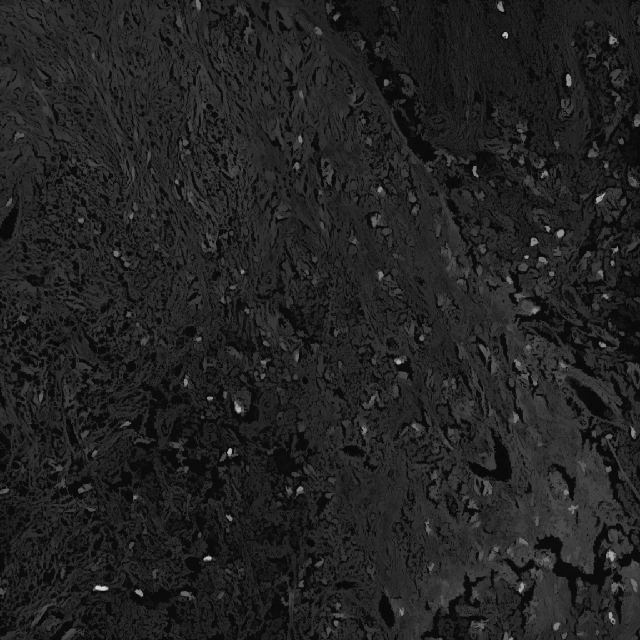}
    \caption{Imagem Reconstruída}
    \label{img:recon-resultado}
\end{minipage}
\end{figure}

\begin{algorithm}
	\label{alg:iwpp-recon}
	\scriptsize
	\Entrada{I, Imagem Máscara}
	\Entrada{J, Imagem Marcadora}
    \Saida{J, Imagem Reconstruída}
	\{\textbf{Fase de Inicialização}\} \\
	\dots \\
	\{\textbf{Fase de Varreduras}\} \\
	\VarreduraIJ{raster}{
		Sendo $p$ o pixel atual \\
		$J(p) \leftarrow (max\{J(q), q \in N_G^+(p) \cup \{p\}\}) \wedge 
		I(p)$ \\
	}
	\VarreduraIJ{anti-raster}{
		Sendo $p$ o pixel atual \\
		$J(p) \leftarrow (max\{J(q), q \in N_G^-(p) \cup \{p\}\}) \wedge 
		I(p)$ \\
		\Se {$\exists q \in N_G^-(p)$ $|$ $J(q) < J(p)$ \textbf{e} $J(q) < 
		I(q)$}{
			fila.insere($p$)
		}
	}
	\{\textbf{Fase de Propagação Irregular}\} \\
	\Enqto {fila.vazia == \textbf{falso}}{
		$p \leftarrow$ fila.retira() \\
		\ParaTodo{q $\in N_G(p)$}{
			\Se{$J(q) < J(p)$ \textbf{e} $I(q)\neq J(q)$}{
				$J(q) \leftarrow min\{J(p),I(q)\}$ \\
				fila.insere($q$)	
			}
		}
	}
	\caption{Reconstrução Morfológica - IWPP}
\end{algorithm} 

No contexto de Bioinformática, uma das aplicações da Reconstrução Morfológica é
destacar setores para possibilitar a identificação e segmentação de células. As
Figuras~\ref{img:recon-original},~\ref{img:recon-mascara},~\ref{img:recon-marcadora}
e~\ref{img:recon-resultado}, exemplificam essa utilização da Reconstrução
Morfológica. A Figura~\ref{img:recon-original} representa a imagem escaneada
com a região do tecido a ser segmentado. A partir dessa imagem escaneada, é
gerada uma versão em tons de cinza para ser utilizada como máscara para a
reconstrução (Figura~\ref{img:recon-mascara}). O ponto de partida para a
identificação de células, a serem segmentadas na região do tecido, utiliza uma
versão da Imagem Máscara (Figura~\ref{img:recon-mascara}) com sua intensidade
reduzida (Figura~\ref{img:recon-marcadora}). Essa redução de intensidade
permite realizar a identificação de picos (ou sementes) que podem ser objetos
ou núcleos no tecido. Dessa forma, a Reconstrução Morfológica irá expandir as
as características dos picos da Imagem Marcadora até os limites permitidos pela
Imagem Máscara gerando, assim, a Imagem Reconstruída
(Figura~\ref{img:recon-resultado}).

\subsection{Transformada de Distância Euclidiana}\label{subsec:tde}
A operação Transformada de Distância calcula o mapa de distâncias $M$ de 
uma imagem binária de entrada $I$, onde para cada pixel $p \in I$, que faz 
parte do objeto (\textit{foreground}), seu valor em $M$ é a menor distância a 
partir de $p$ até o pixel de fundo (\textit{background})  mais próximo da 
imagem.

\begin{figure}[!htb]
\begin{minipage}{.5\textwidth}
  \centering
  \includegraphics[width=.8\linewidth]{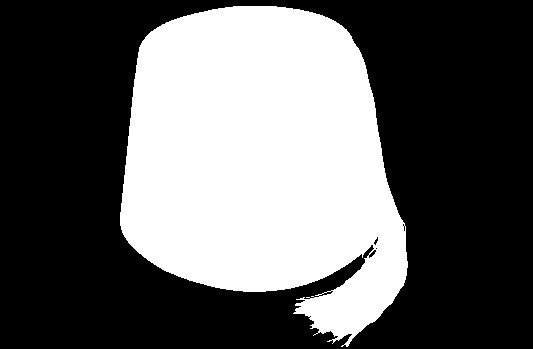}
  \caption{Imagem Original.}
  \label{fig:fez}
\end{minipage}%
\begin{minipage}{.5\textwidth}
  \centering
  \includegraphics[width=.8\linewidth]{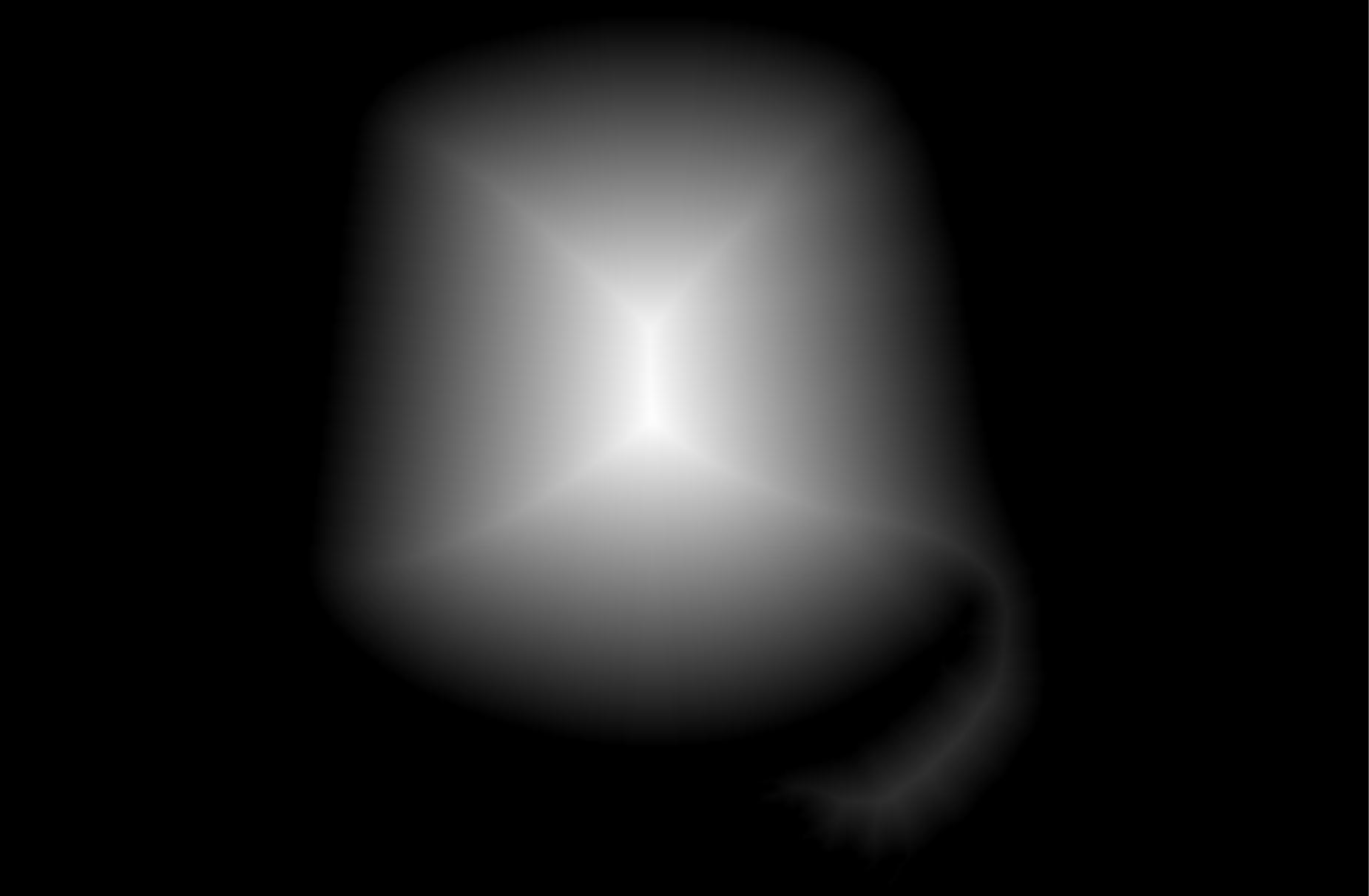}
  \caption{TDE.}
  \label{fig:fezdt}
\end{minipage}
\end{figure}

A métrica de distância mais usual em processamento de imagens é a distância
Euclidiana cuja operação é a Transformada de Distâncias Euclidiana (TDE)
(Figuras~\ref{fig:fez} e~\ref{fig:fezdt}). Essa operação é amplamente utilizada
em diversos outros algoritmos como diagramas de
Voronoi~\cite{rong2007variants}, triangulações de
Delaunay~\cite{preparata2012computational} ou
\textit{watershed}~\cite{vincent1991watersheds}, por exemplo. A TDE possui alto
custo de execução, apesar da simplicidade de seus conceitos.  Com isso,
diversas implementações já foram apresentadas utilizando várias estratégias
para a sua execução, que podem ser classificadas, principalmente, como exatas e
não exatas. A implementação de Danielsson em
1980~\cite{danielsson1980euclidean} foi a primeira proposta de algoritmo TDE,
e, apesar de não exato, as diferenças podem ser rapidamente detectadas e
corrigidas com pós-processamento.

\begin{algorithm}
	\label{TDE}
	\scriptsize
	\Entrada{I, Imagem Máscara}
	\Saida{M, Mapa de distâncias}
	FG: Foreground \{Objeto na imagem\}\\ 
	BG: Background \{Fundo na imagem\}\\
	\{\textbf{Fase de Inicialização}\} \\
	\ParaTodo{$p \in D_I$}{
		\eSe{$p==BG$}{
			$VR(p) \leftarrow p$
		}
		{
			$VR(p) \leftarrow \infty$
		}
		\Se{$I(p)==BG$ \textbf{e} $(\exists q \in N_G(p) \mid I(q) == FG)$}{
			$fila.insere(p)$ \\
		}
	}
	\{\textbf{Fase de Propagação Irregular}\} \\
	\Enqto{fila.vazia() == \textbf{falso}}{
		$p \leftarrow fila.retira(p)$ \\
		\ParaTodo{$q \in N_G(p)$}{
			\Se{$DIST(q,VR(p)) < DIST(q,VR(q))$}{
				$VR(q) = VR(p)$\\
				$fila.insere(q)$
			}
		}
	}
	\ParaTodo{$p \in D_I$}{
		$M(p) = DIST(p,VR(p))$
	}
	\caption{Transformada de Distância Euclidiana - IWPP}
\end{algorithm} 

Na implementação da TDE, vista no Algoritmo~\ref{TDE}, é utilizada uma fila de
pixels IWPP para calcular a distância aproximada de cada vizinhança. Na Fase de
Inicialização, todos os pixels $p$ que fazem parte dos objetos na imagem
(\textit{foreground}) recebem valor infinito (linha 8). A partir desse ponto,
são inseridos na fila pixels $p$ que pertençam ao fundo da imagem e possuam
algum vizinho $q$ que faça parte dos objetos na imagem $I$ (linhas 10 e 11).
Esses pixels formarão a \textit{wavefront} inicial. Na fase \textit{wavefront
propagation}, um pixel é retirado da fila (linha 16) e, para cada vizinho $q$
de $N_g(p)$ em que a distância dele para o fundo seja maior que a distância de
$p$ para o fundo, a propagação irá ocorrer (linhas 17 a 19). Com a propagação,
$q$ será incluído na fila e as iterações continuam até que a fila esvazie.
Alcançada a estabilidade (fila vazia), o mapa de distâncias $M$ é, a partir do
diagrama de Voronoi, calculado pelo algoritmo durante as propagações (linhas 18
e 25).

\subsection{\textit{Fill Holes}}\label{subsec:fillholes}

\textit{Fill Holes} ou \textit{Imfill} é uma técnica utilizada para realizar o
preenchimento de buracos ou áreas em uma imagem binária ou em tons de
cinza~\cite{soille2013morphological}. Um buraco pode ser definido como uma
região de fundo da imagem rodeada por uma borda completamente contornada pelo
objeto da imagem (ou \textit{foreground})~\cite{gonzalez2002digital}.

Em uma imagem binária são realizadas mudanças nos valores dos elementos do
fundo da imagem (\textit{background}) conectados aos valores do objeto com uma
operação de preenchimento, ou inundação (\textit{flood-fill}), até que os limites
do objeto sejam atingidos. Já imagens em tons de cinza é utilizada a mesma
ideia de inundação que modifica a intensidade de áreas escuras contornadas por
áreas mais claras pelas próprias intensidades do contorno.

Uma das estratégias de funcionamento do algoritmo utiliza a Reconstrução
Morfológica como base, onde a entrada é a imagem a ser preenchida. A máscara
a ser utilizada é uma inversão (ou complemento) da própria imagem que irá
limitar o resultado dentro da região de interesse a ser preenchida.

\begin{figure}[h]
	\centering
	\includegraphics[width=\columnwidth]{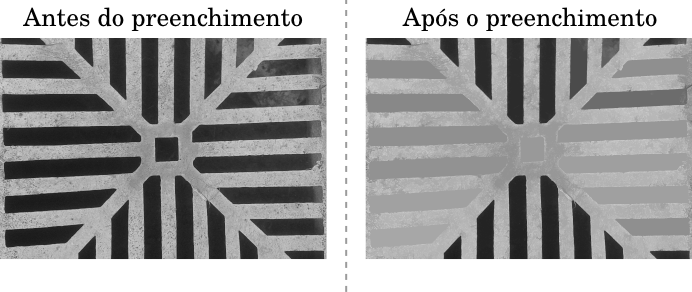}
    \caption{Função \textit{Imfill} Aplicada em Tons de Cinza.}
	\label{fig:fill}
\end{figure}

Esse preenchimento é utilizado no processamento de imagens para homogeneizar
imagens buscando evitar falsa segmentação no uso de outros algoritmos, remover
diferentes intensidades de luminância, ou  facilitar o processamento de outros
algoritmos morfológicos de qualquer natureza. A Figura~\ref{fig:fill}
exemplifica com um objeto com diferentes tonalidades de sombra cuja execução do
algoritmo gera uma imagem mais homogênea do mesmo.

\subsection{Trabalhos Relacionados}\label{sec:relacionados}

As seções subsequentes destacam aspectos e trabalhos relevantes relacionados ao
IWPP (Seção~\nameref{subsubsec:triwpp}), Reconstrução Morfológica
(Seção~\nameref{subsubsec:morphrecon}) e Transformada de Distância Euclidiana
(Seção~\nameref{subsubsec:TD}). 

\subsubsection{IWPP}\label{subsubsec:triwpp}

\textbf{\textit{Breadth-First Search} Eficiente}

Algoritmos IWPP podem ser vistos como algoritmos de varredura em grafos com
suporte a múltiplos pontos de origem. Alguns trabalhos envolvendo
implementações eficientes do \textit{Breadth-First Search} são apresentados
em~\cite{hong2011accelerating} e~\cite{tao2013using}. 

O trabalho apresentado em~\cite{hong2011accelerating} tem como objetivo
desenvolver técnicas para evitar o efeito de desbalanceamento de carga devido a
existência de vértices com número irregular de arestas. Tais técnicas não são
aplicáveis ao contexto do IWPP, uma vez que o número de vértices é constante
por ser baseado no elemento estruturante. 

Já em~\cite{tao2013using} são apresentadas, de maneira isolada, diversas
técnicas para acelerar o BFS utilizando MIC.\ São utilizadas operações de
leitura de arestas aproveitando a largura do vetor com instruções SIMD.\ No
paralelismo em nível de \textit{threads} são utilizados mapas de bits em
conjunto com operações atômicas na exploração da vizinhança dos vértices para
manipular uma fila compartilhada. O uso dessa estratégia busca garantir a não
expansão de vértices iguais por \textit{threads} diferentes, permitindo um maior
controle, e mostrando ser satisfatório a sua implementação. 

Esse trabalho também trás abordagens análogas e adaptáveis para o IWPP
envolvendo tanto a execução no modo ``nativo'', por meio do uso de relaxamento
de dependência durante a expansão de vértices, quanto na execução
\textit{offload}, por meio da computação em tarefas, cujas técnicas
mostraram-se factíveis ao contexto deste trabalho.

\textbf{IWPP em \textit{Graphics Processing Units}}

\begin{figure}[h]
	\centering
	\includegraphics[width=\columnwidth]{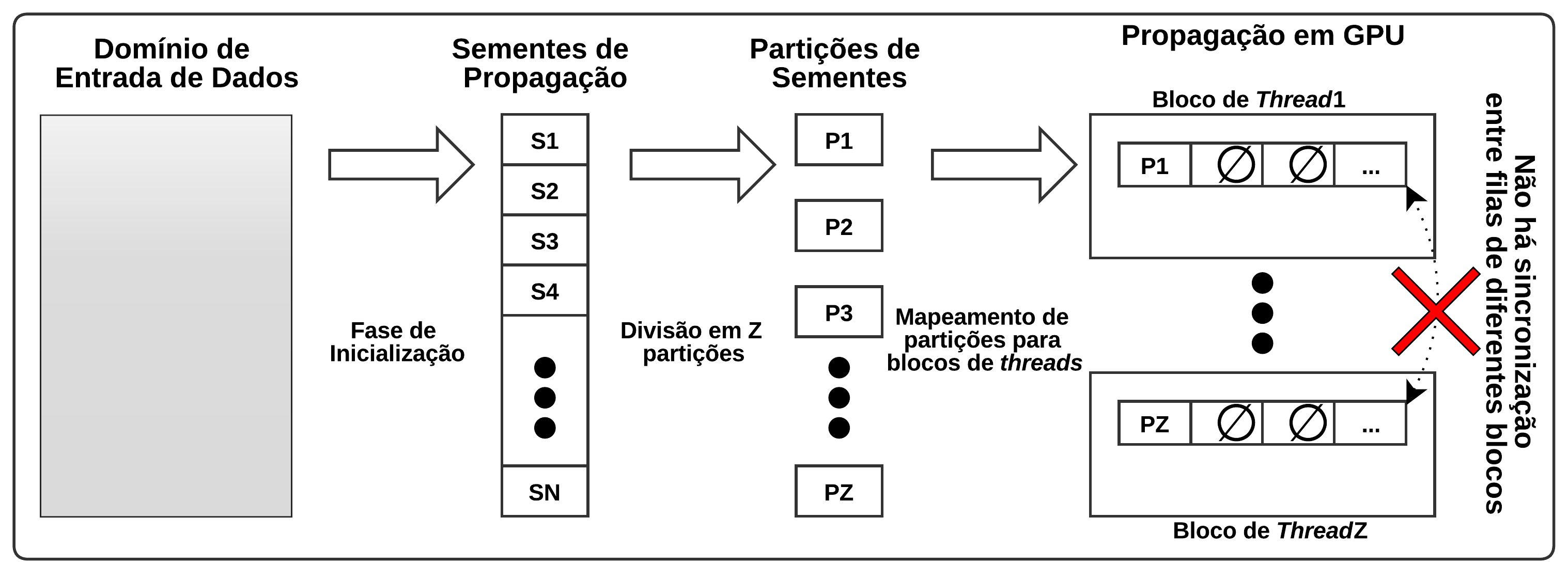}
    \caption{Visão Global da IWPP em GPU.\ Adaptado
    de~\cite{teodoro2013efficient}.}
	\label{fig:iwppgpu}
\end{figure}

Os trabalhos~\cite{teodoro2012fast,teodoro2013efficient,6569804,6267914}
apresentaram uma implementação paralela eficiente para GPUs. A implementação 
da fase de Propagação Irregular pode ser acompanhada no
Algoritmo~\ref{alg:gpuirregular} e a estratégia de implementação completa pode
ser visualizada na Figura~\ref{fig:iwppgpu}, na qual utiliza-se a seguinte
estratégia:

\begin{enumerate}
    \item Durante a inicialização, identifica os elementos ativos e separa-os
        em um conjunto $S$ de sementes;
    \item Divide o conjunto $S$ em $Z$ partições $(P_1,\dots,P_Z)$, onde cada
        partição é mapeada para uma \textit{thread} $t$ na GPU;
    \item Cria-se uma instância independente da fila hierárquica para a
        \textit{thread} $t_i$ que é inicializada com os elementos da partição
        $P_i$;
	\item Utilizam-se filas distintas de entrada e saída de pixels, como forma 
        de melhorar os tempos de leitura e sincronização.
\end{enumerate}

Como cada \textit{thread} possui sua própria fila de elementos a serem
propagados de forma independente, necessita-se que a atualização do pixel
vizinho, quando ocorrer, seja feita de forma atômica para evitar a ocorrência
de \textit{data races}, quando \textit{threads} tentam atualizar o mesmo pixel
com valores diferentes, o que poderia causar resultados não determinísticos ao
algoritmo.

Como a ordem de processamento não pode ser determinada, operações de propagação 
necessitam ser comutativas e atômicas. Para garantir a atomicidade das 
atualizações nos pixels vizinhos, é então necessário custo extra de 
sincronização, dado pela operação \textit{atomicCAS} (\textit{atomic compare 
and swap}) disponível nas placas de vídeo CUDA~\cite{wilt2013cuda} (linha 7 do
Algoritmo~\ref{alg:gpuirregular}).

\begin{algorithm}
	\label{alg:gpuirregular}
	\scriptsize
	\Enqto {fila[idBloco] != $\varnothing$}{
		\Enqto { ( p = fila[idBloco].desenfileira(...)) != $\varnothing$}{
			\ParaTodo{$q \in N_G(p)$} {
				\Repita { Verdadeiro } {
					valorCorrenteQ = I(q) \\
					\Se {CondiçãoDePropagação(I(p),valorCorrenteQ)} {
						valorAntigo = atomicCAS(\&I(q),valorCorrenteQ,op(I(p))) 
						\\
						\Se {valorAntigo == valorCorrenteQ} {
							fila[idBloco].enfileira(q) \\
							terminaLaco \\
						}
						\Senao {
							terminaLaco \\
						}
					}
				}
			}
		}
		fila[idBloco].troca\_entrada\_saida
	}
	\caption{Propagação Irregular em GPU}
\end{algorithm} 

\subsubsection{Reconstrução Morfológica}\label{subsubsec:morphrecon} 

Recentemente, a Reconstrução Morfológica vem sendo amplamente empregada nos
mais diversos tipos de aplicações. Das principais publicações nesse campo,
pode-se citar o trabalho~\cite{vincent1992morphological} que definiu
formalmente a Reconstrução Morfológica aplicada a imagens em tons de cinza. Ele
apresentou diversas aplicações para a reconstrução, outorgando a utilização da
reconstrução baseada em filas de pixels \textit{First In, First Out} (FIFO). 

Outra publicação de relevância foi a~\cite{LaurentR98} que utilizou uma
arquitetura em \textit{cluster} com memória distribuída. A ideia dessa
arquitetura é explorar paralelismo entre pixels próximos durante a
reconstrução, através de estruturas de dados hierárquicas em uma abordagem
assíncrona em imagens de tamanho pequeno. 

Em~\cite{karas2011efficient} e~\cite{jivet2008image}, apesar do foco em
arquiteturas distintas,~\cite{jivet2008image} em \textit{Field-Programmable
Gate Arrays} (FPGAs) e~\cite{karas2011efficient} em GPUs, ambos utilizaram
algoritmos paralelos cujos ganhos com paralelização não foram significativas
quando comparadas com as versões mais eficientes das implementações
sequenciais. Isso aconteceu devido a ambos os trabalhos terem paralelizado uma
versão menos eficiente da reconstrução morfológica. 

Em~\cite{teodoro2014comparative,teodoro2015performance,Teodoro27072015,6970651}
foram realizadas análises mensurando o uso de diversos algoritmos morfológicos
em CPUs, GPUs e MICs. Nesses estudos foram constatados o bom desempenho da
arquitetura \textit{Many Integrated Core} em acessos regulares de dados, porém
com o uso apenas de autovetorização por meio de anotações de diretivas no
código (\texttt{\#pragma simd}, por exemplo).

\subsubsection{Transformada de Distância Euclidiana}\label{subsubsec:TD}

A Transformada de Distância é um instrumento de suma importância no 
processamento de imagens, pois é utilizada em um extensivo número de 
aplicações. A Transformada de Distância Euclidiana possui diversas abordagens 
devido ao seu alto custo computacional e a possibilidade de busca pela 
eficiência através de soluções não determinísticas.

Uma das primeiras implementações apresentadas, equivocadamente tratada como
exata, foi em~\cite{danielsson1980euclidean}. Essa implementação gera um mapa
com os valores absolutos das coordenadas com o fundo da imagem mais próximo, ao
invés de um mapa de distâncias propriamente dito. Essa diferença difere do caso
exato devido a não conexidade\footnote{Espaço topológico que não pode ser
representado como a união de dois ou mais conjuntos abertos disjuntos e
não-vazios.} do diagrama de Voronoi no caso
discreto~\cite{zampirolli2003transformada}.  Apesar dessa inexatidão,
destacou-se por demonstrar a eficiência de algoritmos sequenciais. 

Posteriormente, diversas abordagens foram apresentadas, como
em~\cite{cuisenaire1999fast} que propõe uma Transformada de Distância exata
utilizando uma lista ordenada. Essa implementação realiza uma aproximação
rápida do resultado e utiliza vizinhanças maiores para executar possíveis
correções. 

Outro algoritmo exato utiliza o método de Meijster~\cite{meijster2000general}
que destaca-se por ser ligeiramente mais eficiente que outras abordagens, e por
possuir, também, uma implementação simples.

Implementações mais recentes em~\cite{teodoro2013efficient} mesclam o uso de
GPUs e de CPUs, por meio de uma estratégia utilizando paralelizações em nível
de tarefas, para a execução mútua nesses dispositivos. Também foram utilizadas
estruturas de dados multiníveis (filas), buscando otimizar o uso de memórias
mais rápidas. Esse estudo mostrou significativas performances, principalmente,
para GPUs, alcançando \textit{speedups} notáveis em sua execução. O uso
extensivo desses algoritmos também incluem o uso em Diagramas de
Voronoi~\cite{ye1988signed}, Esqueletos
Morfológicos~\cite{shih1995skeletonization} e
\textit{Watersheds}~\cite{vincent1991watersheds}.

\subsection{Sumário}

Nesta seção foi apresentado o padrão de computação IWPP em algoritmos
morfológicos, o qual pode ser observado em diversas soluções na literatura.
Esses algoritmos são divididos em duas fases: Identificação das Frentes de Onda
e Propagação Irregular.

Em seguida, foram apresentados casos de uso dos algoritmos de Reconstrução
Morfológica, Transformada de Distância Euclidiana e \textit{Fill Holes}
buscando evidenciar o seu funcionamento e as características de IWPP.

Por fim, foram apresentados trabalhos relacionados elencando aspectos e
resultados a serem considerados no desenvolvimento da solução e, também,
utilizados como parâmetro de comparação na apresentação dos resultados.

A próxima seção abordará uma implementação eficiente para o IWPP na
arquitetura \textit{Many Integrated Core}, com foco no
Intel\textsuperscript{\textregistered} Xeon
Phi\textsuperscript{\texttrademark}~\cite{rahman2013xeon}, explorando as
principais características dessa arquitetura.

  \section{IWPP Paralelo Eficiente na Arquitetura \textit{Many Integrated
Core}}\label{sec:IWPPphi} 

Este trabalho tem como objetivo a apresentação de uma abordagem eficiente para
algoritmos da classe \textit{Irregular Wavefront Propagation Pattern} na
arquitetura MIC/. Essa nova abordagem visa a obtenção de um algoritmo IWPP
vetorizável e paralelizável utilizando instruções SIMD/. A consequência de um
algoritmo com as duas características citadas é uma implementação sem o uso de
operações atômicas, que são um elemento limitador na eficiência de algoritmos
no Intel\textsuperscript{\textregistered} Xeon
Phi\textsuperscript{\texttrademark}, pois o mesmo não suporta instruções SIMD
atômicas~\cite{rahman2013xeon}.

\subsection{Algoritmo Proposto}\label{subsec:algoritmoproposto} 
A partir do algoritmo IWPP (mostrado no Algoritmo~\ref{alg:IWPPpadrao}), que
não é vetorizável, a solução apresentada nesta seção propõe um algoritmo com
\textit{data races} benignas~\cite{nasre2013atomic} no momento da propagação,
cuja ocorrência não altera o resultado final. Tal abordagem se dá nos seguintes
passos: 

\begin{enumerate}
    \item A fase de inicialização é similar a apresentada no algoritmo da
        Reconstrução Morfológica (Algoritmo~\ref{alg:iwpp-recon}), que na
        varredura \textit{anti-raster} os elementos são inseridos em uma fila
        chamada \emph{ondaAtual};
    \item Na fase de identificação de elementos recebedores de propagação, ao
        varrer os pixels $p$ contidos em \emph{ondaAtual}, o algoritmo irá
        inserir aqueles pixels $q \in N_G(p)$, cuja condição de propagação é
        verdadeira, em \emph{proximaOnda} (linhas 6 a 11 do
        Algoritmo~\ref{alg:proposal});
    \item Na fase de propagação será utilizada a fila \emph{proximaOnda}, que
        está populada com elementos que receberão alguma atualização, onde o
        pixel $p$ irá receber a respectiva atualização de algum pixel $q \in
        N_G(p)$ (linhas 14 a 18 do Algoritmo~\ref{alg:proposal});
    \item Os elementos que se encontram em \emph{proximaOnda} constituem o
        conjunto de elementos ativos das frentes de onda, então as filas
        \emph{proximaOnda} e \emph{ondaAtual} são trocadas (linhas 19 e 20 do
        Algoritmo~\ref{alg:proposal}) e o algoritmo continua sendo executado
        até que a mesma se encontre vazia.
\end{enumerate}

\begin{algorithm}
    \label{alg:proposal}
    \scriptsize
    \Entrada{$D$: \emph{Conjunto de elementos em um espaço multidimensional}}
    \Saida{$D$: \emph{Conjunto estável com todas as propagações realizadas}}
    \{\textbf{Inicialização}\}\\
    \dots \\
    \{\textbf{Fase de Propagação Irregular}\}\\
    \Enqto{$ondaAtual.vazia() =$ falso} {
        \{\textbf{Identificação de Elementos Recebedores de Propagação}\}\\
        \ParaTodo{$p \in ondaAtual$} {
            \ParaTodo{$q \in N_G(p)$} {
                \Se{$condiçãoDePropagação$($D(q), D(p)$) $=$ verdadeiro} {
                    $proximaOnda.insere(q)$\\
                }
            }
        }
    \{\textbf{Propagação}\}\\
    \ParaTodo{$p \in proximaOnda$} {
        \ParaTodo{$q \in N_G(p)$} {
            $D(p) \leftarrow Max$/$Min(D(q), D(p))$\\
        }
    }
    $ondaAtual \leftarrow proximaOnda$\\
    $proximaOnda \leftarrow \varnothing$\\
}
\caption{Novo IWPP proposto não-vetorizado}
\end{algorithm} 

\begin{figure}[!htb]
\hspace*{-3mm}
\begin{minipage}{.5\textwidth}
  \centering
  \includegraphics[width=.5\linewidth]{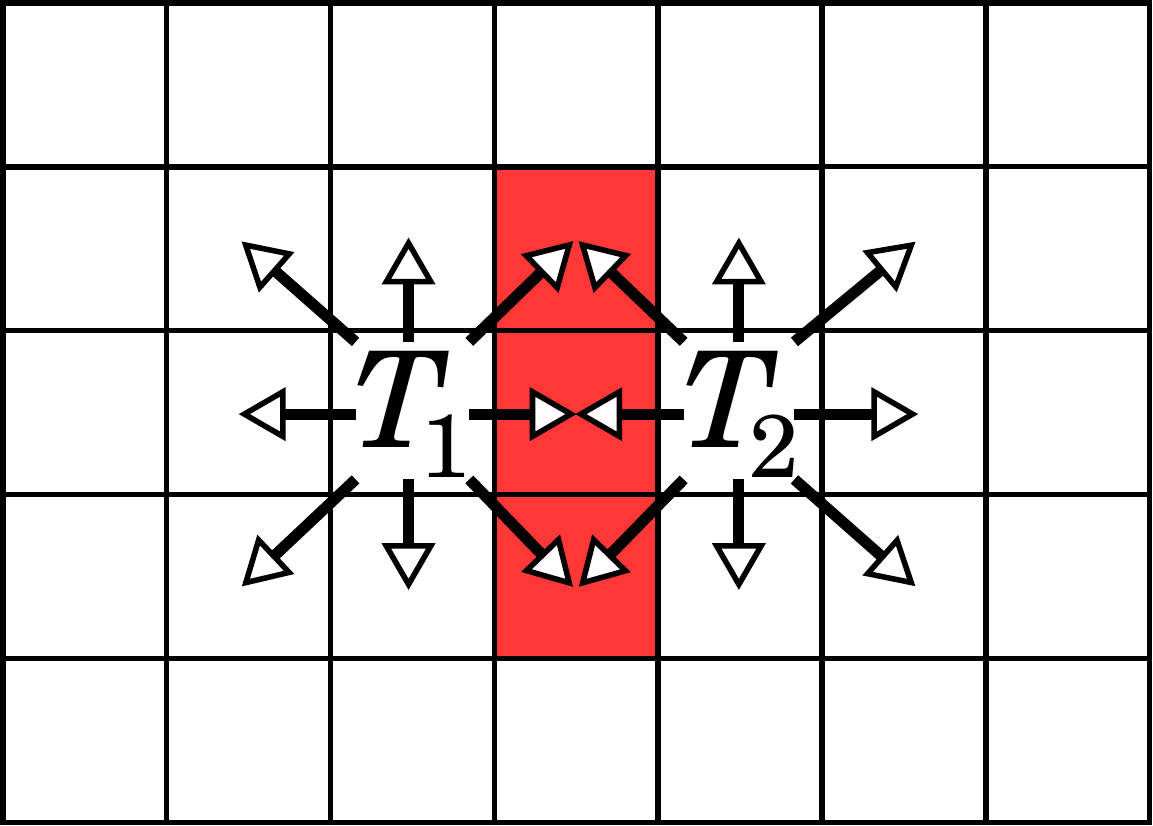}
  \caption{Exemplo de Propagação Paralela Oriunda de Elementos Diferentes.}
  \label{fig:corrida1}
\end{minipage}
\hspace*{3mm}
\begin{minipage}{.5\textwidth}
  \centering
  \includegraphics[width=.5\linewidth]{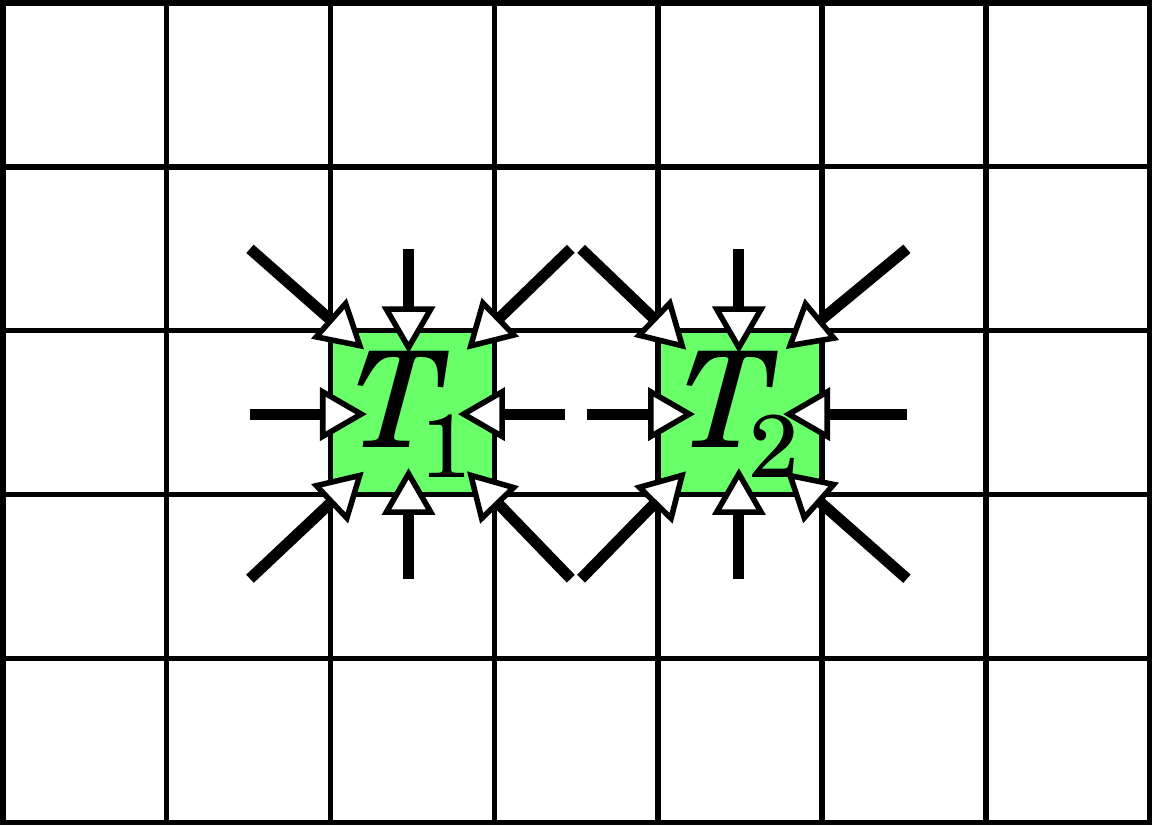}
  \caption{Exemplo de Propagação Paralela Utilizando a Abordagem Proposta.}
  \label{fig:corrida2}
\end{minipage}
\end{figure}

As modificações propostas nesse algoritmo objetivam que elementos alterem
unicamente seus próprios valores na Fase de Propagação Irregular. Na
implementação apresentada no Algoritmo~\ref{alg:iwpp-recon}, um elemento pode
modificar a sua vizinhança, e a versão paralela desse algoritmo exige o uso de
operações atômicas durante a verificação da condição de propagação e a
propagação, pois \textit{threads} com elementos de posições diferentes e
valores diferentes podem tentar modificar uma mesma posição (vizinho) ao mesmo
tempo, ocasionando uma condição de corrida (ou \textit{data race} prejudicial).
Um exemplo dessa possibilidade de modificação de uma determinada posição por
elementos diferentes, inseridos na estrutura de dados de \textit{threads}
diferentes pode ser visualizado na Figura~\ref{fig:corrida1}. Já a abordagem
proposta neste trabalho visa eliminar esse caso de escrita de valores
diferentes em uma mesma posição, fazendo com que cada elemento modifique apenas
o seu próprio conteúdo, como pode ser visualizado no exemplo da
Figura~\ref{fig:corrida2}.

\subsection{Implementação Vetorial do Algoritmo Proposto}\label{subsec:implementacaovetorial}

Utilizando a abordagem apresentada na seção anterior, uma variante vetorial
pode ser implementada (conforme apresentado no Algoritmo~\ref{alg:vectorized}).
As fases desse algoritmo, identificação de elementos recebedores de propagação
e propagação, são detalhadas nas próximas seções.

\begin{algorithm}
    \label{alg:vectorized}
    \scriptsize
    \Entrada{$D$: \emph{Conjunto de elementos em um espaço multidimensional}}
    \Saida{$D$: \emph{Conjunto estável com todas as propagações realizadas}}
    $vet_{shift} \leftarrow$ \emph{Constante de distâncias de endereços da vizinhança}\\
    \{\textbf{Inicialização}\}\\
    \dots\\
    \{\textbf{Escaneamento}\}\\
    \dots\\
    \{\textbf{Fase \textit{wavefront propagation}}\}\\
    \Enqto{$ondaAtual.vazia() == falso$} {
        \{\textbf{Identificação de Elementos Recebedores de Propagação}\}\\
        \ParaTodo{$p \in ondaAtual$} {
            $vet_p \leftarrow $ Extração de elementos ativos\\
            $vet_{enderecos} \leftarrow vetAdd(vet_p, vet_{shift})$\\
            $vet_{vizinhos} \leftarrow gather(d, vet_{enderecos})$\\
            $mascara_{condicao} \leftarrow vetorCondicaoDePropagacao(vet_p,vet_{vizinhos})$\\
            $vet_{prefixsum} \leftarrow prefixSum(mascara_{condicao})$\\
            $ondaproxima.insere(vet_{enderecos}, mascara_{condicao},vet_{prefixSum})$\\
        }
        \{\textbf{Propagação}\}\\
        \ParaTodo{$q \in proximaOnda$} {
            $vet_q \leftarrow $ Extração de elementos\\
            $vet_{enderecos} \leftarrow vetAdd(vet_q, vet_{shift})$\\
            $vet_{vizinhos} \leftarrow gather(d, vet_{enderecos})$\\
            $d(q) \leftarrow (Max/Min)Reducao(vet_q, vet_{vizinhos})$\\
        }
        $ondaAtual \leftarrow proximaOnda$\\
        $proximaOnda \leftarrow \varnothing$\\
    }
\caption{IWPP vetorizado}
\end{algorithm}

\subsubsection{Identificação de Elementos Recebedores de
Propagação}\label{subsubsec:insercaovetorialfila}

Esta fase é dividida em três passos principais, os quais são: Leitura de Pixels
Vizinhos (linhas 10 a 12), Contagem de Elementos e \textit{Prefix sum} (linha
13), e Inserção Vetorial na Fila (linhas 14 e 15).

\textbf{Passo 1: Leitura de Pixels Vizinhos}

Para um pixel $p$ retirado da fila de entrada, é realizado a verificação da
condição de propagação com cada um dos pixels de sua vizinhança. A leitura dos
pixels vizinhos $q$ vetorial é feita pelo seguinte raciocínio:
\begin{itemize}
    \item Dado o endereço na imagem da posição do pixel $p$, retirado da fila
        de entrada, as distâncias entre os pixels $q \in N_G(p)$ em relação a
        $p$ são constantes. Com isso, é possível calcular a posição da
        vizinhança de $p$ realizando a soma de seu endereço ($vet_p$) com um
        vetor\footnote{Vetor indica um registrador vetorial de 512 bits, para
        fins de simplificação.} $vet_{shift}$ de deslocamentos referentes à
        distância de $q$ e seus vizinhos (linha 11). O resultado nos dá um
        vetor $vet_{enderecos}$ com os endereços dos pixels vizinhos a serem
        lidos.
    \item De posse do vetor $vet_{enderecos}$ contendo os endereços a serem
        buscados na imagem, o Intel\textsuperscript{\textregistered} Xeon
        Phi\textsuperscript{\texttrademark} dispõe de uma instrução de leitura
        chamada \textit{gather} que carrega até 16 (dezesseis) inteiros de
        posições irregulares, como pode ser observado na
        Figura~\ref{fig:gather}. Dessa forma, é possível carregar eficientemente
        os vizinhos de $p$ da imagem máscara $I$ e da imagem marcadora $J$
        utilizando essa instrução.

\begin{figure}[t]
	\centering
	\includegraphics[width=\columnwidth]{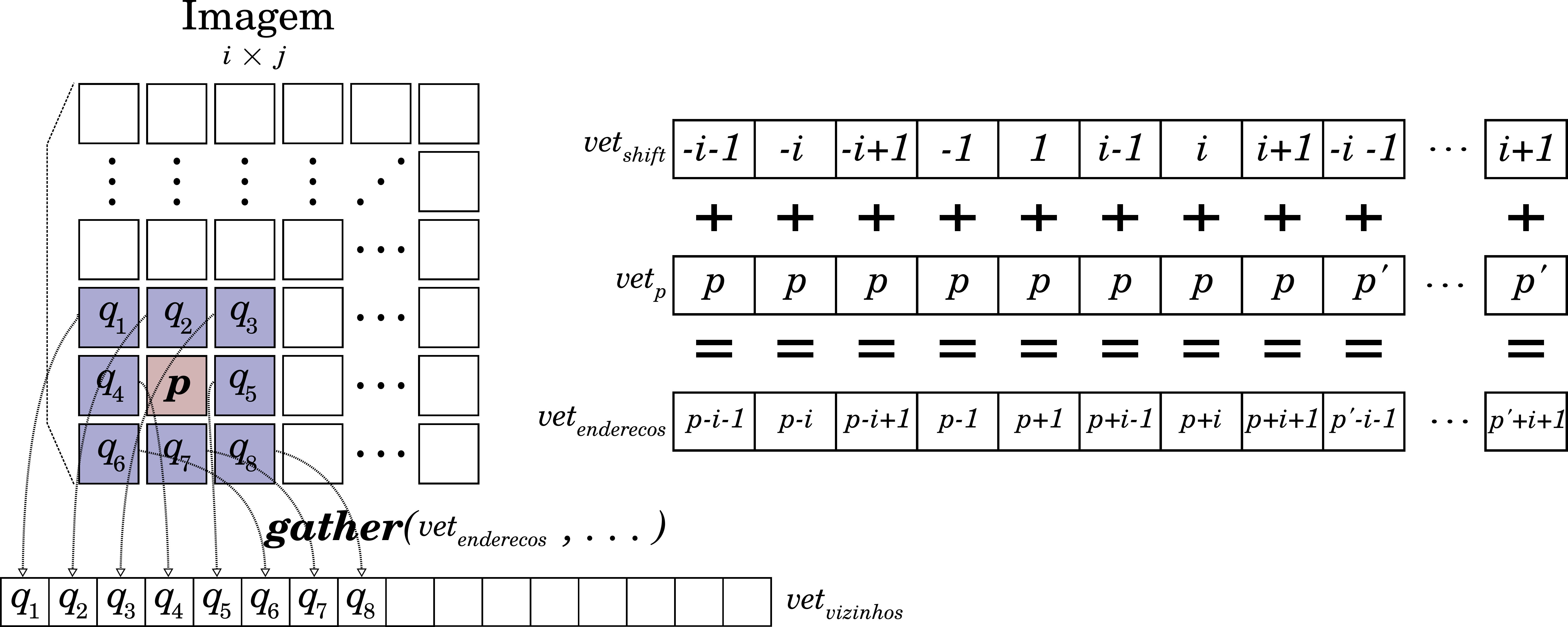}
	\caption{Cálculo de Endereços e Operação \textit{Gather} de um Pixel $p$.}
	\label{fig:gather}
\end{figure}
\end{itemize}

\begin{figure}[!htb]
	\centering
	\includegraphics[width=.7\columnwidth]{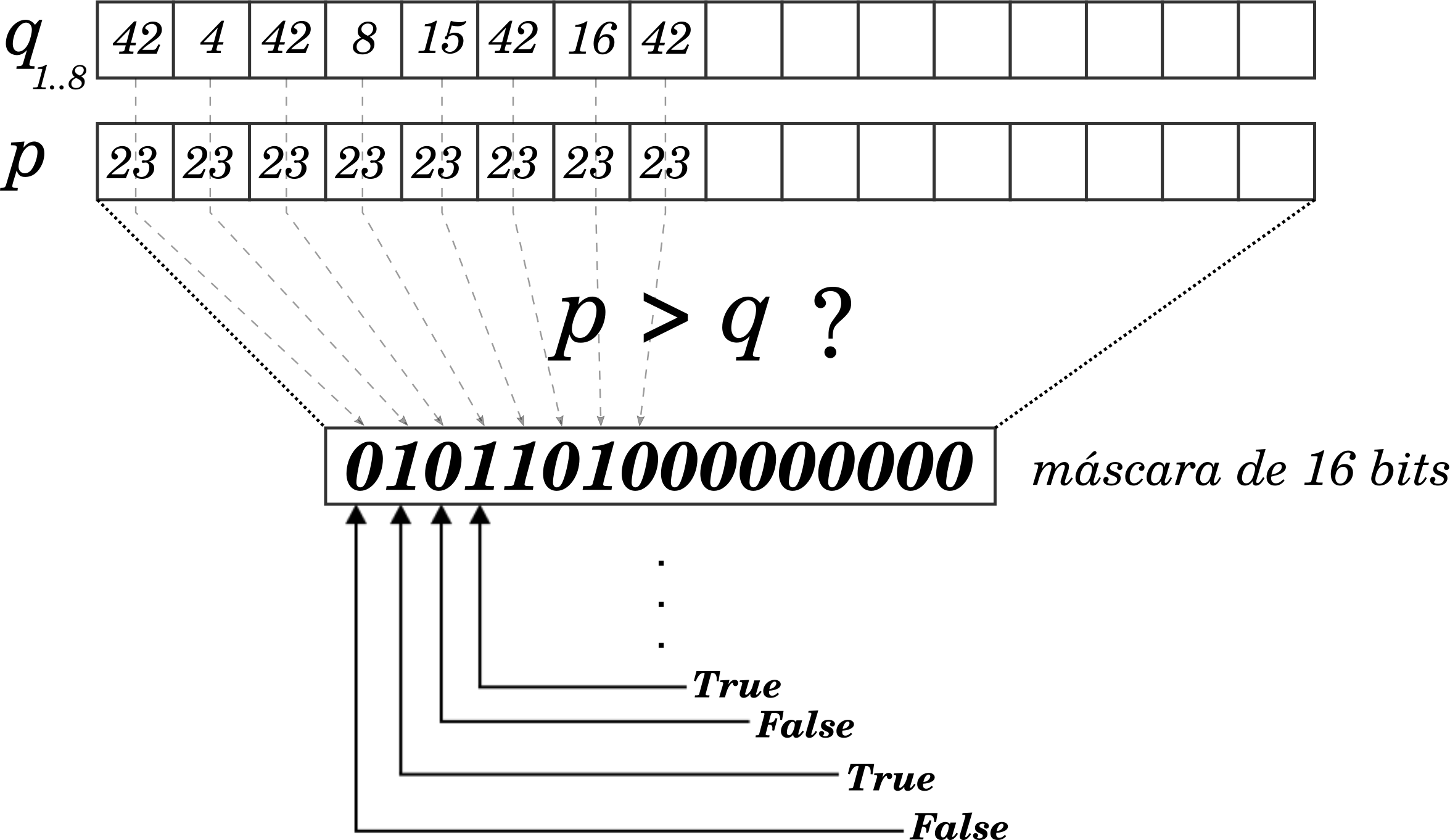}
	\caption{Verificação da Condição de Propagação Vetorial.}
	\label{fig:condicaopropagacao}
\end{figure}

A partir dos pixels carregados para o vetor, são feitas as respectivas
verificações acerca da condição de propagação.  Cabe ressaltar nesta seção que
operações de comparação vetoriais geram máscaras binárias com valor 1 (um) para
posições onde a condição de propagação é verdadeira
(Figura~\ref{fig:condicaopropagacao}), que são utilizadas na contagem de
elementos no passo de inserção na fila. 

\textbf{Passo 2: Contagem de Elementos e \textit{Prefix Sum}}

Após descobertos os elementos que deverão ser inseridos na fila, por meio da
verificação da condição de propagação e a máscara de 16 bits gerada como
resposta, necessita-se inserir na fila somente os elementos que possuem o valor
1 na sua posição da máscara de resposta. Esta etapa é executada em dois passos:
a contagem dos elementos e o cálculo do vetor \textit{prefix sum}.

Para a contagem de elementos, utiliza-se uma instrução
\textit{popcnt}~\cite{rahman2013xeon} em que, dado um valor inteiro como
entrada (a máscara de resposta da verificação da condição de propagação), ela
retorna a quantidade de bits cujo valor é $1$ em sua representação binária.

Já o \textit{Prefix sum} é a soma cumulativa dos prefixos $\{y_0, y_1, y_2,
\dots\}$ de uma determinada entrada $\{x_0, x_1, x_2,
\dots\}$~\cite{blelloch1990prefix}. Por meio desse vetor de prefixos e a
máscara de respostas gerada pela verificação da condição de propagação é
possível identificar e inserir esses elementos na fila.

A implementação utiliza um vetor de tamanho fixo (512 bits), que é suficiente
para operar cada um dos bits da máscara recebida. A estratégia é executada nos
seguintes passos: 

\begin{enumerate}
    \item Dado uma máscara de 16 bits, cada bit dessa máscara é distribuído
        para cada uma das $16$ posições do registrador vetorial $vet_{r1}$ de
        $512$ bits (Figura~\ref{fig:prefixexpansao}).

\begin{figure}[!htb]
	\centering
	\includegraphics[width=.75\columnwidth]{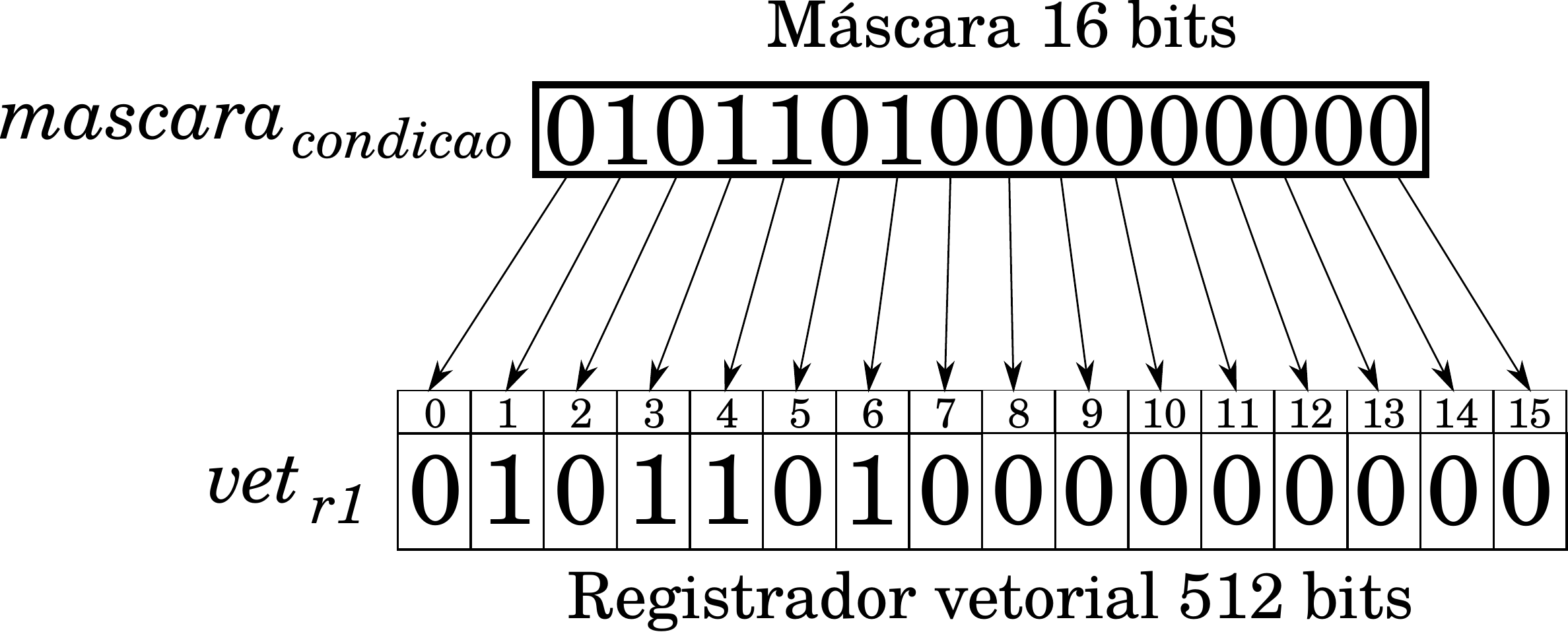}
    \caption{Exemplo de Distribuição de Máscara em Registrador Vetorial.}
	\label{fig:prefixexpansao}
\end{figure}

    \item O registrador $vet_{r2}$ recebe uma permutação dos valores de
        $vet_{r1}$, deslocando todos os elementos em uma posição à direita
        (Figura~\ref{fig:perm01}). É então realizada a soma de $vet_{r1}$ e
        $vet_{r2}$ (Figura~\ref{fig:soma01}). O resultado dessa soma
        elemento-a-elemento é salvo no próprio $vet_{r1}$.

\begin{figure}[!htb]
\begin{minipage}{.5\textwidth}
  \centering
  \includegraphics[width=1\linewidth]{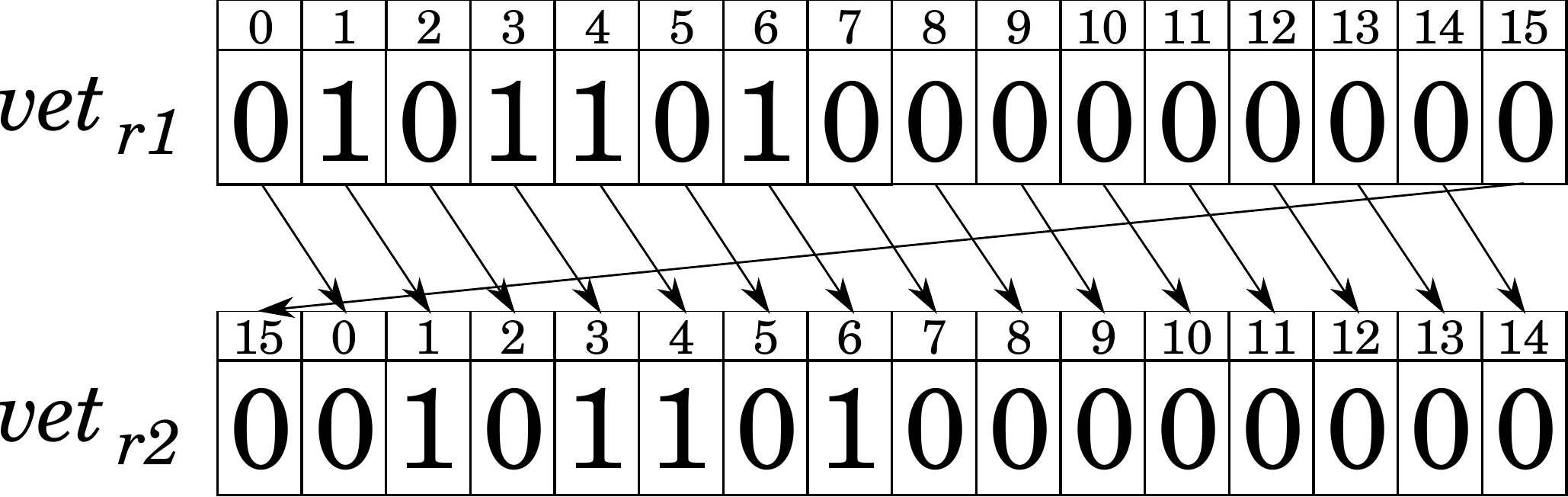}
  \caption{Permutação 01.}
  \label{fig:perm01}
\end{minipage}%
\begin{minipage}{.5\textwidth}
  \centering
  \includegraphics[width=1\linewidth]{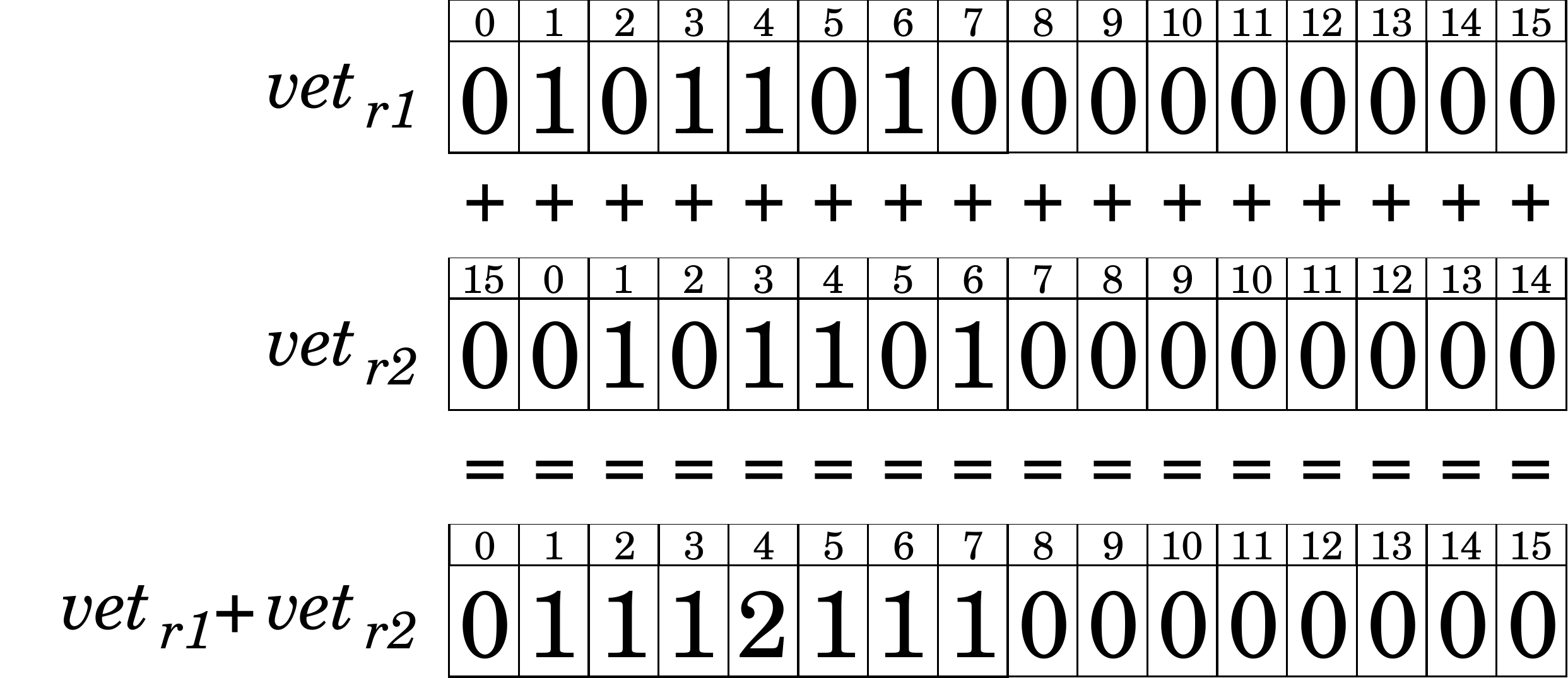}
  \caption{Soma Vetorial 01.}
  \label{fig:soma01}
\end{minipage}
\label{fig:prefix01}
\end{figure}

\item Para completar o cálculo do \textit{prefix sum} nas oito primeiras
    posições do vetor, são aplicadas mais duas vezes o passo anterior, como
    única ressalva de modificação das permutações dos valores. Em cada passo
    serão permutadas duas (Figuras~\ref{fig:perm02} e~\ref{fig:soma02}) e
    quatro posições (Figuras~\ref{fig:perm04} e~\ref{fig:soma04}) à direita.

\begin{figure}[!htb]
\centering
\begin{minipage}{.5\textwidth}
  \centering
  \includegraphics[width=1\linewidth]{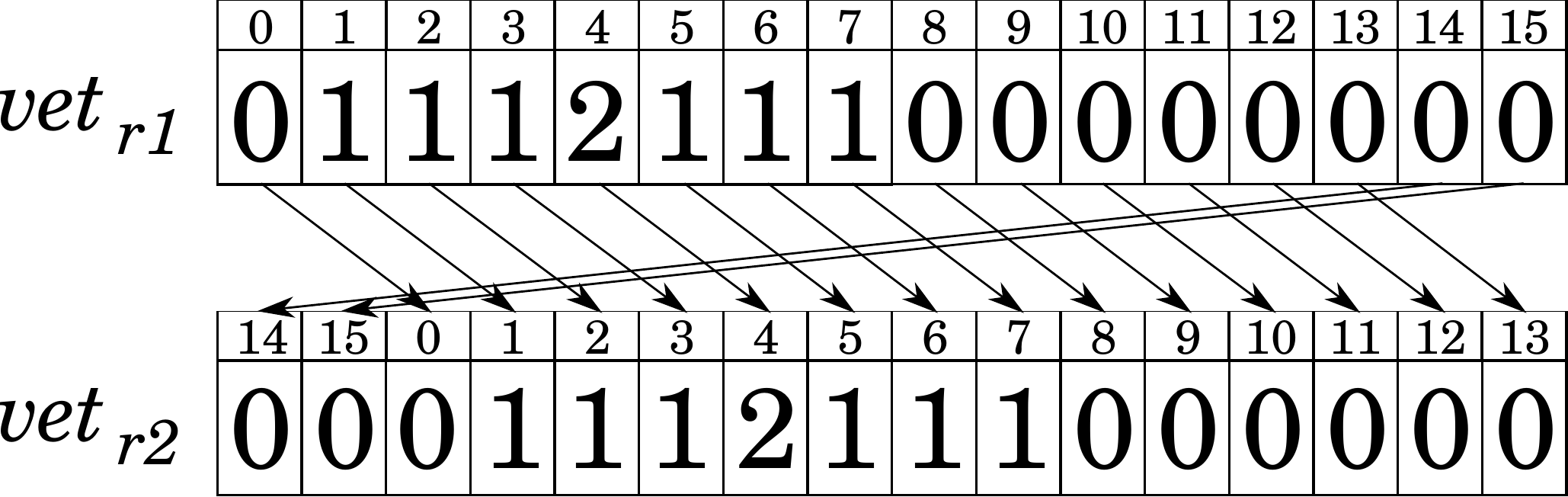}
  \caption{Permutação 02.}
  \label{fig:perm02}
\end{minipage}%
\begin{minipage}{.5\textwidth}
  \centering
  \includegraphics[width=1\linewidth]{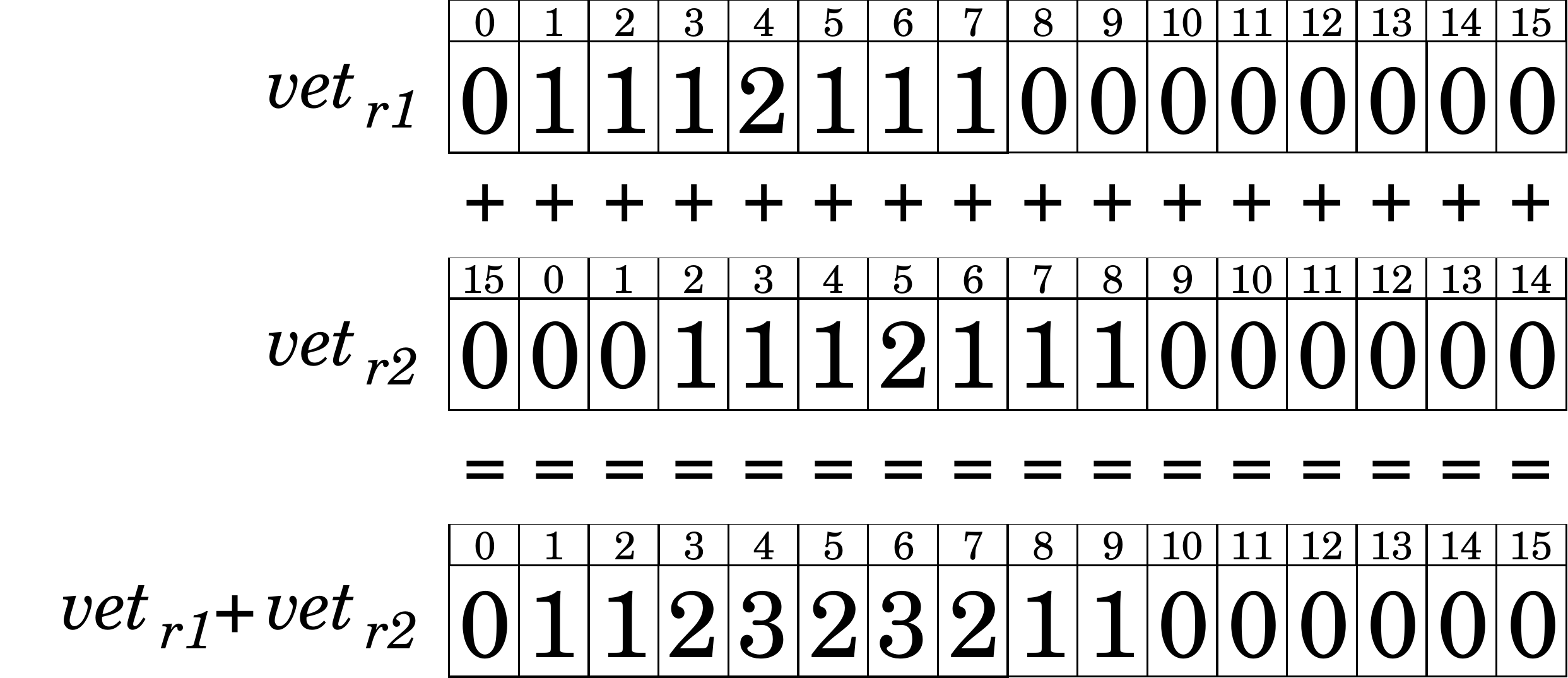}
  \caption{Soma Vetorial 02.}
  \label{fig:soma02}
\end{minipage}
\end{figure}
\begin{figure}[!htb]
\begin{minipage}{.5\textwidth}
  \centering
  \includegraphics[width=1\linewidth]{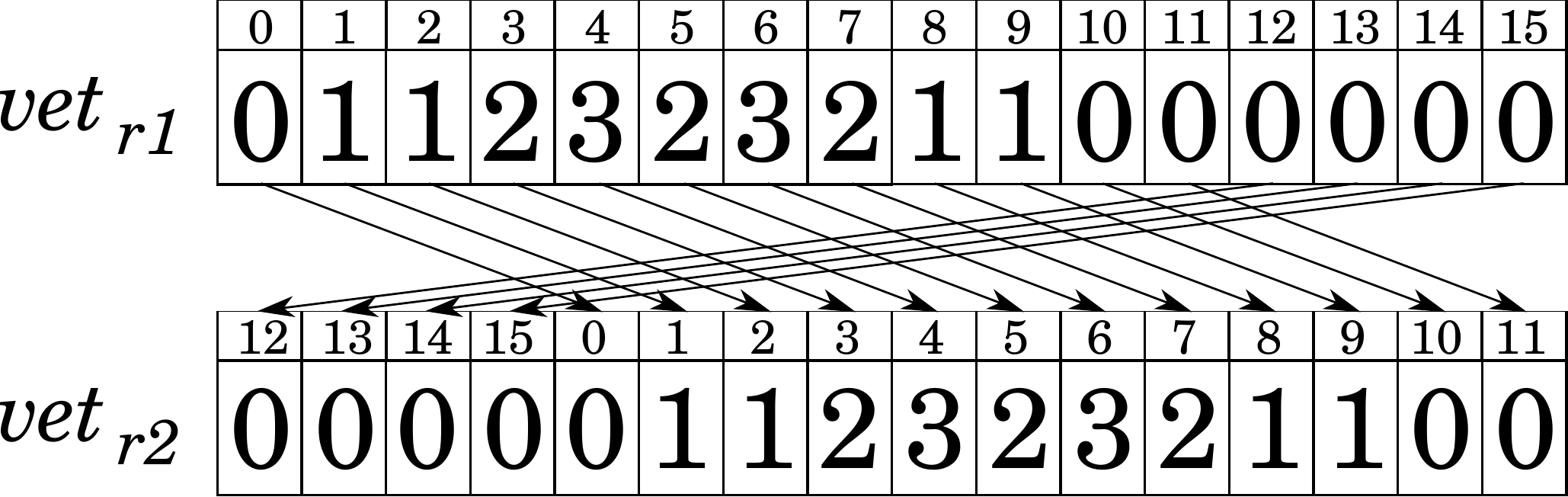}
  \caption{Permutação 03.}
  \label{fig:perm04}
\end{minipage}%
\begin{minipage}{.5\textwidth}
  \centering
  \includegraphics[width=1\linewidth]{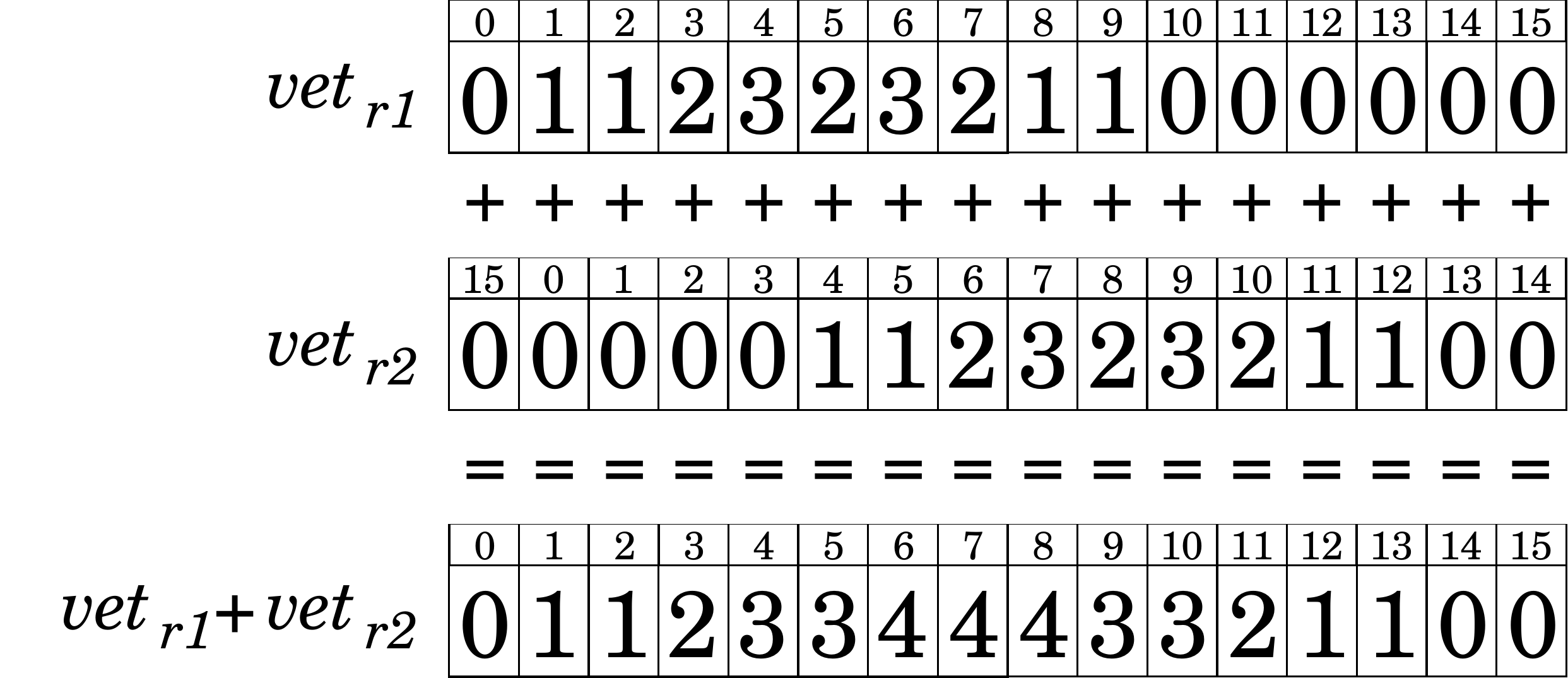}
  \caption{Soma Vetorial 03.}
  \label{fig:soma04}
\end{minipage}
\end{figure}
\item Os valores contidos na primeira metade do registrador $vet_{r1}$ são o 
\textit{prefix sum}, referente a máscara de entrada, e seu conteúdo é retornado 
como resultado, conforme apresentado na Figura~\ref{fig:prefixresultado}.
\begin{figure}[!htb]
	\centering
	\includegraphics[width=.75\linewidth]{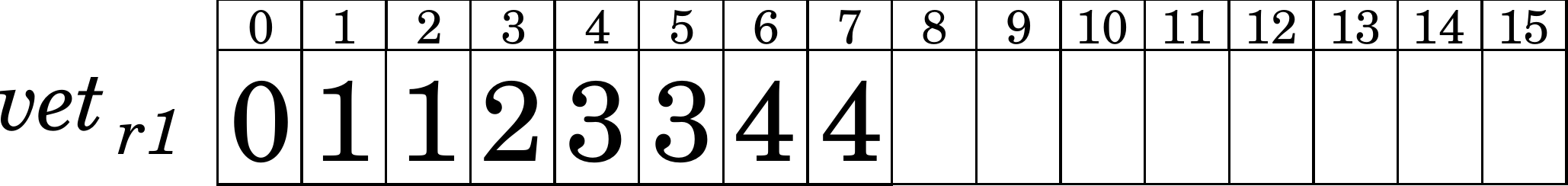}
	\caption{Registrador Retornado como Resposta.}
	\label{fig:prefixresultado}
\end{figure}
\end{enumerate}

Há, na literatura~\cite{harris2007parallel}, implementações do \textit{prefix
sum} que exploram eficientemente o processamento dessa operação para grandes
entradas de dados. Porém, como o tamanho da entrada é um valor fixo (512 bits),
essas versões \textit{work-efficient}, que foram concebidas para grandes
entradas de dados, não melhoram o desempenho nesse caso. Dessa forma, é
utilizada uma adaptação da versão paralela~\cite{harris2007parallel} para
funcionar de forma vetorial e operando com os pixels vizinhos $G$ de um pixel.

\textbf{Passo 3: Inserção Vetorial na Fila}

\begin{figure}[h]
	\centering
	\includegraphics[width=\columnwidth]{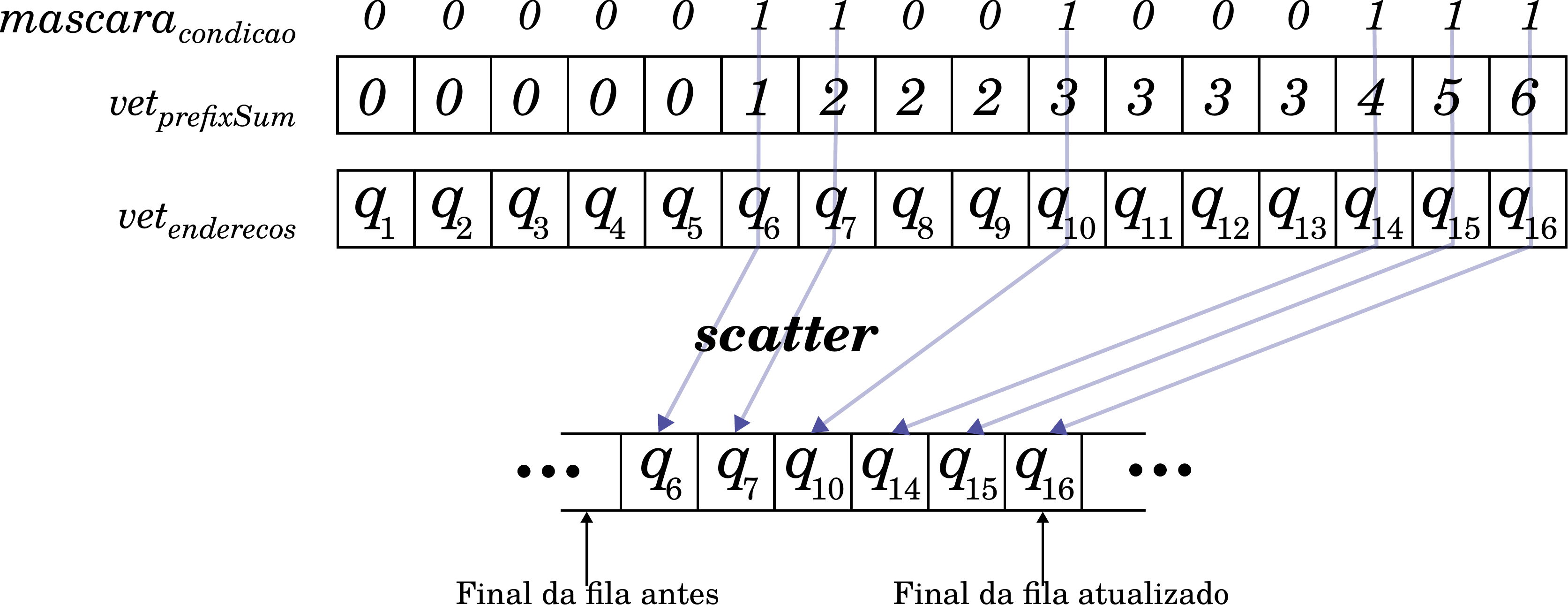}
	\caption{\textit{Scatter} e Atualização da Quantidade de Elementos.}
\label{fig:scatter}
\end{figure}

O último passo da Identificação de Elementos Recebedores de Propagação integra
na função \textit{scatter} a utilização do vetor de endereços calculados, a
máscara de valores a serem inseridos na fila de saída e o vetor contendo o
\textit{prefix sum} das posições a serem inseridas na fila (linha 15 do
Algoritmo~\ref{alg:vectorized}). Para cada posição da máscara cujo valor é $1$,
o elemento correspondente a mesma posição no vetor $vet_{enderecos}$ é inserido
ao final da fila, na posição apontada pelo vetor $vet_{\textit{prefixSum}}$,
como pode ser visualizado na Figura~\ref{fig:scatter}.

Após a inserção, a variável responsável por controlar o final da fila é, então,
acrescida da quantidade de elementos adicionados.

\subsubsection{Propagação}\label{subsubsec:propagacao}

Durante a fase de Propagação, o pixel $q$ verifica qual pixel $r \in N_G(q)$
ele irá receber a atualização. Nessa estratégia, a função verifica pela
atualização de dois pixels simultaneamente, como forma de aproveitar a largura
do vetor disponível. Sua execução ocorre nos seguintes passos:

\begin{itemize}
    \item São consumidos da fila de saída dois pixels $p_1$ e $p_2$;
    \item Utilizando o mesmo método apresentado na
        Seção~\nameref{subsubsec:insercaovetorialfila}, os vizinhos $r_{pn}$, onde $n$
        varia de $\{1..8\}$ para cada um dos elementos de uma determinada
        vizinhança, são carregados para um registrador vetorial
        (\textit{gather}), de forma que a vizinhança de cada um dos pixels
        ocupe metade do vetor com a seguinte organização:
\begin{center}
$ [r_{11}\quad r_{12}\quad r_{13}\quad r_{14}\quad r_{15}\quad 
r_{16}\quad r_{17}\quad r_{18}\quad r_{21}\quad r_{22}\quad r_{23}\quad 
r_{24}\quad r_{25}\quad r_{26}\quad r_{27}\quad r_{28}] $
\end{center}

    \item São geradas máscaras para cada uma das vizinhanças $q$ de um pixel
        $p$, para os pixels $q$ cuja condição de propagação é verdadeira;

    \item A partir dessas máscaras, é calculado o valor de atualização a partir
        da vizinhança, sendo ele o pixel $r$ que irá propagar a $q$. As
        atualizações são feitas individualmente para cada um dos pixels;

    \item O procedimento continua até que toda a fila de saída seja consumida.

\end{itemize}

Durante a execução da Propagação, existe a possibilidade de um mesmo elemento
$q$ ser inserido diversas vezes em uma mesma iteração da fila
\emph{proximaOnda}. Quando isso acontece, é preciso avaliar um cenário
específico no qual o mesmo pixel é processado concorrentemente, ou seja $p_1$ é
igual a $p_2$, ocasionando um \textit{data race} no momento da atualização em
$D(q)$. Contudo, para o processamento desses dois elementos - que são a réplica
de um único elemento - ocorrer, as vizinhanças desses dois elementos serão
lidas por uma única instrução \textit{gather}, fazendo com que as duas
vizinhanças possuam sempre os mesmos elementos lidos. Dessa forma, este cenário
gera um \textit{data race} benigno e não afeta o resultado da
aplicação~\cite{narayanasamy2007automatically}.

\subsection{Implementação Paralela do Algoritmo Proposto}\label{subsec:implementacaoparalela}

A partir da versão vetorizada (apresentado no Algoritmo~\ref{alg:vectorized}),
a estratégia de paralelização se aplica diretamente no processamento
independente dos elementos ativos que serão inseridos na fila, após as
varreduras \textit{raster} e \textit{anti-raster}. 

\begin{figure}[!htb]
	\centering
	\includegraphics[width=.95\linewidth]{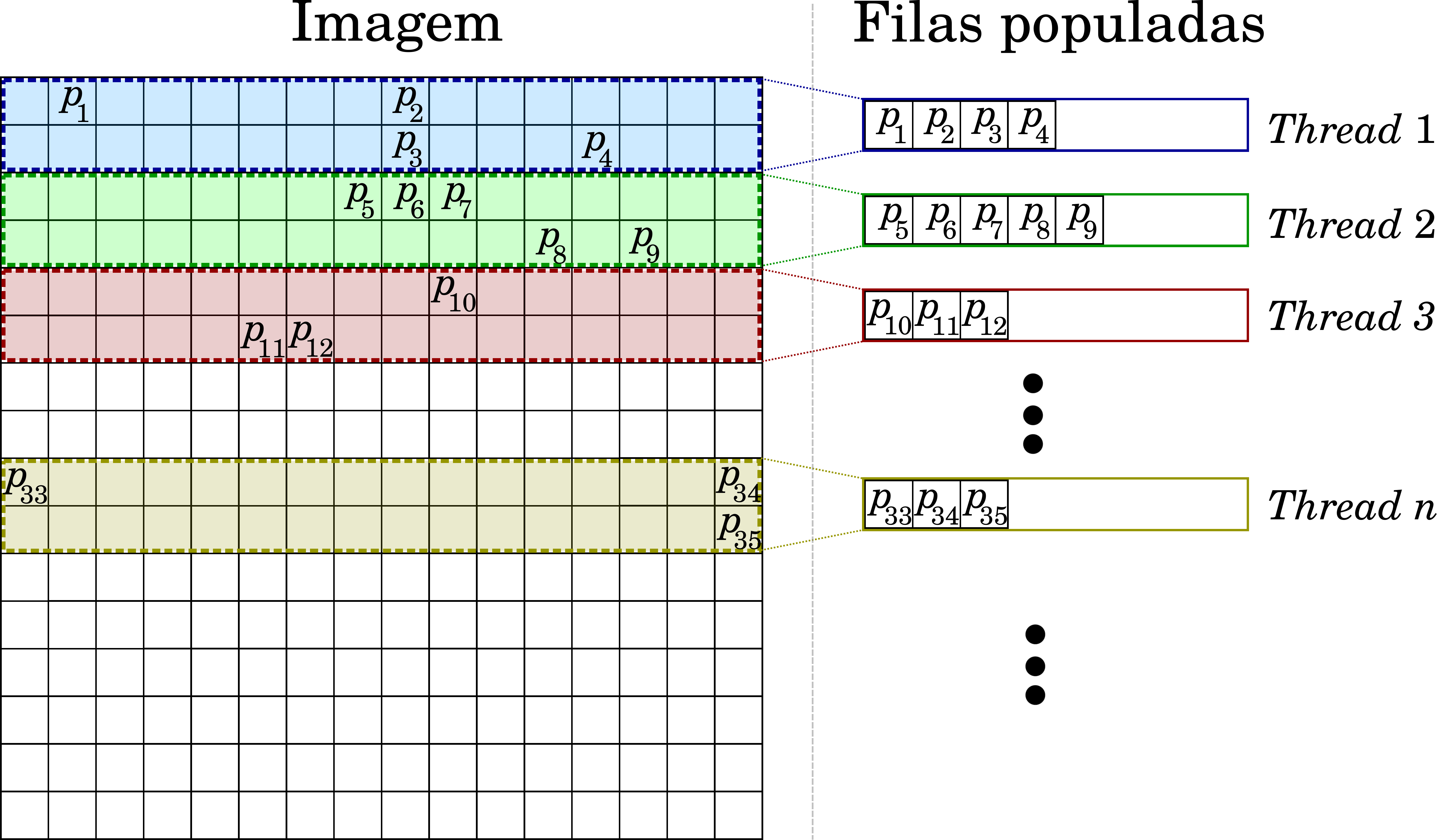}
	\caption{Divisão de Elementos Ativos entre \textit{Threads}.}
\label{fig:threadsfila}
\end{figure}

A última das varreduras realiza a distribuição das iterações, a partir das
linhas da imagem, entre as \textit{threads} disponíveis. Cada uma das
\textit{threads} irá percorrer um determinado segmento da imagem, identificando
elementos ativos que formarão as ondas de propagação de cada \textit{thread},
como pode ser visto na Figura~\ref{fig:threadsfila}. 

Assim, nessa modificação, cada \textit{thread} passa a possuir suas próprias
filas (ondaAtual e ondaProxima) a serem utilizadas na fase de Propagação com um
conjunto específico de frentes de onda (sementes) e, a partir deles, serão
realizadas as propagações de forma independente entre as \textit{threads},
utilizando as estratégias já descritas na
Seção~\ref{subsec:implementacaovetorial}. 

Apesar das ondas de propagação iniciais de cada \textit{thread} serem
independentes e iniciarem em posições distintas, durante o processamento (Fase
de Propagação Irregular) as frentes de onda de diferentes \textit{threads}
podem cruzar umas com as outras. Como consequência, é possível que diferentes
\textit{threads} insiram um mesmo elemento $q$ em seus conjuntos distintos de
elementos ativos. Neste caso, um \textit{data race} pode ocorrer na atualização
de $D(q)$ por múltiplas \textit{threads}. 

\begin{figure}[!htb]
	\centering
	\includegraphics[width=.95\linewidth]{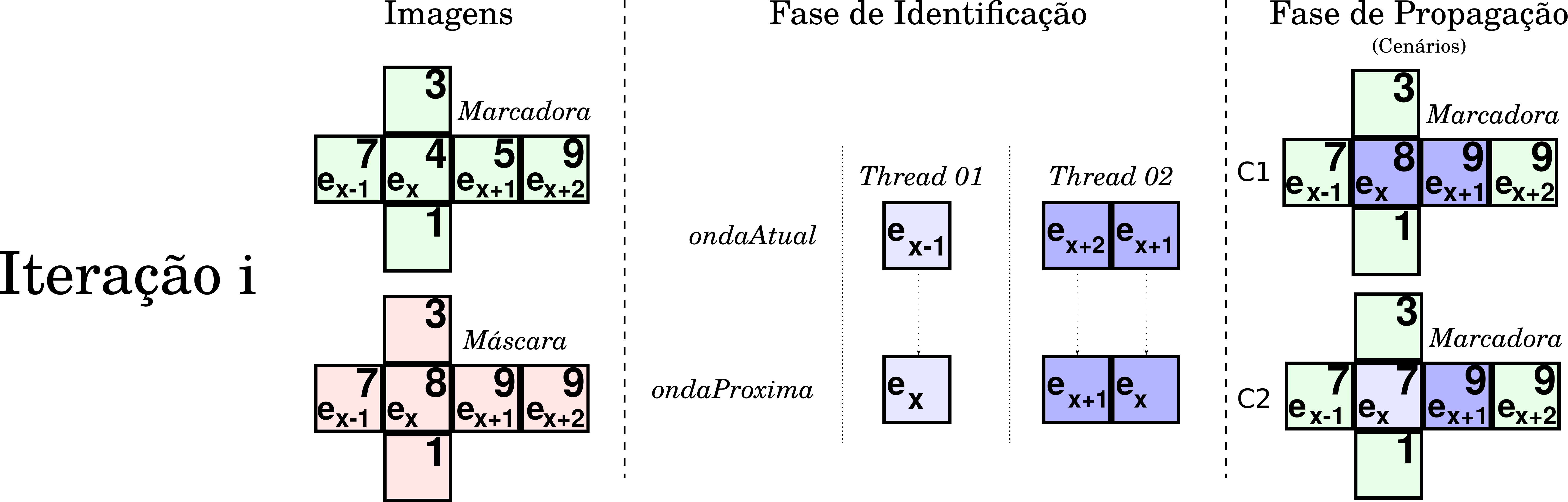}
    \caption{Exemplo de Identificação de Elementos Iguais por \textit{Threads}
    Diferentes.}
\label{fig:racecondition01}
\end{figure}

A consequência do caso descrito no parágrafo anterior é exemplificado na
Figura~\ref{fig:racecondition01} para a Reconstrução Morfológica. Neste
exemplo, a \textit{Thread} $01$ possui o elemento $e_{x-1}$ em sua fila
\emph{ondaAtual} na iteração $i$, enquanto a \textit{Thread} $02$ possui os
elementos $e_{x+2}$ e $e_{x+1}$. É notável que na iteração $i$ ambas as
\textit{threads} possuem elementos que satisfazem a condição de propagação e
irão inserir o elemento $e_{x}$ e suas respectivas filas \emph{proximaOnda}, na
Fase de Identificação. Dada essa possibilidade, durante a Fase de Propagação, a
leitura da vizinhança de $e_x$ pode desencadear dois cenários ($C1$ e $C2$)
distintos.

\begin{figure}[!htb]
	\centering
	\includegraphics[width=.95\linewidth]{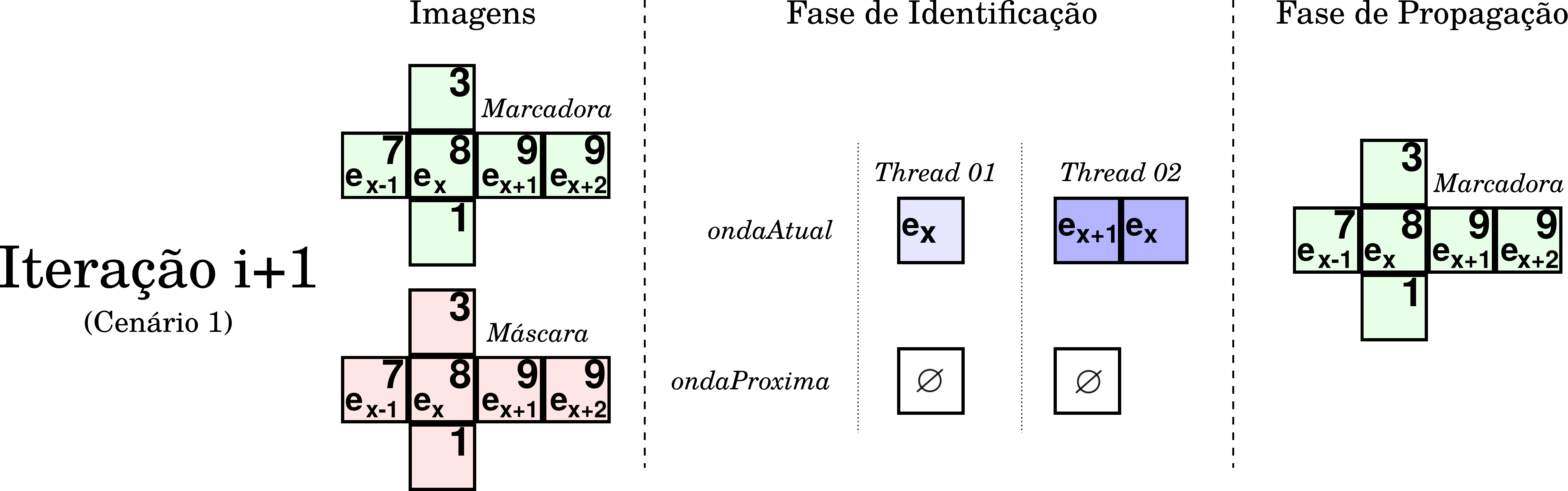}
    \caption{Cenário 1: Estabilidade Alcançada na Iteração $i$.}
\label{fig:racecondition02}
\end{figure}

No primeiro cenário ($C1$), a \textit{thread} 01 verifica a vizinhança de $e_x$
somente após a \textit{thread} 02 escrever a propagação do elemento $e_{x+1}$
e, nesse caso, a estabilidade será alcançada já na iteração $i$ onde os valores
dos elementos serão finais (Figura~\ref{fig:racecondition02}).

No segundo cenário ($C2$), a execução ocorre na seguinte ordem: 

\begin{itemize}
    \item A \textit{thread} 01 realiza a leitura da vizinhança de $e_x$;
    \item A \textit{thread} 02 processa leitura da vizinhança e propagação dos
        elementos $e_{x+1}$ e $e_{x}$;
    \item A \textit{thread} 01 realiza a propagação de $e_x$.
\end{itemize}

Neste cenário, o maior valor da vizinhança lido pela \textit{thread} 01 ($7$) é
menor do que o maior valor da vizinhança lido pela \textit{thread} 02
(9)\footnote{O valor é atualizado para 8 devido aos limites de reconstrução dados
pela Máscara.}. Dessa forma, mesmo possuindo um valor maior após a
propagação executada pela \textit{thread} 02, $e_x$ irá terminar a iteração com
um valor menor oriundo da leitura da \textit{thread} 01.

\begin{figure}[!htb]
	\centering
	\includegraphics[width=.95\linewidth]{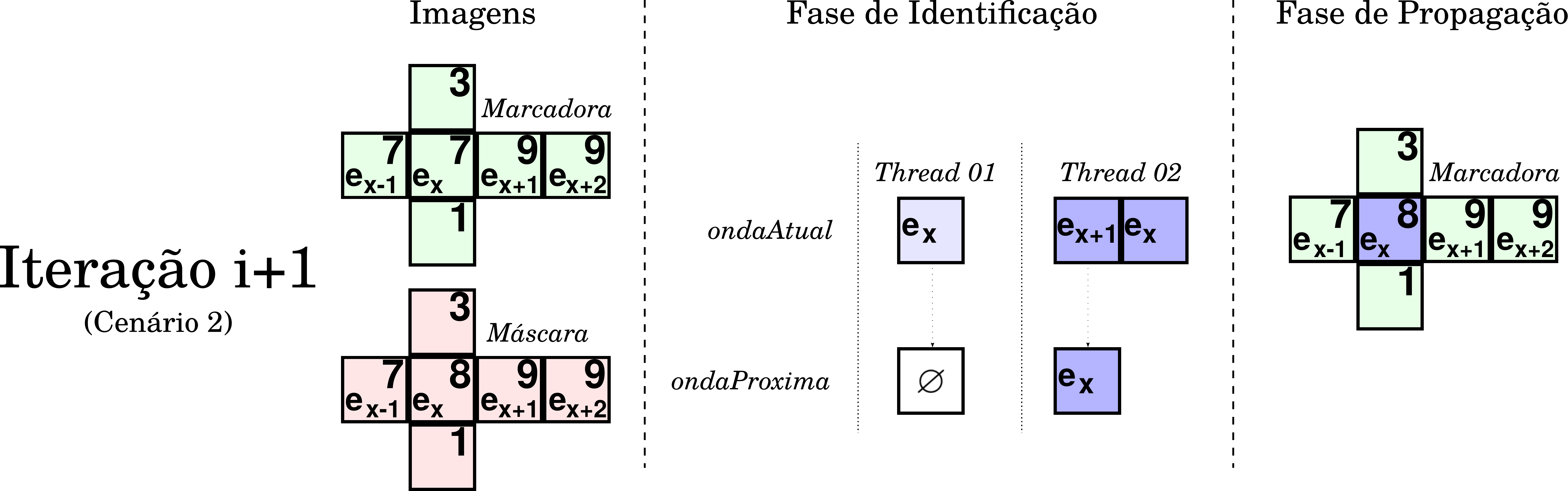}
	\caption{Cenário 2: Estabilidade Alcançada na Iteração $i+1$.}
\label{fig:racecondition03}
\end{figure}

Apesar dos cenários diferenciados, como na iteração $i$ a \textit{thread} 01
inseriu o elemento $e_x$ e a \textit{thread} 02 inseriu os elementos $e_x$ e
$e_{x+1}$, na iteração $i+1$ esses elementos constituirão as frentes de onda e
estarão na fila \emph{ondaAtual}. Dessa forma, durante a Fase de Identificação
a \textit{thread} 02 irá inserir $e_x$ na fila \emph{proximaOnda}, e a
propagação resultará em uma imagem igual aquela gerada pelo Cenário 1, quando a
propagação alcançou a estabilidade.

Por consequência dessa rechecagem de estados, observada no Cenário 2, o
\textit{data race} identificado na implementação paralela proposta é benigno e
classificado como dupla checagem~\cite{narayanasamy2007automatically}, não
influenciando no resultado final.

\subsection{Uso de Diferentes Estruturas de Dados}\label{subsec:estruturas}

Como já apontado pela Seção~\ref{subsec:implementacaoparalela}, na implementação
paralela do algoritmo proposto, apesar das \textit{threads} serem disparadas
em diferentes segmentos, elas trabalham simultaneamente em toda a imagem onde
ondas de propagação podem atravessar e influenciar umas as outras.  Dessa
forma, além da ocorrência de \textit{data races} durante a Fase de Propagação,
outra consequência do uso de paralelismo é o crescimento da quantidade de
elementos a serem propagados a cada iteração, ou a quantidade de elementos
inseridos na fila para processamento. 

Esse fato ocorre em razão dos elementos poderem ser inseridos na estrutura de
dados mais de uma vez ao longo do tempo, devido a origem das ondas de
propagação serem díspares e ainda serem oriundos de \textit{threads}
diferentes. Os reprocessamentos provenientes desses casos citados tendem a
aumentar, à medida em que se aumenta ou o nível de paralelismo explorado ou o
número de \textit{threads}. Assim, o aumento do \textit{throughput} (elementos
processados por unidade de tempo) utilizando paralelismo não reflete na mesma
proporção no tempo de execução por causa desse reprocessamento.

\begin{figure}[!htb]
	\centering
	\includegraphics[width=.95\linewidth]{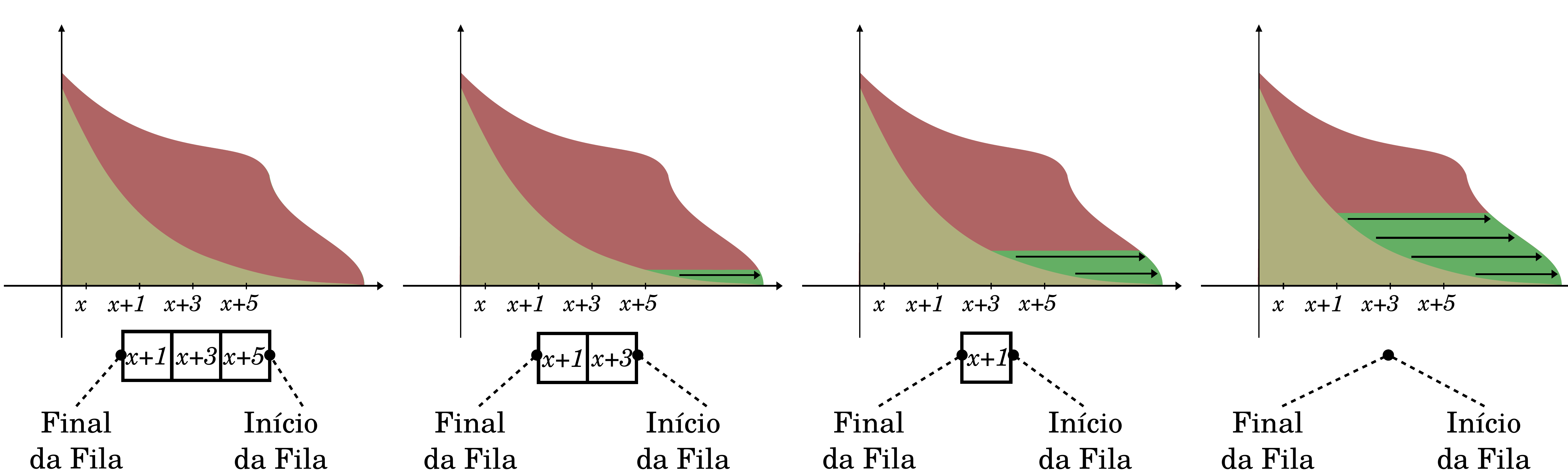}
	\caption{Exemplo de Propagação Utilizando Fila FIFO.}
\label{fig:1dfifo}
\end{figure}

A Figura~\ref{fig:1dfifo} mostra um exemplo desses reprocessamentos. Nesse caso, a
estrutura de dados é uma fila FIFO que possui três valores inseridos em ordem
crescente. Da forma como estão inseridos, inicialmente, o elemento $x + 5$
propaga o seu valor. Em seguida, o elemento $x + 3$ realiza a propagação em uma
camada superior a do valor $x + 5$, e por último, o elemento $x + 1$ também
realiza a propagação superior aos outros dois elementos. É possível observar
que a ordem de inserção dos elementos, por causa da fila FIFO, provoca uma
propagação gradativa das ondas de propagação até que seja alcançado o valor
final dessas propagações.

Uma forma de lidar com este problema é processar os elementos na Fase de
Propagação utilizando uma ordem diferente daquela dada pela Fase de
Identificação e inserção na fila, na tentativa de reduzir o número de
operações executadas.

Uma vez que a operação almejada por algoritmos IWPP é a maximização ou a
minimização de valores em determinada posição, executar primeiro as propagações
cujos valores estão mais próximos do valor final, possibilita a redução do
processamento, já que este cenário pode evitar que ondas de propagação com valores
mais distantes do final sejam executadas e sobrepostas em seguida por ondas com 
valores mais próximos do almejado.

Uma das estruturas de dados que se adequam ao IWPP, que já utiliza inicialmente
a fila FIFO, agregando a ordenação e a extração dos elementos de acordo com seu
conteúdo, é a fila de prioridades~\cite{van1976design}. Essa fila utiliza uma
estrutura \textit{heap}\cite{Musser:2001:STR:374262} que mantém as ondas de
propagação ordenadas de acordo com seus respectivos valores, permitindo a
redução de reprocessamento ocasionado pelos eventos citados anteriormente.

\begin{figure}[!htb]
	\centering
	\includegraphics[width=\linewidth]{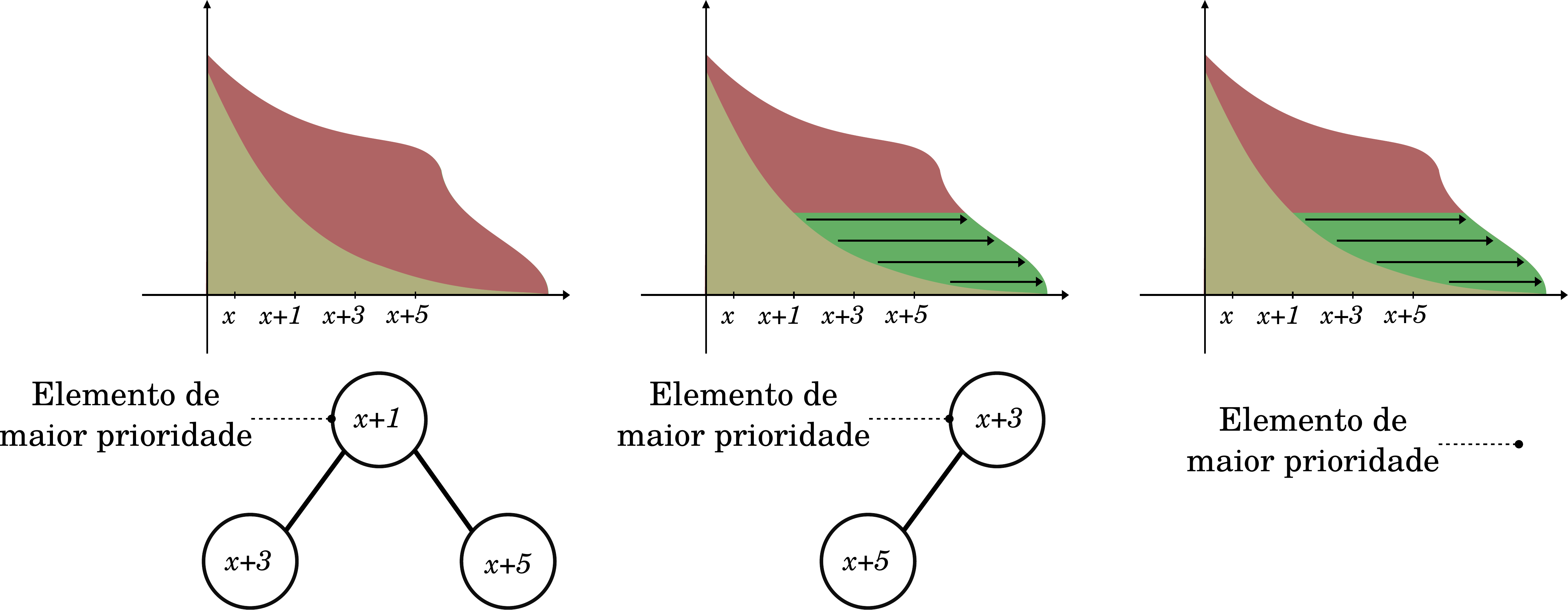}
	\caption{Exemplo de Propagação Utilizando Fila de Prioridades.}
\label{fig:1dheap}
\end{figure}

A Figura~\ref{fig:1dheap} demonstra o funcionamento da fila de prioridades. A
estrutura possui seus elementos ordenados utilizando uma \textit{heap}, e a
extração retira da fila o elemento com maior valor de dentro da estrutura para
realizar a propagação (elemento $x + 1$). Assim, quando os elementos com
valores inferiores são retirados da estrutura (elementos $x + 3$ e $x + 5$),
nenhuma propagação é realizada economizando em processamento.

\subsection{Execução Cooperativa em Processadores Heterogêneos}
\label{subsec:cooperativa}

Como forma de aproveitar os recursos disponíveis, e pela possibilidade das
imagens utilizadas neste estudo poderem possuir um tamanho superior a memória
disponível no Intel\textsuperscript{\textregistered} Xeon
Phi\textsuperscript{\texttrademark}, uma estratégia para execução cooperativa
em processadores heterogêneos é utilizada.

\begin{figure}[!htb]
\centering
\begin{minipage}{.45\textwidth}
  \centering
  \includegraphics[width=.9\linewidth]{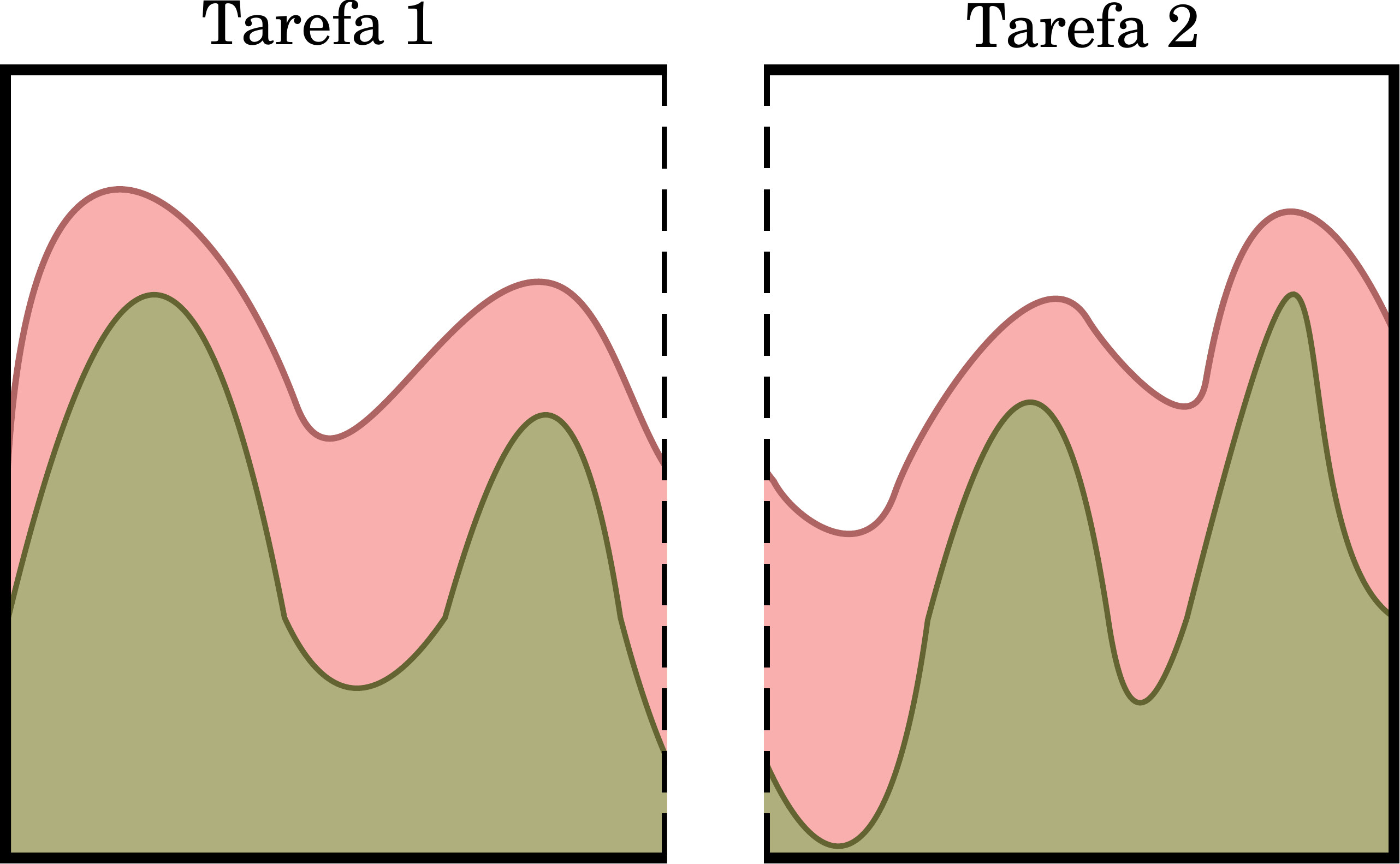}
  \caption{Divisão de Uma Imagem em Duas Tarefas.}
  \label{fig:2tiles01}
\end{minipage}%
\hspace*{10mm}
\begin{minipage}{.45\textwidth}
  \centering
  \includegraphics[width=.9\linewidth]{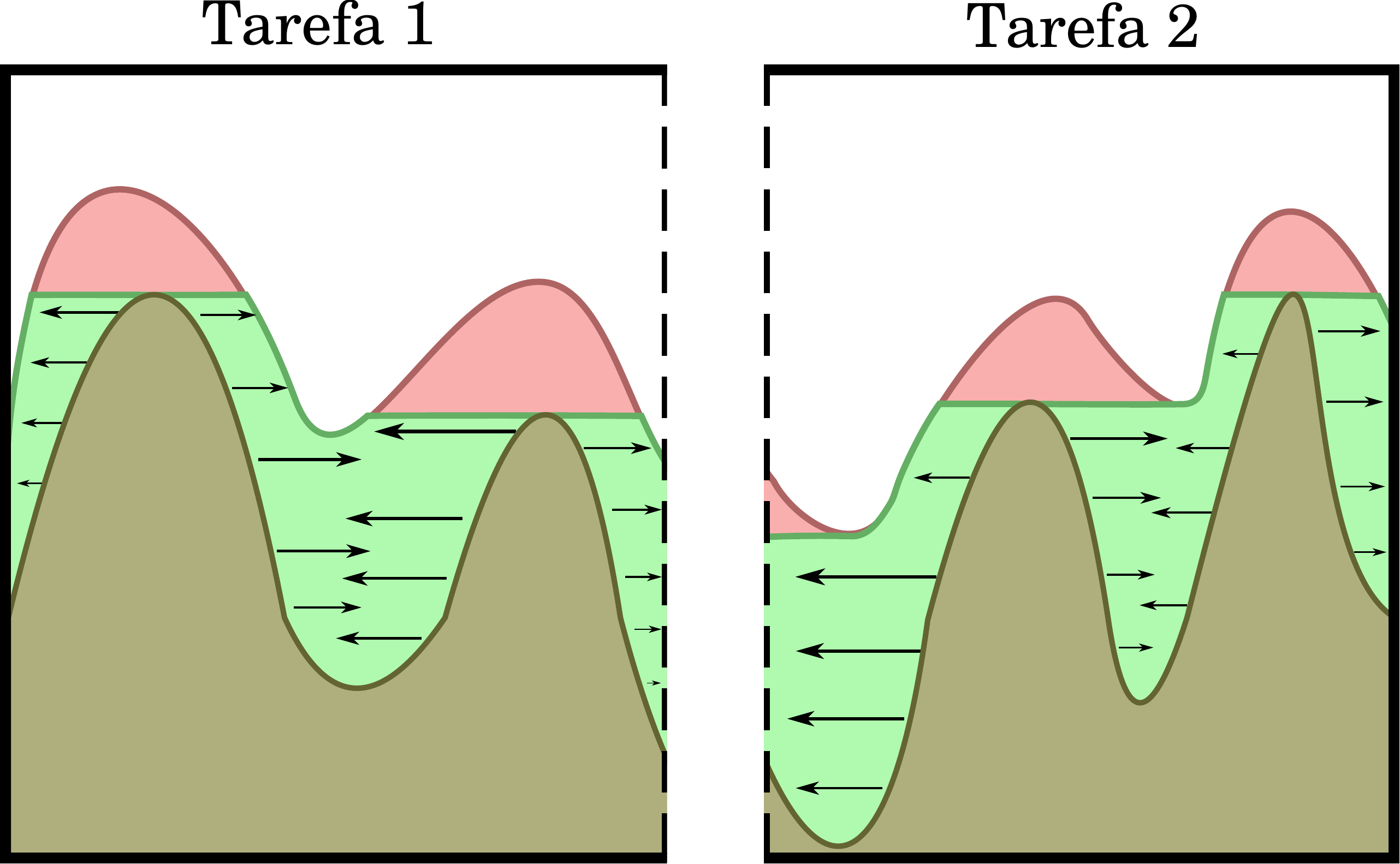}
  \caption{Reconstrução Morfológica Aplicada em Cada Tarefa.}
  \label{fig:2tiles02}
\end{minipage}
\end{figure}

Em um computador, além da própria CPU, podem existir mais de um
Intel\textsuperscript{\textregistered} Xeon Phi\textsuperscript{\texttrademark}
e até GPUs que podem ser assinaladas para execução paralela entre os
processadores disponíveis para o processamento de uma imagem. Dessa forma, é
possível empregar uma estratégia que divida a entrada a ser processada em
tarefas. A Figura~\ref{fig:2tiles01} demonstra um exemplo de divisão de uma
imagem em duas tarefas.

Utilizando a Reconstrução Morfológica como exemplo, em cada uma das tarefas é
executada individualmente esse algoritmo, gerando uma versão reconstruída que
pode ser visualizada na Figura~\ref{fig:2tiles02}.

\begin{figure}[!htb]
 \centering
 \includegraphics[width=.6\linewidth]{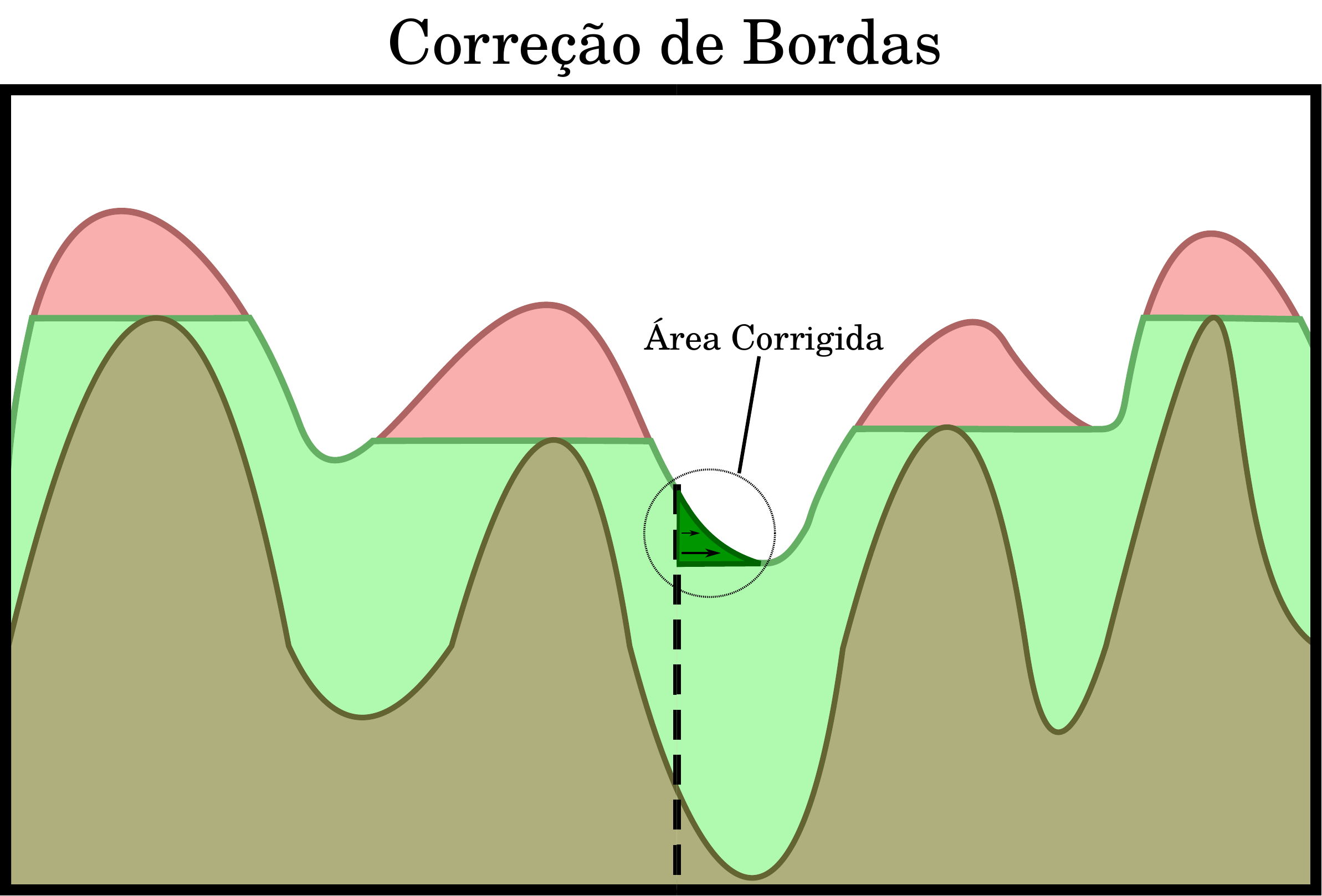}
 \caption{Reconstrução Morfológica Após a União das Imagens com a Área de
 Fronteira Corrigida.}
 \label{fig:2tiles03}
\end{figure}

Após o processamento das tarefas de maneira independente, existem propagações
que podem migrar das fronteiras de uma imagem (tarefa) para outra. Dessa forma,
após a união das imagens processadas individualmente como tarefas, um
processamento para a correção da propagação dessas fronteiras é realizado. A
região corrigida ou reprocessada para o exemplo da Reconstrução Morfológica
pode ser visualizada na Figura~\ref{fig:2tiles03}.

\begin{figure}[!htb]
 \centering
 \includegraphics[width=1\linewidth]{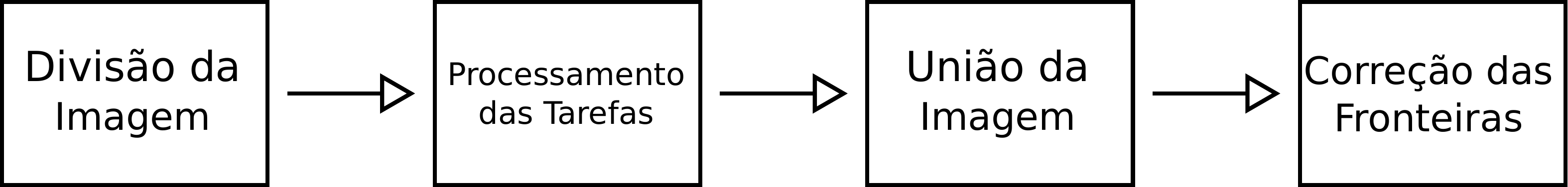}
 \caption{Área Processada pela Correção de Bordas.}
 \label{fig:tarefas}
\end{figure}

A implementação dessa estratégia pode ser observada na Figura~\ref{fig:tarefas}
que constitui-se de quatro estágios: Divisão da Imagem, Processamento de Tarefas,
União da Imagem e Correção das Fronteiras. A Divisão da Imagem consiste na
criação das tarefas a serem processadas, dada uma imagem como entrada. Essas
tarefas passam a fazer parte de uma fila de tarefas a serem processadas na fase
de Processamento de Tarefas. Nessa fase, os dispositivos disponíveis para
executar algum processamento (GPUs, MICs e CPU) irão consumir, de forma
independente, tarefas da fila até que a mesma se encontre vazia. Na fase de
União da Imagem todas as tarefas já foram processadas e esses pedaços da imagem
assinados para cada tarefa são montados em sua posição original. E por último,
na fase Correção de Fronteiras, é executado um processamento em CPU
identificando elementos ativos nas fronteiras dos pedaços das imagens, agora
unidas, e a propagação das ondas relacionadas a essas fronteiras.

\subsection{Sumário}

Nesta seção foi apresentada uma abordagem eficiente para algoritmos da
classe \textit{Irregular Wavefront Propagation Pattern}. Inicialmente, para
permitir a aplicação de estratégias de alto desempenho, foram propostas
modificações no algoritmo IWPP separando a propagação em duas fases: Fase de
Identificação e Fase de Propagação.

A partir desse novo algoritmo, foram aplicadas estratégias de vetorização na
identificação dos elementos recebedores de propagação, realizando leitura da
vizinhança de um elemento e a inserção de elementos na fila utilizada na fase
de Propagação do algoritmo. Para a realização da inserção vetorial, também foi
apresentada a função \textit{prefix sum} vetorial que permite identificar a
posição dos elementos a serem inseridos na estrutura de dados (fila) utilizada.

Posteriormente, foi empregada uma estratégia paralela na versão vetorizada do
algoritmo. Essa estratégia permite o uso de várias \textit{threads} por meio da
distribuição de segmentos da imagem durante a fase de Inicialização. Após a
Inicialização, as \textit{threads} executam a fase de Propagação e trabalham
por toda a imagem realizando modificações de forma independente entre as
\textit{threads}.

Em seguida, foi relatado o uso de uma fila de prioridades na implementação
paralela. A finalidade do seu uso é ordenar os elementos inseridos na Fase de
Identificação pelo seu conteúdo, buscando realizar primeiro a propagação de
elementos que estão mais próximos de seu valor final.

Por fim, foi apresentada uma estratégia para a execução cooperativa em
processadores heterogêneos. Essa estratégia permite processar imagens que
ultrapassem o tamanho da memória disponível no
Intel\textsuperscript{\textregistered} Xeon
Phi\textsuperscript{\texttrademark}, realizando a divisão dela em tarefas que
podem ser processadas de forma independente, a partir de uma fila de tarefas e
uso de correção das fronteiras dos pedaços dessas imagens.

  \section{Análise dos Resultados}
Nesta seção será apresentada uma análise dos resultados alcançados pelas
estratégias desenvolvidas na Seção~\ref{sec:IWPPphi}. As seções a seguir
detalham os ambientes de desenvolvimento e testes (Seção~\ref{subsec:ambiente}), as
configurações dos experimentos realizados (Seção~\ref{subsec:config}) e os
resultados alcançados (Seção~\ref{subsec:resultados}).

\subsection{Ambiente de Desenvolvimento e Testes}
\label{subsec:ambiente}

Os testes e análises foram realizados em três computadores com as seguintes
configurações: 
    
\begin{enumerate}
    \item Um SGI C1104 com Sistema Operacional CentOS 6.5, e as seguintes
        configurações: 
	\begin{itemize}
        \item 02 (dois) processadores Intel\textsuperscript{\textregistered}
            Xeon\textsuperscript{\textregistered} 8-\textit{Core} 64-bit E5-processors
            com 2.6GHz de frequência; 
        \item 20MB de memória cache L3; 
        \item 64GB de memória RAM DDR3-1866MHz; 
        \item 01 (um) coprocessador Intel\textsuperscript{\textregistered}
            Xeon Phi\textsuperscript{\texttrademark} 7120P 1.33GHz; 
        \item HD SATA de 1TB, 10000 RPM.
    \end{itemize}

    \item Um computador Dell C8220z com Sistema Operacional CentOS 6.3, e
        as seguintes configurações: 
	\begin{itemize}
        \item 02 (dois) processadores Intel\textsuperscript{\textregistered}
            Xeon\textsuperscript{\textregistered} 8-\textit{Core} 64-bit
            E5-processors com 2.7GHz de frequência; 
        \item 20MB de memória cache L3; 
        \item 256KB de memória cache L2; 
        \item 32GB de memória RAM DDR3-1600MHz; 
        \item 01 (um) coprocessador Intel\textsuperscript{\textregistered}
            Xeon Phi\textsuperscript{\texttrademark} SE10P 1.1GHz;
        \item 01 (uma) placa gráfica NVIDIA Tesla K20 (GK110);
        \item HD SATA de 250GB, 7500 RPM.
    \end{itemize}
    
    \item Um computador com Sistema Operacional CentOS 6.5, e as seguintes
        configurações: 
	\begin{itemize}
        \item 02 (dois) processadores Intel\textsuperscript{\textregistered}
            Xeon\textsuperscript{\textregistered} 8-\textit{Core} 64-bit
            E5-processors com 2.7GHz de frequência; 
        \item 20MB de memória cache L3; 
        \item 256KB de memória cache L2; 
        \item 32GB de memória RAM DDR3-1600MHz; 
        \item 02(dois) coprocessadores Intel\textsuperscript{\textregistered}
            Xeon Phi\textsuperscript{\texttrademark} SE10P 1.1GHz. 
    \end{itemize}
\end{enumerate}

Os algoritmos desenvolvidos foram comparados com outras implementações para CPU
e GPU.\ Para isso, foram utilizadas as versões mais eficientes conhecidas na
literatura para CPU~\cite{vincent1993morphological} e
GPU~\cite{teodoro2013efficient}~\cite{teodoro2012fast}. 

A GPU utilizada foi a NVIDIA Tesla K20, e cada um dos dispositivos
utilizados nos testes tiveram sua largura de banda testada de acordo com o tipo
de acesso aos dados (regular ou randômico). No acesso regular, a largura de
banda foi mensurada utilizando o \textit{STREAM benchmark}~\cite{McCalpin1995}.
Para o acesso randômico foi desenvolvido um \textit{benchmark} simplificado
com acesso randômico paralelo de 10 milhões de elementos em uma imagem
$4K\times4K$. Os resultados podem ser visualizados na
Tabela~\ref{tab:comparative} que contém os \textit{benchmarks} realizados e as
informações relevantes para os dispositivos testados.

\begin{table}[!htb]
\caption{Características dos Processadores.}
\begin{center}
    \begin{tabular}{l l l l}
    \hline
                                                    & K20 GPU   & SE10P  & 7120P     \\ \hline 
        \hline
        Número de núcleos                           & 2496      & 61     & 61        \\ \hline
        \textit{Clock} por núcleo (MHz)             & 706       & 1100   & 1238      \\ \hline
        Largura de banda - Acesso regular (GB/s)    & 148       & 160    & 177       \\ \hline
        Largura de banda - Acesso randômico (MB/s)  & 895       & 399    & 438       \\ \hline
    \end{tabular}
\end{center}
\label{tab:comparative}
\end{table}

\subsection{Configuração dos Experimentos}
\label{subsec:config}
\begin{figure}[!htb]
    \centering
    \includegraphics[width=\columnwidth]{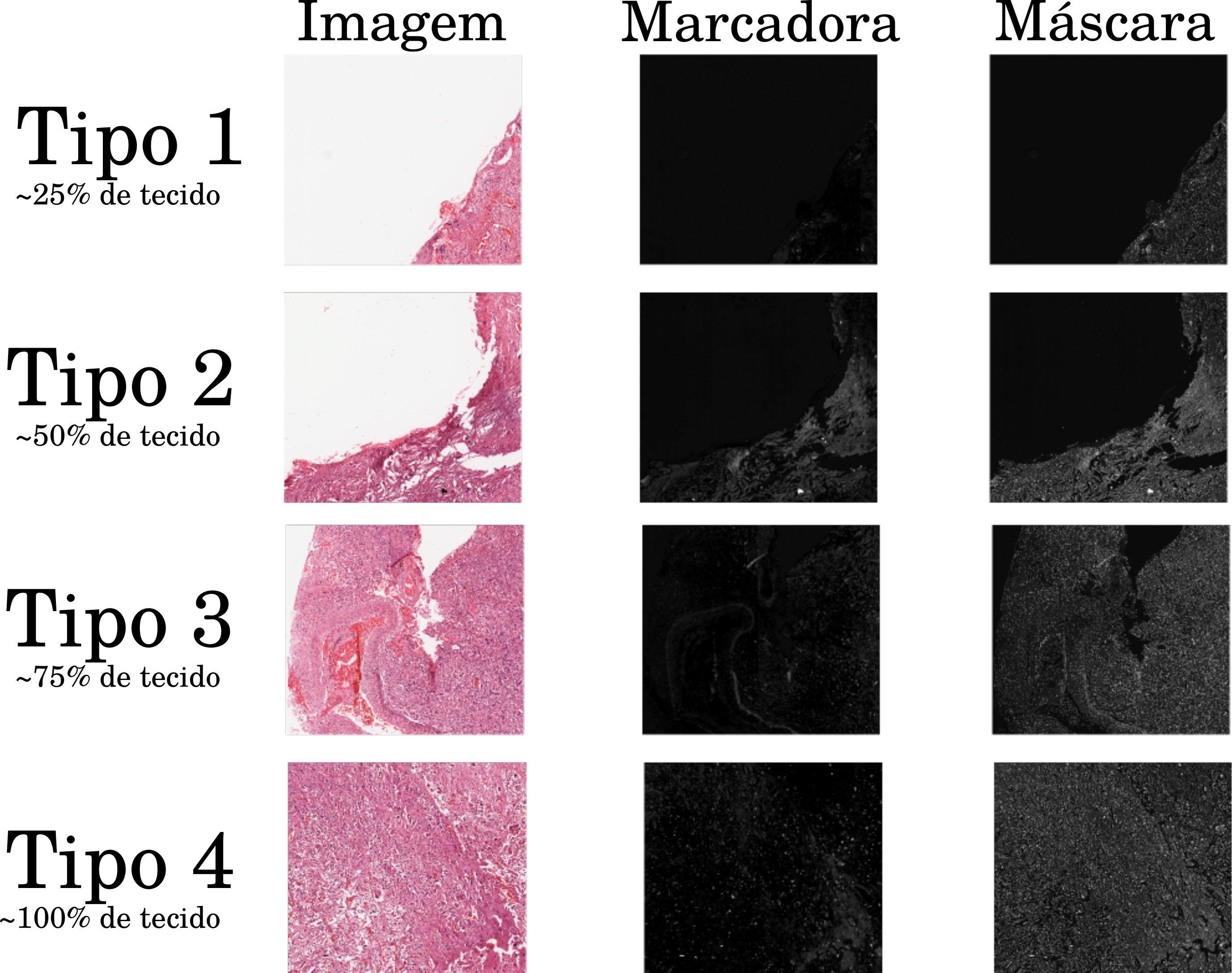}
    \caption{Tipos Diferentes de Cobertura.}
    \label{fig:tissue}
\end{figure}

O desempenho da implementação do IWPP foi avaliado utilizando a Reconstrução
Morfológica e a função \textit{Fill Holes}. Para ambos os algoritmos, suas
configurações foram variadas em termos de conectividade (4 e 8), tamanho da
imagem de entrada e o tipo ou grau de cobertura. Os testes foram executados em
imagens coletadas para pesquisa do \textit{In Silico Brain Tumor Research
Center} (ISBTRC)\cite{saltz2010multi} da \textit{Emory University}. Essas
imagens foram variadas quanto ao tamanho de entrada de $4K\times4K$ até
$32K\times32K$, e foram diferidas em tipo de cobertura e percentual aproximado
de tecido na imagem (Figura~\ref{fig:tissue}). 

Em todas as amostras, o algoritmo foi executado dez vezes para cada uma das
configurações com cada tipo de cobertura. Para essas execuções, o coeficiente
de variação não foi superior a $1\%$.

\subsection{Resultados Experimentais}
\label{subsec:resultados}

Esta seção apresenta os resultados obtidos da execução dos testes e uma
discussão acerca da análise dos mesmos.

\subsubsection{Impacto da Vetorização para o Desempenho}
Nesta seção foi avaliado o ganho de vetorização em relação a versão
não-vetorizada do algoritmo. Os experimentos foram executados no
Intel\textsuperscript{\textregistered} Xeon Phi\textsuperscript{\texttrademark}
SE10P com a Reconstrução Morfológica e o \textit{Imfill} utilizando fila FIFO e
as imagens de entrada com tamanho $4K\times4K$ variando seu percentual de
cobertura.

\begin{figure}[!htb]
\label{grafico:vetorizacao}
\begin{center}
\begin{tikzpicture}[scale=1]
    \begin{axis}[
        ybar=10pt,
        xmin = 10,
        xmax = 115,
        ymin=0,
        ymax=7,
        grid=both,
        width  = 12cm,
        height = 5cm,
        bar width=20pt,
        ylabel={\textbf{\textit{Speedup} (vs.\ não-vetorizado)}},
        xlabel={\textbf{Cobertura ($\%$)}},
        nodes near coords,
        legend image code/.code={%
                    \draw[#1, draw=none] (0cm,-0.1cm) rectangle (0.8cm,0.1cm);
                }, 
        xtick = data,
        legend columns=2,
        legend style={at={(.5,1.35)},anchor=north},
        cycle list name ={my chart colors},
        enlarge y limits={value=0.2,upper}
    ]
    \addplot [black,  postaction={pattern color=blue, pattern=north east lines}] 
                     table[x=cover,y=recon]{graficos/01-vetorizacao.dat};
    \addlegendentry{Reconstrução Morfológica}
    \addplot [draw=black,black,  postaction={pattern color=red,    
    pattern=grid}] 
                     table[x=cover,y=fill]{graficos/01-vetorizacao.dat};
    \addlegendentry{\textit{Imfill}}
    \end{axis}
\end{tikzpicture}
\end{center}
\caption{Ganho de Vetorização Variando a Porcentagem de Cobertura.}
\end{figure}
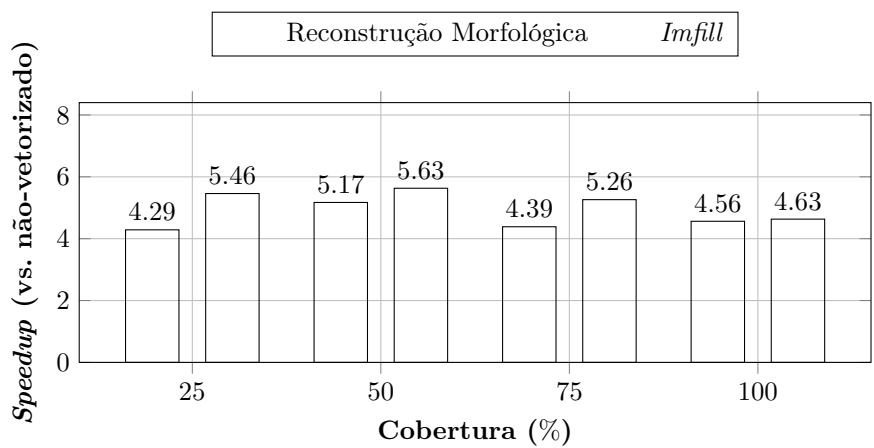

A Figura~\ref{grafico:vetorizacao} mostra o gráfico de \textit{speedups} do
algoritmo vetorizado em relação ao algoritmo sem vetorização.

Como pode ser visto na Figura~\ref{grafico:vetorizacao}, os algoritmos
alcançaram significativos \textit{speedups} para todas as configurações de
entrada e o \textit{Imfill} obteve melhores resultados em termos de vetorização
do que a Reconstrução Morfológica. A melhor configuração obteve
\textit{speedup} de $5.63\times$ para imagens com $50\%$ de cobertura. A
cobertura da imagem é um indicativo da quantidade de trabalho a ser processado
durante a execução do algoritmo, porém, devido a irregularidade do IWPP,
imagens com coberturas menores podem executar menos ou mais trabalho que
imagens de mesmo tamanho com coberturas maiores, ocasionando essa oscilação nos
\textit{speedups}, que podem ser observadas no desempenho de imagens com $75\%$
e $100\%$ de cobertura, que obtiveram ganhos inferiores aos de $50\%$. 

Dadas as instruções vetoriais disponíveis e utilizadas no
Intel\textsuperscript{\textregistered} Xeon
Phi\textsuperscript{\texttrademark}, a estratégia mostrou-se eficiente.
Contudo, os ganhos em desempenho são menores que a proporção de aumento da
quantidade de elementos que são processados em um vetor de $512$ bits, quando
comparado com a versão não vetorizada. Na implementação são utilizados inteiros
de 32 bits para manipular os dados e as instruções SIMD podem processar até 16
elementos por vez. Isso ocorre porque o código vetorizado não é uma tradução
direta do código não-vetorizado, e sua versão demanda mais instruções que o
código original, em virtude da sua natureza irregular, como pode ser visto nos
detalhes da implementação. A inserção vetorial na fila, por exemplo, requer o
uso da função \textit{prefix sum} que inexiste na versão não-vetorizada.

\subsubsection{Resultados da Paralelização do Algoritmo Vetorizado}
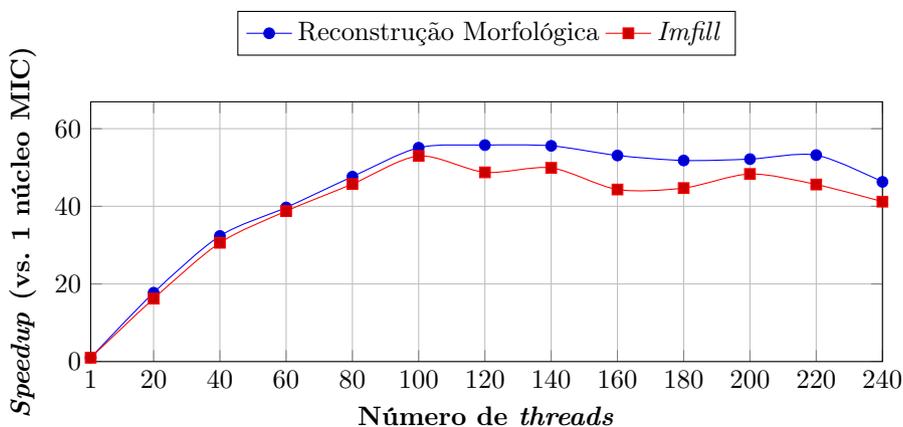
\begin{figure}[!htb]
    \begin{center}
       \begin{tikzpicture}[scale=1]
       \begin{axis}[
       xmin = 1,
       xmax = 240,
       ymin = 0,
       grid=both,
       width  = 12cm,
       height = 5cm,
       ylabel={\textbf{\textit{Speedup} (vs. 1 núcleo MIC)}},
       xlabel={\textbf{Número de \textit{threads}}},
       every node near coord/.style=above left,
       xtick = data,
       legend columns=2,
       legend style={at={(.5,1.35)},anchor=north},
       enlarge y limits={value=0.2,upper}
       ]
       \addplot+[color=blue, smooth, every node near coord/.style=black] 
       table[x=threads,y=recon]{graficos/02-escalabilidade.dat};
       \addlegendentry{Reconstrução Morfológica}
       \addplot+[color=red, smooth, every node near coord/.style=black] 
       table[x=threads,y=fill80]{graficos/02-escalabilidade.dat};
       \addlegendentry{\textit{Imfill}}
       \end{axis}
       \end{tikzpicture}
    \end{center}
\caption{Análise de Escalabilidade Variando a Quantidade de \textit{Threads}.}
\label{grafico:escalabilidade}
\end{figure}

A Figura~\ref{grafico:escalabilidade} apresenta o gráfico de \textit{speedups}
alcançados ao variar a quantidade de \textit{threads} utilizadas para a
paralelização apresentada na Seção~\ref{subsec:implementacaovetorial}. Os
\textit{speedups} são calculados tomando como referência a execução do
respectivo algoritmo no MIC utilizando 1 \textit{thread}.

Nessa implementação, o algoritmo atingiu \textit{speedups} de até $55.7\times$
para a configuração utilizada (imagens de $8K\times8K$).  Apesar de o
coprocessador suportar $240$ \textit{threads} de hardware, o ganho acima do
número de núcleos do coprocessador (60) é limitado. Além disso, não é possível
perceber ganhos significativos com o aumento de \textit{threads} para valores
superiores a 120.

\subsubsection{Impacto do Percentual de Tecido de Cobertura}

\begin{figure}[!htb]
\begin{center}
\begin{tikzpicture}[scale=1]
    \begin{axis}[
        ybar=10pt,
        xmin = 15,
        xmax = 110,
        ymin=0,
        grid=both,
        width  = 12cm,
        height = 5cm,
        bar width=20pt,
        ylabel={\textbf{Speedup (vs. 1 núcleo CPU)}},
        xlabel={\textbf{Tecido de Cobertura ($\%$)}},
        nodes near coords,
        legend image code/.code={%
                    \draw[#1, draw=none] (0cm,-0.1cm) rectangle (0.8cm,0.1cm);
                },  
        xtick = data,
        legend columns=2,
        legend style={at={(.5,1.35)},anchor=north},
        cycle list name ={my chart colors},
        enlarge y limits={value=0.2,upper}
    ]
    \addplot [black,  postaction={pattern color=blue, pattern=north east lines}] 
                     table[x=cover,y=recon]{graficos/03-cobertura.dat};
    \addlegendentry{Reconstrução Morfológica}
    \addplot [draw=black,black,  postaction={pattern color=red, pattern=grid}] 
                     table[x=cover,y=fill]{graficos/03-cobertura.dat};
    \addlegendentry{\textit{Imfill}}
    \end{axis}
\end{tikzpicture}
\end{center}
\caption{Variação do Percentual de Tecido de Cobertura.}
\label{grafico:cobertura}
\end{figure}
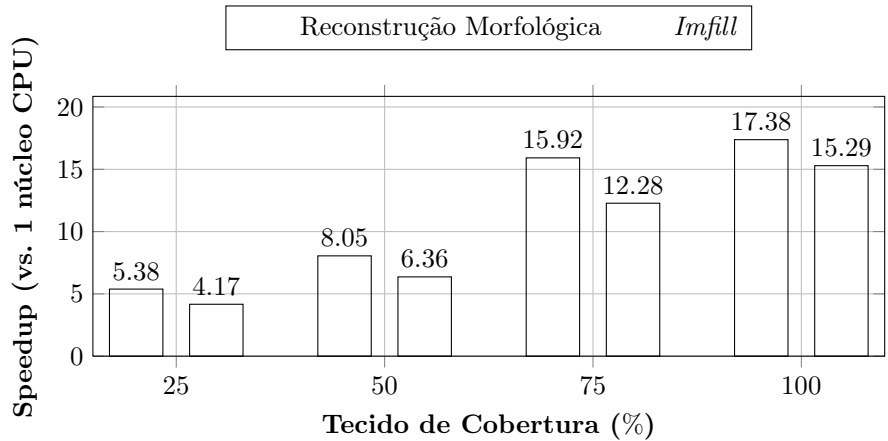

A Figura~\ref{grafico:cobertura} contém o gráfico do impacto do tipo de
cobertura para o algoritmo da Reconstrução Morfológica e do algoritmos
\textit{Fill Holes}. Neste experimento foram utilizadas imagens com
$4K\times4K$ onde a execução do algoritmo é comparada com a execução sequencial
em CPU.\

Ambos os algoritmos apresentaram melhorias quando comparados com a versão
sequencial de CPU.\ Como pode ser observado no gráfico, maior quantidade de
tecido a ser processado resulta em maior \textit{speedup}. Isso ocorre devido
ao aumento da quantidade de processamento e redução da porcentagem de
\textit{overhead} necessário ao início do processamento.

Outro fato observado são os \textit{speedups} maiores por parte da Reconstrução
Morfológica. Isso ocorre porque a função \textit{Imfill} utiliza a própria
inversa da imagem a ser processada como máscara, resultando em um número muito
maior de propagações. Dessa forma, é requerido um número maior de escaneamentos
na inicialização do algoritmo, e o ganho na Fase de Propagação Irregular em
relação à execução do algoritmo é reduzida.

\subsubsection{Impacto do Tamanho da Imagem}

Esta seção avalia o impacto do tamanho da entrada para a execução do IWPP.\
Para esses testes foi escolhido o algoritmo da Reconstrução Morfológica
utilizando a fila FIFO.\

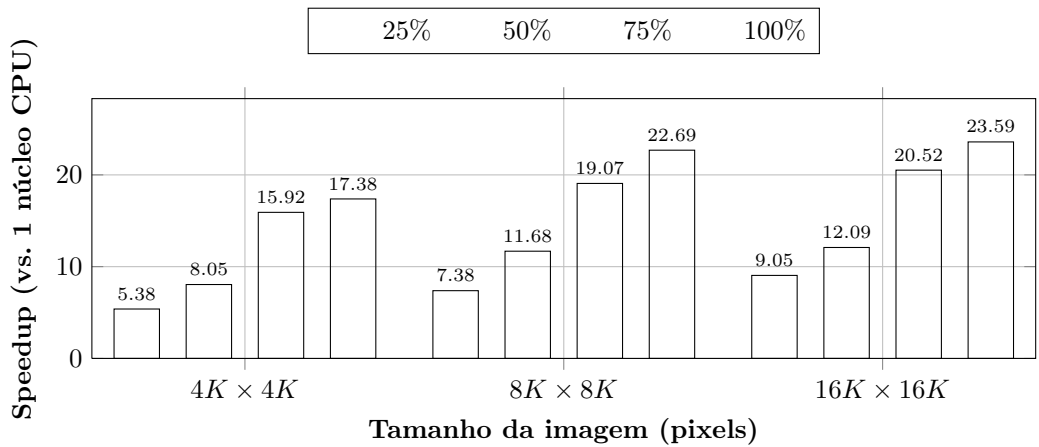
\begin{figure}[!htb]
\begin{center}
\begin{tikzpicture}[scale=1]
    \begin{axis}[
        ybar=10pt,
        xmin = -2,
        xmax = 72,
        ymin=0,
        grid=both,
        width  = 14cm,
        height = 5cm,
        bar width=17pt,
        ylabel={\textbf{Speedup (vs. 1 núcleo CPU)}},
        xlabel={\textbf{Tamanho da imagem (pixels)}},
        nodes near coords,
        legend image code/.code={%
                    \draw[#1, draw=none] (0cm,-0.1cm) rectangle (0.8cm,0.1cm);
                },  
        xtick = data,
        xticklabels from table={graficos/04-tamanhos.dat}{size},
        xticklabel style={align=center},
        legend columns=4,
        legend style={at={(.5,1.35)},anchor=north},
        cycle list name ={my chart colors},
        enlarge y limits={value=0.2,upper},
        every node near coord/.append style={font=\scriptsize}
    ]
    \addplot [black,  postaction={pattern color=blue, pattern=horizontal lines}] 
                     table[x=nr,y=25]{graficos/04-tamanhos.dat};
    \addlegendentry{25\%}
    \addplot [black,  postaction={pattern color=green,    pattern=grid}] 
                     table[x=nr,y=50]{graficos/04-tamanhos.dat};
    \addlegendentry{50\%}
    \addplot [black,  postaction={pattern color=red,      pattern=north east lines}] 
                     table[x=nr,y=75]{graficos/04-tamanhos.dat};
    \addlegendentry{75\%}
    \addplot [black,  postaction={pattern color=black,   pattern=crosshatch dots}] 
                     table[x=nr,y=100]{graficos/04-tamanhos.dat};
    \addlegendentry{100\%}

    \end{axis}
\end{tikzpicture}
\end{center}
\caption{Variação do Tamanho das Imagens de Entrada para a Reconstrução
Morfológica.}
\label{grafico:tamanhos}
\end{figure}

Os resultados apresentados na Figura~\ref{grafico:tamanhos} demonstraram o
aumento do \textit{speedup} ao passo em que a imagem de entrada também aumenta.
Essa melhoria no \textit{speedup} é de $1.45\times$ quando as imagens são
aumentadas de $4K\times4K$ para $16K\times16K$. A melhoria apresentada por esse
experimento é resultado de uma maior utilização do paralelismo que permite um
melhor uso do coprocessador e reduz o custo de computação e transferência de
dados entre CPU e o coprocessador.

\subsubsection{Avaliação de Diferentes Coprocessadores e Estrutura de Dados}

Esta seção compara a performance do IWPP no
Intel\textsuperscript{\textregistered} Xeon Phi\textsuperscript{\texttrademark}
SE10P, 7120P, a GPU NVIDIA K20 e a CPU Intel\textsuperscript{\textregistered}
Xeon\textsuperscript{\textregistered} 8-\textit{Core} E5. Ambos os
coprocessadores possuem a mesma quantidade de núcleos (61), porém os mesmos se
diferem no \textit{clock}, como já apresentado na Tabela~\ref{tab:comparative}
(1.1GHz e 1.33GHz). Para ambos os processadores, também foram avaliados o uso
da fila de prioridades que ordena as ondas de propagação de acordo com a
intensidade dos elementos inseridos na Estrutura de Dados.

\begin{figure}[!htb]
\begin{center}
\begin{tikzpicture}[scale=1]
    \begin{axis}[
        ybar=5pt,
        xmin = -2,
        xmax = 42,
        ymin=0,
        grid=both,
        width  = 14cm,
        height = 5cm,
        bar width=13pt,
        ylabel={\textbf{\textit{Speedup} (vs. 1 núcleo CPU)}},
        xlabel={\textbf{Tamanho da imagem}},
        nodes near coords,
        legend image code/.code={%
                    \draw[#1, draw=none] (0cm,-0.1cm) rectangle (0.8cm,0.1cm);
                },  
        xtick = data,
        xticklabels from table={graficos/05-estruturas.dat}{size},
        xticklabel style={align=center},
        legend columns=3,
        legend style={at={(.5,1.5)},anchor=north},
        cycle list name ={my chart colors},
        enlarge y limits={value=0.2,upper},
        every node near coord/.append style={font=\scriptsize}
    ]
    \addplot [black,  postaction={pattern color=black, pattern=vertical lines}] 
                     table[x=nr,y=10]{graficos/05-estruturas.dat};
    \addlegendentry{8-CPU (FIFO)}
    \addplot [black,  postaction={pattern color=blue, pattern=horizontal lines}] 
                     table[x=nr,y=20]{graficos/05-estruturas.dat};
    \addlegendentry{SE10P MIC (FIFO)}
    \addplot [black,  postaction={pattern color=green,    pattern=grid}] 
                     table[x=nr,y=35]{graficos/05-estruturas.dat};
    \addlegendentry{7120P MIC (FIFO)}
    \addplot [black,  postaction={pattern color=red,      pattern=north east lines}] 
                     table[x=nr,y=50]{graficos/05-estruturas.dat};
    \addlegendentry{K20 GPU (FIFO)}
    \addplot [black,  postaction={pattern color=black,   pattern=crosshatch dots}] 
                     table[x=nr,y=65]{graficos/05-estruturas.dat};
    \addlegendentry{SE10P MIC (\textit{Heap})}
    \addplot [black,  postaction={pattern color=black,   pattern=north west lines}] 
                     table[x=nr,y=80]{graficos/05-estruturas.dat};
    \addlegendentry{7120P MIC (\textit{Heap})}

    \end{axis}
\end{tikzpicture}
\end{center}
\caption{Avaliação de Múltiplos Coprocessadores e Uso de Diferentes Tipos de
Estrutura de Dados.}
\label{grafico:estruturas}
\end{figure}
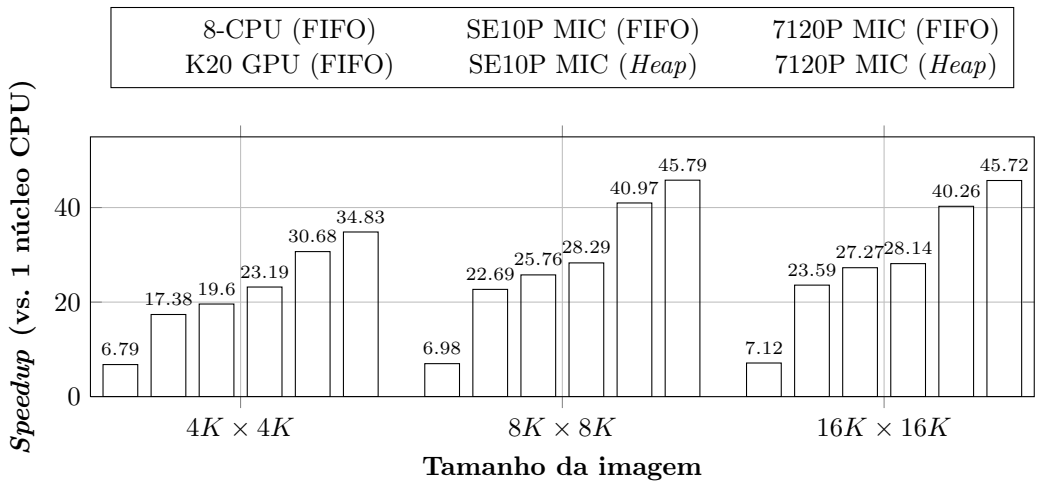

Utilizando a execução sequencial em  CPU como base, a
Figura~\ref{grafico:estruturas} apresenta o gráfico comparativo de
desempenho. Como pode ser observado, todas as versões avaliadas apresentaram
ganho em relação a versão utilizada como base. 

A execução do Intel Phi 7120P apresentou ganho de $1.14\times$ em relação ao
modelo SE10P, o que pode ser explicado pela diferença entre o \textit{clock} de
cada núcleo dos coprocessadores. 

Já a versão de GPU apresentou melhor \textit{speedup} do que as melhores
execuções no Intel Phi utilizando filas FIFO, devido a diferença de largura de
banda entre ela e os coprocessadores ($28.29\times$). Algoritmos da classe IWPP
possuem acesso intensivo de dados de forma regular na Fase de Inicialização, e
irregular na Fase de Propagação Irregular. Como pode ser observado na
Tabela~\ref{tab:comparative}, o Intel Phi é mais eficiente que a GPU K20 no
acesso regular, porém a GPU possui largura de banda $2.04\times$ mais rápida
quando o acesso é irregular.

Por fim, os testes utilizando uma fila de prioridades, onde os elementos
inseridos estão ordenados pela intensidade de seus valores, resultaram na mais
significativa melhoria deste experimento. A execução no 7120P, utilizando
imagens de tamanho $16K\times16K$, alcançou \textit{speedup} $1.62\times$ mais
rápido do que a melhor execução em GPU.\ Ainda essa execução, utilizando fila
de prioridades, resultou em um ganho decorrente da redução de $20\times$ menos
elementos processados durante a Fase de Propagação Irregular do que a versão
com fila FIFO do mesmo modelo de coprocessador. Isso se deve a ordenação
propiciada pela fila de prioridades que ocasiona em propagações de elementos
cujos valores estão mais próximos de seu valor final, como já explicado na
Seção~\ref{subsec:estruturas}.

\subsubsection{Avaliação da Execução Cooperativa em Processadores Heterogêneos}

Esta seção retrata o uso da execução cooperativa em processadores heterogêneos,
de acordo com a estratégia apresentada na Seção~\ref{subsec:cooperativa}. Os
testes foram realizados utilizando o algoritmo da Reconstrução Morfológica, e
com as seguintes configurações:

\begin{itemize}
    \item Versão serial em 1 \textit{core} de CPU utilizada como base;
    \item Versão CPU \textit{multithread} executada em 16 \textit{cores};
    \item 1 MIC;\
    \item 1 MIC $+$ versão CPU \textit{multithread};
    \item 2 MICs;
    \item 2 MICs $+$ versão CPU \textit{multithread}.
\end{itemize}

Visando um melhor aproveitamento dos recursos de processamento disponíveis,
aquelas configurações que combinam o uso de CPUs e
Intel\textsuperscript{\textregistered} Xeon
Phi\textsuperscript{\texttrademark} utilizaram uma estratégia de divisão do
tamanho da imagem (ou tarefa) a ser processada. Essa divisão aproveita
informações anteriores dos \textit{speedups} individuais de cada um dos
dispositivos, e se dá da seguinte forma: Dado o \textit{speedup}
$S_{CPU}$ da execução \textit{multithread} e o \textit{speedup} $S_{MIC}$
da execução no Intel\textsuperscript{\textregistered} Xeon
Phi\textsuperscript{\texttrademark}, a porcentagem do tamanho da imagem $P_{CPU}$
que a versão \textit{multithread} deverá processar é a seguinte:

\begin{equation}
    P_{CPU} = \frac{100 * S_{CPU}}{S_{CPU} + S_{MIC}}
\end{equation}

e a porcentagem do tamanho da imagem $P_{MIC}$ que o
Intel\textsuperscript{\textregistered} Xeon Phi\textsuperscript{\texttrademark}
deverá processar é:

\begin{equation}
    P_{MIC} = \frac{100 * S_{MIC}}{S_{CPU} + S_{MIC}}
\end{equation}

Dessa forma, a Tabela~\ref{tab:tiles} apresenta as porcentagens empregadas em
cada um dos dispositivos utilizados nos testes de uso combinado de MICs e CPUs.

\begin{table}[!htb]
\caption{Porcentagem de Divisão das Tarefas por Dispositivo.}
\begin{center}
    \begin{tabular}{c c c c c c c c c}
        \hline
        Dispositivos & & \multicolumn{2}{c}{1 MIC +
    CPU (\textit{multithread})} & & \multicolumn{3}{c}{2 MIC + CPU
    (\textit{multithread})} &  \\
        \hline
        \hline
        Tamanho & & MIC (\%) & CPU (\%) & & & MIC (\%) & CPU (\%) &  \\
        \hline
        \hline
        $16K\times16K$ & & 68 & 32 & & & 82 & 18 & \\ \hline
        $24K\times24K$ & & 71 & 29 & & & 84 & 16 & \\ \hline
        $32K\times32K$ & & 70 & 30 & & & 82 & 18 & \\ \hline
    \end{tabular}
\end{center}
\label{tab:tiles}
\end{table}

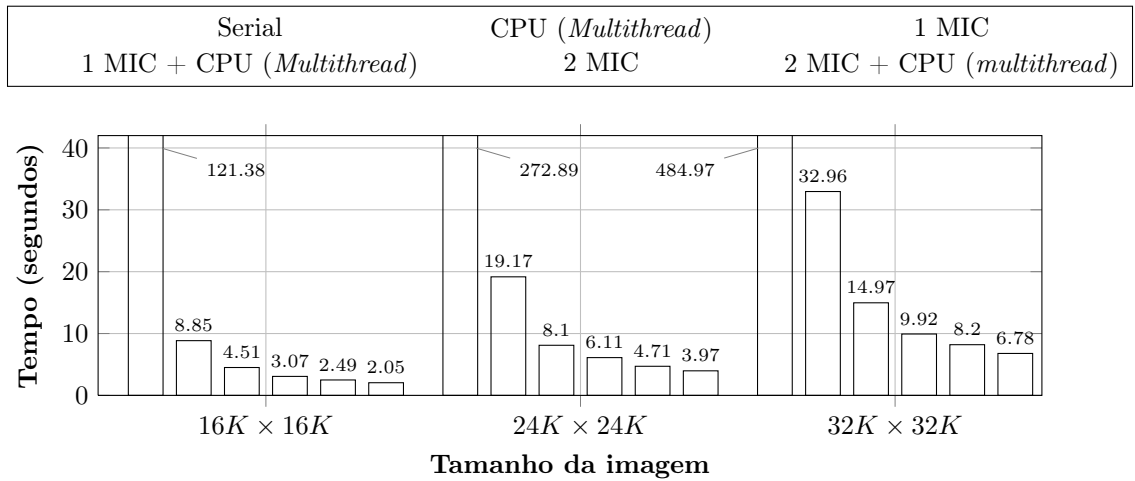
\begin{figure}[!htb]
\begin{center}
\begin{tikzpicture}[scale=1]
    \begin{axis}[
        ybar = 5pt,
        xmin = -3,
        xmax = 42,
        ymin = 0,
        ymax = 35,
        point meta = rawy,
        grid = both,
        width  = 14cm,
        height = 5cm,
        bar width = 13pt,
        ylabel={\textbf{Tempo (segundos)}},
        xlabel={\textbf{Tamanho da imagem}},
        nodes near coords,
        legend image code/.code={%
                    \draw[#1, draw=none] (0cm,-0.1cm) rectangle (0.8cm,0.1cm);
                },  
        xtick = data,
        xticklabels from table={graficos/06-cooperativo.dat}{size},
        xticklabel style={align=center},
        legend columns=3,
        legend style={at={(.5,1.5)},anchor=north},
        cycle list name ={my chart colors},
        enlarge y limits={value=0.2,upper},
        every node near coord/.append style={font=\scriptsize}
    ]
    \addplot [black,  postaction={pattern color=black, pattern=vertical lines}] 
                     table[x=nr,y=base]{graficos/06-cooperativo.dat};
                     \addlegendentry{Serial}
    \addplot [black,  postaction={pattern color=blue, pattern=horizontal lines}] 
                     table[x=nr,y=1cpu]{graficos/06-cooperativo.dat};
                     \addlegendentry{CPU (\textit{Multithread})}
    \addplot [black,  postaction={pattern color=green,    pattern=grid}] 
                     table[x=nr,y=1mic]{graficos/06-cooperativo.dat};
                     \addlegendentry{1 MIC}
    \addplot [black,  postaction={pattern color=black,   pattern=crosshatch dots}] 
                     table[x=nr,y=1cpu1mic]{graficos/06-cooperativo.dat};
                     \addlegendentry{1 MIC + CPU (\textit{Multithread})}
    \addplot [black,  postaction={pattern color=red,      pattern=north east lines}] 
                     table[x=nr,y=2mic]{graficos/06-cooperativo.dat};
                     \addlegendentry{2 MIC}
    \addplot [black,  postaction={pattern color=black,   pattern=bricks}] 
                     table[x=nr,y=1cpu2mic]{graficos/06-cooperativo.dat};
                     \addlegendentry{2 MIC + CPU (\textit{multithread})}

    \node at (axis cs:0,40) [pin={-10:\scriptsize 121.38},inner sep=0pt] {};
    \node at (axis cs:15,40)[pin={-10:\scriptsize 272.89},inner sep=0pt,scale=0.3] {};
    \node at (axis cs:28.5,40) [pin={-170:\scriptsize 484.97},inner sep=0pt,scale=0.3] {};

    \end{axis}
\end{tikzpicture}
\end{center}
\caption{Avaliação da Execução Cooperativa em Processadores Heterogêneos.}
\label{grafico:cooperativo}
\end{figure}

Assim, a Figura~\ref{grafico:cooperativo} analisa as execuções das
configurações citadas anteriormente.

Como pode ser notado, o uso combinado de dispositivos se demonstrou
proveitoso em todas as configurações do experimento, alcançando uma redução
média de até $50\%$ dos tempos de execução dos dispositivos individuais.

A execução utilizando 2 Intel\textsuperscript{\textregistered} Xeon
Phi\textsuperscript{\texttrademark} alcançou \textit{speedups} próximos a
$1.8\times$, comparado a execução utilizando apenas um desses dispositivos. A
justificativa para esse valor de redução do tempo abaixo de linear remete ao
tempo de transferência dessas imagens (\textit{offload transfer}) para os
dispositivos que, mesmo divididas a uma fração de seu tamanho, continuam
possuindo um custo alto de inicialização para cada um dos dispositivos. Na
arquitetura MIC, o uso de \textit{offload} inclui um \textit{overhead} de
inicialização, gerência e transferência de dados, e chamada de função na
aplicação~\cite{newburn2013offload}. 

Por último, o uso combinado de CPU e 2 MICs resultou em um decrescimento de
tempo médio próximo a $18\%$ em relação ao uso somente de 2 MICs. Esse valor
representa a proporção das tarefas entregues ao dispositivo mais lento (CPUs),
quando comparados em relação ao tempo de execução da configuração de maior
\textit{speedup} utilizando um mesmo tipo de dispositivo (execução em 2 MICs).

  \section{Conclusão}%

Neste trabalho, foi proposta e avaliada uma solução eficiente para o
\textit{Irregular Wavefront Propagation Pattern} na arquitetura \textit{Many
Integrated Core}. Esta solução explorou o uso de instruções \textit{Single
Instruction, Multiple Data} SIMD que permitiu alcançar bons resultados para a
solução.

Inicialmente, o algoritmo do IWPP foi reprojetado para permitir uma
implementação eficiente sem o uso de operações atômicas. Essa versão separa a
fase \textit{Wavefront Propagation} em duas fases (Identificação de Elementos
Recebedores de Propagação e Propagação), fazendo com que elementos passem a
modificar apenas suas próprias posições.

Por meio dessa versão reprojetada, a implementação vetorial explora a largura
do vetor de $512$ bits dos Intel\textsuperscript{\textregistered} Xeon
Phi\textsuperscript{\texttrademark} durante a leitura e a propagação de
elementos vizinhos, a execução de uma função \textit{prefix sum} com permutação
dos valores dentro do próprio registrador vetorial e a inserção vetorial de
elementos na estrutura de dados utilizada durante a propagação. Essa estratégia
alcançou \textit{speedups} de até $5.63\times$ para ambos os algoritmos
implementados (Reconstrução Morfológica e \textit{Imfill}), demonstrando a
viabilidade do uso de instruções vetoriais ao se desenvolver aplicações para o
Intel\textsuperscript{\textregistered} Xeon
Phi\textsuperscript{\texttrademark}.

A versão paralela do algoritmo vetorizado realiza uma segmentação do trabalho
inicial das \textit{threads} entre as linhas da imagem. Apesar dessa
segmentação inicial, durante a fase de propagação, as \textit{threads} podem
realizar propagações por toda a imagem, gerando \textit{data races} comprovados
benignos devido a rechecagem de estados utilizado no algoritmo reprojetado.
Testes nessa versão paralelizada demonstraram boa escalabilidade, atingindo
\textit{speedups} de $55.7\times$ quando comparados a execução de um núcleo do
Intel\textsuperscript{\textregistered} Xeon
Phi\textsuperscript{\texttrademark}. Ao variar o percentual de tecido de
cobertura, aumenta-se a quantidade de processamento distribuído a cada
\textit{thread} que reduz o \textit{overhead} vinculado ao início do
processamento. Os testes para essa variação resultaram em \textit{speedups} que
variaram de $5.38\times$ em imagens com $25\%$ de tecido de cobertura a
$17.38\times$ em imagens com $100\%$ de tecido de cobertura na Reconstrução
Morfológica, e de $4.17\times$ em imagens com $25\%$ de tecido de cobertura a
$15.29\times$ em imagens com $100\%$ de tecido de cobertura na \textit{Imfill}.
Avaliações referentes ao tamanho da entrada demonstraram um aumento de
$1.45\times$ quando imagens são aumentadas de $4K\times4K$ para $16K\times16K$,
sendo resultado de uma maior utilização do paralelismo do coprocessador.

Como consequência do uso da estratégia paralela e pela possibilidade de
elementos recomputados aumentarem de acordo com o nível de paralelismo, foi
empregada uma fila de prioridades que mantiveram as ondas de propagação
ordenadas de acordos com seus respectivos valores, ocasionando a redução da
quantidade desse reprocessamento. Além disso, foram feitos testes relacionados
ao uso dessa estrutura comparando dois tipos de estruturas de dados (fila FIFO
e fila de prioridades) e dois modelos de coprocessador com a implementação mais
rápida conhecida na literatura em GPU e uma versão \textit{multicore} para
CPU.\ Os resultados mostraram que as versões do
Intel\textsuperscript{\textregistered} Xeon Phi\textsuperscript{\texttrademark}
utilizando fila FIFO alcançaram bons \textit{speedups} em relação à versão
\textit{multicore} em CPU (cerca de $2.72\times$ mais rápido), porém não mais
que a implementação em GPU utilizando fila FIFO que foi $1.06\times$ mais
rápido. Já o uso de fila de prioridades alcançou \textit{speedups} de
$1.62\times$ mais rápido do que a implementação em GPU, confirmando a redução
de elementos recomputados devido ao uso de paralelismo ocasionada pela
propagação de elementos cujos valores estão mais próximos de seu valor final. 

Por último, para possibilitar o processamento de imagens que superem a
capacidade de memória do Intel\textsuperscript{\textregistered} Xeon
Phi\textsuperscript{\texttrademark}, e como forma de aproveitar os dispositivos
disponíveis no computador, foi implementada uma estratégia para execução
cooperativa em processadores heterogêneos. Essa estratégia realiza a divisão do
processamento em estágios. Nesses estágios, a imagem é divida em tarefas que
são entregues aos dispositivos para processamento. Após esse processamento, a
imagem é novamente unida e, então, é executada uma correção das fronteiras da
imagem que não foram totalmente processadas em virtude da divisão. Essa
estratégia demonstrou-se exequível com os tempos de execução, utilizando 2
Intel\textsuperscript{\textregistered} Xeon Phi\textsuperscript{\texttrademark}
mais a implementação \textit{multithread} em CPU, reduzindo o tempo médio de
execução para $56\%$ do tempo de processamento utilizando apenas 1
Intel\textsuperscript{\textregistered} Xeon
Phi\textsuperscript{\texttrademark}.

\subsection{Trabalhos Futuros}

As implementações desenvolvidas neste trabalho utilizam apenas o paradigma de
programação de memória compartilhada, não aproveitando a possibilidade de uso
em sistemas distribuídos para uma mesma imagem utilizada como entrada. Como
trabalho futuro, deseja-se propor uma solução em memória distribuída que
aproveite de maneira eficiente os recursos disponíveis em tempo de execução.

Além disso, é possível aproveitar as estratégias apresentadas nesse algoritmos
do IWPP reprojetado e implementar outros algoritmos que também tenham seus
princípios de funcionamento baseados nele (Transformação Watershed, Esqueletos
por Zona de Influência, etc).

Por fim, o barramento do Intel\textsuperscript{\textregistered} Xeon
Phi\textsuperscript{\texttrademark} não permite um controle detalhado do
barramento como forma de melhorar a coerência de cache, permitindo um controle
da memória interna de cada núcleo do coprocessador. Assim, é válido um estudo
detalhado para aumentar a coerência de memória, através de outras estratégias
como \textit{prefetching} de instruções~\cite{jeffers2013intel}.

  \bibliographystyle{abbrv}
  \bibliography{bibliografia}

\end{document}